\documentclass[12pt,a4paper]{article}
\usepackage{amsmath,amssymb,cite,graphicx,mathtools,rotating}
\usepackage[usenames]{color}
\usepackage[bookmarksopen,colorlinks=true,linkcolor=dark_green,citecolor=dark_red,urlcolor=dark_red,linktocpage=false]{hyperref}

\definecolor{dark_blue}{rgb}{0,0,0.6}
\definecolor{dark_green}{rgb}{0,0.4,0}
\definecolor{dark_red}{rgb}{0.6,0,0}

\usepackage[height=22.5cm,width=16.5cm,centering]{geometry}

\def\thefootnote{\fnsymbol{footnote}}

\renewcommand{\thefootnote}{\fnsymbol{footnote}}
\newcommand\Order{\mathop{\mathcal{O}}}
\newcommand\unit[1]{\,\mathrm{#1}}

\begin{document}
%%%%%%%%%%%%%%%%%%%%%%%%%%%%%%%%%%%%%%%%%%%%%%%%%%

%%%%%%%%%%%%%%%%%%%%%%%%%%%%%%%%%%%%%%%%%%%%%%%%%%
\begin{titlepage}

\begin{center}

\hfill CTPU-PTC-19-18 \\
\hfill DESY 19-109 \\
\hfill UT-19-13

\vskip 2cm

{\fontsize{18pt}{0pt} \bf
Fingerprint matching of beyond-WIMP dark matter:
}
\\[3ex]
{\fontsize{18pt}{0pt} \bf
neural network approach
}

\vskip 2cm

{\large
Kyu Jung Bae$^{a}$, %Respected Prof.
Ryusuke Jinno$^{a,b}$,
Ayuki Kamada$^{a}$,
and Keisuke Yanagi$^{c}$
}

\vskip 0.7cm

\begin{tabular}{ll}
$^{a}$ &\!\! {\em Center for Theoretical Physics of the Universe, Institute for Basic Science (IBS),} \\
&{\em Daejeon 34126, Korea} \\[.3em]
$^{b}$ &\!\! 
{\em Deutsches Elektronen-Synchrotron DESY, 22607 Hamburg, Germany} \\[.3em]
$^{c}$ &\!\! {\em Department of Physics, University of Tokyo, Bunkyo-ku, Tokyo 113-0033, Japan}
\end{tabular}

\vskip 2cm

\begin{abstract}
Galactic-scale structure is of particular interest since it provides important clues to dark matter properties and its observation is improving.
Weakly interacting massive particles (WIMPs) behave as cold dark matter on galactic scales, while beyond-WIMP candidates suppress galactic-scale structure formation.
Suppression in the linear matter power spectrum has been conventionally characterized by a single parameter, the thermal warm dark matter mass.
On the other hand, the shape of suppression depends on the underlying mechanism.
It is necessary to introduce multiple parameters to cover a wide range of beyond-WIMP models.
Once multiple parameters are introduced, it becomes harder to share results from one side to the other.
In this work, we propose adopting neural network technique to facilitate the communication between the two sides. 
To demonstrate how to work out in a concrete manner, we consider a simplified model of light feebly interacting massive particles.
\end{abstract}

\end{center}
\end{titlepage}

\tableofcontents
\thispagestyle{empty}

\renewcommand{\thepage}{\arabic{page}}
\setcounter{page}{1}
\renewcommand{\thefootnote}{$\diamondsuit$\arabic{footnote}}
\setcounter{footnote}{0}
%%%%%%%%%%%%%%%%%%%%%%%%%%%%%%%%%%%%%%%%%%%%%%%%%%

\newpage
\setcounter{page}{1}

%%%%%%%%%%%%%%%%%%%%%%%%%%%%%%%%%%%%%%%%%%%%%%%%%%
\section{Introduction}
\label{sec:Intro}
\setcounter{equation}{0}
%%%%%%%%%%%%%%%%%%%%%%%%%%%%%%%%%%%%%%%%%%%%%%%%%%

Dark matter (DM) is an essential component for the Universe to form the current shape.
Its existence and abundance are probed by gravitational observations such as galaxy rotation curves,
bullet cluster collision, and cosmic microwave background (CMB) anisotropy.
On the other hand, we have not seen any DM signal by any non-gravitational interactions,
and thus we still do not know the identity of DM: what it is and how it is produced.
One intriguing possibility is that DM consists of a new particle, which provides a clue to physics beyond the standard model (SM) (see Ref.~\cite{Feng:2010gw} for a review).

One of the early attempts is a weakly interacting massive particle (WIMP) (see Refs.~\cite{Arcadi:2017kky, Roszkowski:2017nbc} for recent reviews).
In this direction, much efforts have been devoted at the large hadron collider (LHC) (for example, mono-jet searches~\cite{Aaboud:2017phn, Sirunyan:2017jix}) and at direct/indirect detection searches~\cite{Aprile:2018dbl, Cui:2017nnn, Amole:2019fdf, Ahnen:2016qkx}.
However, no firm signals have been reported yet.
It may motivate us to consider beyond-WIMP scenarios that can be probed by cosmological/astrophysical observations.%
\footnote{
We refer readers to Ref.~\cite{Buckley:2017ijx} for a recent review of gravitational probes of DM properties.
}
WIMPs behave as cold dark matter (CDM) on galactic scales.
They are in good agreement with many independent observations such as CMB anisotropy~\cite{Aghanim:2018eyx} and galaxy clustering~\cite{Alam:2016hwk}.
On the other hand, their predictions of galactic-scale structure are in debate.
On galactic scales, there have been issues that are difficult to explain in CDM (small-scale issues).%
\footnote{
Prominent examples are the missing satellite problem~\cite{Klypin:1999uc, Moore:1999nt, Zavala:2009ms, Papastergis:2011xe, Klypin:2014ira}, core-cusp problem~\cite{Flores:1994gz, Moore:1994yx, Moore:1999gc, Oh:2015xoa}, and too-big-to-fail problem~\cite{BoylanKolchin:2011de, BoylanKolchin:2011dk, Ferrero:2011au, Tollerud:2014zha, Garrison-Kimmel:2014vqa, Papastergis:2014aba}.
We refer readers to Ref.~\cite{Bullock:2017xww} for a recent review and further details.
State-of-the-art hydrodynamical simulations have been demonstrating that astrophysical processes also play an important role~\cite{Governato:2009bg, Governato:2012fa, Zolotov:2012xd, Brooks:2012vi, Munshi:2012nt, Pontzen:2014lma, Madau:2014ija, Sawala:2014xka, Chan:2015tna, Tollet:2015gqa, Sawala:2015cdf, Dutton:2015nvy, Wetzel:2016wro, Maccio:2016egb, Fattahi:2016nld, Brooks:2017rfe, Verbeke:2017rfd, Dutton:2018nop}.
There have also been reports that small-scale issues persist even in state-of-the-art hydrodynamical simulations~\cite{Oman:2015xda, Pawlowski:2015qta, 2016A&A...591A..58P, Oman:2016zjn, Sales:2016dmm, Trujillo-Gomez:2016pix, 2018arXiv180604143G, Bose:2018oaj, 2018arXiv181004186B}.
To our best knowledge, it is still controversial if astrophysical processes fully resolve the small-scale issues.
}
Alternatives to CDM may explain small-scale issues: warm dark matter (WDM)~\cite{AvilaReese:2000hg, Bode:2000gq, Knebe:2001kb, Zavala:2009ms, Lovell:2011rd, Papastergis:2011xe, Kamada:2013sh, Klypin:2014ira, Trujillo-Gomez:2016pix}; fuzzy DM~\cite{Hu:2000ke, Woo:2008nn, Marsh:2010wq, Marsh:2013ywa, Schive:2014dra, Schive:2014hza, Marsh:2015wka, Schwabe:2016rze, Hui:2016ltb, Zhang:2016uiy, Du:2018zrg}; and long-lasting DM interaction with primordial plasma or free-streaming light particles~\cite{Boehm:2001hm, Sigurdson:2003vy, Profumo:2004qt, vandenAarssen:2012ag, Aarssen:2012fx, Kamada:2013sh, Boehm:2014vja, Buckley:2014hja, Schewtschenko:2014fca, Cyr-Racine:2015ihg, Vogelsberger:2015gpr, Schewtschenko:2015rno, Binder:2016pnr, Bringmann:2016ilk, Kamada:2017oxi}.

\begin{figure}
\begin{center}
\includegraphics[width=0.9\columnwidth]{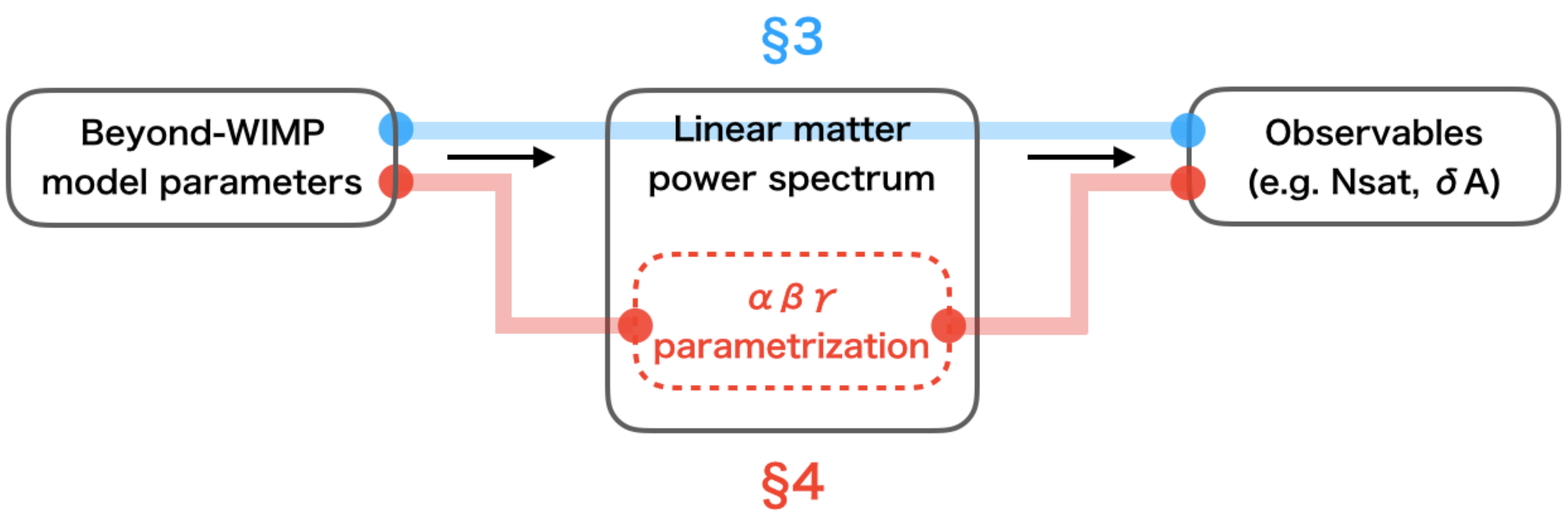}
\caption {\small
Sketch of the proposal of this paper.
}
\label{fig:Schematic}
\end{center}
\end{figure}

On the other hand, impacts on galactic-scale structure formation depend on beyond-WIMP scenarios.
Free-streaming of light WDM particles smears out the primordial density contrast.
Quantum pressure of fuzzy DM prevents DM from gravitational clustering.
Pressure of radiation to which DM couples involves DM in acoustic oscillation rather than gravitational clustering.
Such effects are reflected in the linear matter power spectrum, which one can obtain by following evolution of the primordial density contrast.
Generally by performing a suit of simulations with the resulting linear matter power spectrum, one can obtain observable quantities, which can be directly compared with cosmological/astrophysical observations.
In summary, we need to work out the following procedure on a model-by-model basis: 
\begin{itemize}
\item[]
\begin{center}
Model $\to$ Linear matter power spectrum $\to$ Observables.
\end{center}
\end{itemize}
See the blue flow in Fig.~\ref{fig:Schematic}.
The whole procedure requires interdisciplinary expertise from particle phenomenology to (computational) astrophysics.
Moreover, each step often requires a dedicated calculation.
In particular, simulations in the last step are often too time-consuming to repeat.

One can work out each step independently by parametrizing the linear matter power spectrum.
See the red flow in Fig.~\ref{fig:Schematic}.
A single parameter has been adopted conventionally: the thermal WDM mass $m_{\rm WDM}$.%
\footnote{
An underlying model may be light gravitino~\cite{Moroi:1993mb, Pierpaoli:1997im}.
WDM particles are thermalized in the early Universe and decouple from thermal plasma at some point.
}
On the other hand, a single parameter is not enough to cover a wide range of beyond-WIMP scenarios.
For this purpose, Ref.~\cite{Murgia:2017lwo} introduces the 3-parameter ($\{ \alpha, \beta, \gamma \}$) characterization of the linear matter power spectrum.
On one side, one (likely particle physicist) can construct a map of model parameters onto $\{ \alpha, \beta, \gamma \}$.
On the other side, one (likely astrophysicist) can provide observational constraints on $\{ \alpha, \beta, \gamma \}$, as indeed done for the Lyman-$\alpha$ forest data in Ref.~\cite{Murgia:2018now}.
By combining results from the two sides, one can obtain observational constraints on a given beyond-WIMP scenario.
Nevertheless, once multiple parameters are introduced, it becomes hard to share results from one side to the other.

In this respect, we propose building ready-to-use networks: 
one maps model parameters onto $\{ \alpha, \beta, \gamma \}$; and another maps $\{ \alpha, \beta, \gamma \}$ onto observables.
One can use these networks to examine models without repeating the aforementioned time-consuming procedure.
Ideally, it would be the most efficient if one obtained analytic maps, but in reality, it is hard to establish such analytic maps.
Thus, a numerical method is helpful to develop such effective maps.
For this purpose, we adopt neural network technique.

To be concrete, in this paper, we consider a feebly interacting massive particle (FIMP)~\cite{Hall:2009bx} (see Ref.~\cite{Bernal:2017kxu} for a recent review).
Light (keV-scale) FIMPs, which are produced through the freeze-in mechanism, are a compelling example of WDM.
Even in FIMP models, the shape of suppression in the linear matter power spectrum depends on production processes such as 2-body decay, 3-body decay, and 2-to-2 scattering~\cite{Shaposhnikov:2006xi, Gorbunov:2008ui, Boyanovsky:2008nc, Adulpravitchai:2015mna, McDonald:2015ljz, Roland:2016gli, Heeck:2017xbu} (see Ref.~\cite{Bae:2017dpt} for a comprehensive discussion).%
\footnote{
We refer readers to Refs.~\cite{Dodelson:1993je, Colombi:1995ze, Abazajian:2005gj, Venumadhav:2015pla} for sterile neutrino DM.
Sterile neutrinos are produced through mixing with active neutrinos.
We also refer readers to Ref.~\cite{Kaplinghat:2005sy} for superWIMPs.
SuperWIMPs are produced by the decay of WIMPs long after the WIMP freeze-out.
If the WIMP decay occurs close after the WIMP freeze-out, one may need to take into account the momentum distribution function of WIMPs~\cite{Petraki:2007gq, Merle:2014xpa, Bezrukov:2014qda, Merle:2015oja, Konig:2016dzg}.
In this paper, we do not consider these possibilities, although they may be FIMPs in a broad sense.
}
Thus 3-parameter characterization rather than conventional single-parameter characterization is required to cover a wide range of FIMP models.
By taking a simplified FIMP model, we demonstrate how one can work out the simplified procedure.
We also provide the obtained neural networks through the {\tt arXiv} website: one is a map of ``model parameters $\to$ $\{ \alpha, \beta, \gamma \}$" and the others are ``$\{ \alpha, \beta, \gamma \}$ $\to$ observables".

The organization of this paper is following. In Sec.~\ref{sec:rev}, we overview the conventional procedure to place constraints on FIMPs and describe the simplified procedure with the $\{ \alpha, \beta, \gamma \}$ parametrization.
In Sec.~\ref{sec:model}, we introduce a simplified FIMP model.
Our FIMP model shares many common aspects with a broad class of FIMP models.
The basic production process is $2$-body decay.
We take into account late-time entropy production after freeze-in (case A) and also freeze-in production through $2$-to-$2$ scattering (case B).
In Sec.~\ref{sec:NN}, we introduce neural network technique and work out the simplified procedure.
We compare the constraints from the simplified procedure and those from the conventional procedure.
Sec.~\ref{sec:sum} is devoted to the summary.
In Appendix~\ref{app:analytic}, we compare our constraints to those obtained through an analytic map from the conventional thermal WDM mass.
In Appendix~\ref{app:precision}, we examine precision of the neural networks in detail.
In Appendix~\ref{app:howto}, we explain how to use the neural networks we provide.

%%%%%%%%%%%%%%%%
%%%%%%%%%%%%%%%%

%%%%%%%%%%%%%%%%%%%%%%%%%%%%%%%%%%%%%%%%%%%%%%%%%%
\section{Procedure for FIMP DM as an example}
\label{sec:rev}
\setcounter{equation}{0}
%%%%%%%%%%%%%%%%%%%%%%%%%%%%%%%%%%%%%%%%%%%%%%%%%%

As we described in introduction, to study galactic-scale structure formation of beyond-WIMP scenarios, generically one has to take a 2-step procedure on a model-by-model basis:
\begin{itemize}
\item[]
\begin{center}
Model $\to$ Linear matter power spectrum $\to$ Observables
\end{center}
\end{itemize}
(corresponding to the blue flow in Fig.~\ref{fig:Schematic}).
In the case of FIMP, the first step of ``Model $\to$ Linear matter power spectrum''
actually consists of two steps:
\begin{itemize}
\item[]
\begin{center}
Model $\to$ DM phase space distribution $\to$ Linear matter power spectrum.
\end{center}
\end{itemize}
To follow the two steps, one first needs to construct the collision term of the Boltzmann equation and
integrate it to obtain the phase space distribution of the DM species.
Then one has to follow evolution of the primordial density contrast with the obtained phase space distribution, 
possibly by using public cosmological Boltzmann solvers such as {\tt CLASS}~\cite{Blas:2011rf, Lesgourgues:2011rh}.
In the following we overview this conventional procedure more specifically.

%%%%%%%%%%%%%%%%%%%%%%%%%%%%%%%%%%%%%%%%%%%%%%%%%%
\subsection{Model $\to$ DM phase space distribution $\to$ Linear matter power spectrum}
%%%%%%%%%%%%%%%%%%%%%%%%%%%%%%%%%%%%%%%%%%%%%%%%%%

We define the DM phase space distribution $f_\chi(t, p)$ as a function of the cosmic time $t$ and the physical momentum $p$, such that the DM number density is given by $n_\chi = g_\chi\int d^3p/(2\pi)^3 f_\chi(t, p)$, where $g_\chi$ is the spin degrees of freedom.
We assume that the DM phase space distribution $f_{\chi}$ is much smaller than unity.
We then obtain the phase space distribution at a late cosmic time $t_{f}$ by integrating the collision term as
\begin{align}
  \label{eq:integ-boltz}
  f_{\chi}(t_f, p) = \int_{t_i}^{t_f} dt \, \frac{1}{E_\chi} C \left( t, \frac{a(t_f)}{a(t_i)} p \right) \,,
\end{align}
where $t_i$ is the reheating time and $a(t)$ is the cosmic scale factor.
Given a squared matrix element of a specific production process, one obtains a semi-analytic expression of the corresponding collision term (see Ref.~\cite{Bae:2017dpt} for expressions).

FIMP production is most efficient when the heaviest particle in the process becomes non-relativistic (freeze-in mechanism).
After that, FIMP particles free-stream and the phase space distribution is invariant as a function of the comoving momentum $q \equiv p/T_\chi$, where $T_\chi$ is the effective DM temperature (see Sec.~\ref{subsubsec:CaseA} for a specific expression of $T_\chi$).
Thus we use $f(q) = f(t, p)$ to characterize the distribution.
Practically, we fit the obtained phase space distribution of DM by
\begin{align}
  \label{eq:fit-1}
q^{2} f(q)
&= 
\sum_{i=1}^{N}c_i \, q^{a_i} e^{- b_i q} \,,
\end{align}
where $(a_i, b_i, c_i)$ are fitting parameters and $i$ runs for different production processes.%
\footnote{
One may wonder if we can work out ``Model $\to$ DM phase space distribution'' and ``DM phase space distribution $\to$ Linear matter power spectrum'' separately by using $(a, b, c)$.
On one side, one can report constraints on $(a, b, c)$.
On the other side, one can calculate $(a, b, c)$ as a function of model parameters.
It is worth investigating this possibility somewhere else.
}
We plug the fitting function into the public cosmological Boltzmann solver {\tt CLASS}~\cite{Blas:2011rf, Lesgourgues:2011rh} to obtain the linear matter power spectrum $P(k)$ as a function of the wavenumber $k$.
We use the cosmological parameters from ``Planck 2015 TT, TE, EE+lowP" in Ref.~\cite{Ade:2015xua}.
Practically, we use the {\tt CLASS} fluid approximation of non-cold DM.

%%%%%%%%%%%%%%%%%%%%%%%%%%%%%%%%%%%%%%%%%%%%%%%%%%
\subsection{Linear matter power spectrum $\to$ Observables}
%%%%%%%%%%%%%%%%%%%%%%%%%%%%%%%%%%%%%%%%%%%%%%%%%%

Galactic-scale structure places constraints on the linear matter power spectrum $P (k)$, or, the transfer function that is defined by 
\begin{align}
T^{2} (k) 
\equiv
\frac{P (k)}{P_{\rm CDM} (k)} \,.
\end{align}
It generically requires a suit of time-consuming simulations to obtain constraints on FIMP DM.
We may simplify this step by using semi-analytic models and/or somehow converting the conventional thermal WDM mass $m_{\rm WDM}$.

In the conventional thermal WDM model, WDM particles follow the Fermi-Dirac distribution with two spin degrees of freedom with temperature $T_{\mathrm{WDM}}$.
The relic abundance is expressed by $m_{\mathrm{WDM}}$ and $T_{\mathrm{WDM}}$ as
\begin{align}
  \label{eq:thermal-wdm-abundance}
  \Omega_{\mathrm{WDM}}h^2
  =
  \left( \frac{m_{\mathrm{WDM}}}{94\unit{eV}}\right)
  \left( \frac{T_{\mathrm{WDM}}}{T_\nu} \right)^3
  =
  7.5
  \left( \frac{m_{\mathrm{WDM}}}{7 \, {\rm keV}} \right)
  \left( \frac{106.75}{g_*^{\mathrm{WDM}}} \right)\,.
\end{align}
For a given WDM mass, the temperature is determined such that the relic abundance reproduces the observed DM density.
Note that for a keV-scale mass, somewhat large entropy production after decoupling is required for $\Omega_{\mathrm{WDM}} h^2\simeq 0.12$.
On the other hand, FIMP DM has a different thermal history and thus different temperature and does not follow the Fermi-Dirac distribution.
Thus reported lower bounds on $m_{\rm WDM}$ is not directly applicable to FIMP DM.

In this paper we consider the number of satellite galaxies~\cite{Maccio:2009isa, Polisensky:2010rw, Lovell:2013ola, Kennedy:2013uta, Horiuchi:2013noa, Schneider:2014rda} and Lyman-$\alpha$ forest~\cite{Viel:2005qj, Seljak:2006qw, Viel:2006kd, Viel:2007mv, Viel:2013apy, Baur:2015jsy, Yeche:2017upn, Irsic:2017ixq, Armengaud:2017nkf, Kobayashi:2017jcf, Zhang:2017chj, Nori:2018pka, Leong:2018opi, Bose:2018juc} as observables.%
\footnote{
Other used probes include the delay of the reionization~\cite{Barkana:2001gr, Yoshida:2003rm, Schultz:2014eia, Lapi:2015zea, Tan:2016xvl, Lopez-Honorez:2017csg, Lovell:2017eec}, the counts of high-$z$ gamma-ray bursts~\cite{Mesinger:2005ah, deSouza:2013hsj}, the faint end of luminosity function of high-$z$ galaxies~\cite{Pacucci:2013jfa, Schultz:2014eia, Lapi:2015zea, Schive:2015kza, Menci:2016eww, Menci:2016eui, Lovell:2017eec, Ni:2019qfa}, the flux anomaly of quadrupole lens systems~\cite{Miranda:2007rb, Inoue:2014jka, Kamada:2016vsc, Kamada:2017icv, Birrer:2017rpp, Gilman:2017voy, Vegetti:2018dly, Rivero:2018bcd}, and the redshifted 21\,cm signal~\cite{Sitwell:2013fpa, Sekiguchi:2014wfa, Safarzadeh:2018hhg, Schneider:2018xba, Lidz:2018fqo, Lopez-Honorez:2018ipk, Nebrin:2018vqt, Chatterjee:2019jts}.
The counts of lensed distant supernovae~\cite{Pandolfi:2014rea} and direct collapse black holes~\cite{Dayal:2017yhb} are suggested for a future use.
We also refer readers to Ref.~\cite{Governato:2002cv, Gao:2007yk, Herpich:2013yga, Governato:2014gja, Maio:2014qwa, Colin:2014sga, Gonzalez-Samaniego:2015sfp, Lovell:2016fec, Wang:2016rio, Villanueva-Domingo:2017lae, Bozek:2018ekc, Bremer:2018nuf, Fitts:2018ycl, Lovell:2018gap, Maccio:2019hby} for hydrodynamical simulation results differentiating WDM and CDM in galaxy formation.
}
Our analysis, which follows Refs.~\cite{Schneider:2016uqi, Murgia:2017lwo}, uses a semi-analytic model for the number of satellite galaxies and converts the reported lower bound on $m_{\rm WDM}$ for the Lyman-$\alpha$ forest.

%%%%%%%%%%%%%%%%%%%%%%%%%%%%%%%%%%%%%%%%%%%%%%%%%%
\subsubsection*{Number of satellite galaxies}
%%%%%%%%%%%%%%%%%%%%%%%%%%%%%%%%%%%%%%%%%%%%%%%%%%

One compares the predicted number of satellite galaxies $N_{\rm sat}$ in simulated Milky Way-size or M31-size haloes with the observed one.
If the predicted number is smaller than the observed one, such FIMPs are excluded.
This constraint may be conservative when one counts all the subhalos above a certain mass, since some of subhalos may not host galaxies bright enough to be detected.

We evaluate the number of satellite galaxies $N_{\rm sat}$ from the linear matter power spectrum in our FIMP model
as follows.
Ref.~\cite{Schneider:2014rda} develops a semi-analytic formula of the subhalo mass function in the conventional thermal WDM model.
The formula uses the conditional mass function~\cite{Lacey:1993iv} based on the extended Press-Schechter approach~\cite{Press:1973iz} and the halo model (see Ref.~\cite{Cooray:2002dia} for a review).
The formula adopts the top-hat filter function in the Fourier space (sharp-$k$ filter) to reproduce results of 
$N$-body simulations in the conventional thermal WDM model:
\begin{align}
  \label{eq:nsat}
  \frac{d N_{\rm sat}}{d \ln M} = \frac{1}{C_{n}}\frac{1}{6 \pi^{2}} \frac{M_{0}}{M}
  \frac{P (1 / R)}{R^{3 }\sqrt{2 \pi (S - S_0)}} \,,
\end{align}
where quantities with and without the subscript ``0'' denote those of the host halo and subhalo, respectively.
For example, $M$ ($M_{0}$) is the subhalo (host halo) mass.
The variance $S$ is given by the linear matter power spectrum as
\begin{align}
  \label{eq:variance}
  S = \frac{1}{2 \pi^{2}} \int_{0}^{1/R} dk~ k^{2} P(k) \,.
\end{align}
The filter scale $R$ is related with the mass as
\begin{align}
  \label{eq:mass-radius}
  M = \frac{4 \pi}{3} \rho_{m} \, (c R)^3 \,,
\end{align}
with the matter mass density at present $\rho_{m}$.
Following Ref.~\cite{Schneider:2014rda}, we adopt $c = 2.5$ and $C_{n}=44.5$.
We use $M_{0} = 1.7 \times 10^{12} \, h^{-1} \mathrm{M_\odot}$ as the Milky-Way mass, where $h$ is the dimensionless Hubble constant.
With these values, the number of satellite galaxies above $M = 10^{8} \, h^{-1}\mathrm{M_\odot}$ is $N_{\rm sat} = 159$, which is consistent with the result of the Aquarius simulation~\cite{Springel:2008cc}.
$M > 10^{8} \, h^{-1}\mathrm{M_\odot}$ roughly corresponds to the lower bound on the maximal circular velocity of $V_{\rm max} > 10$\,km/s.

We estimate the observed number of satellites above $M = 10^{8} \, h^{-1} \mathrm{M_\odot}$ as $N_{\rm sat}^{\rm obs} = 63$ (11 classical dwarf galaxies and $3.5 \times 15$ ultra-faint dwarf galaxies).%
\footnote{
Classical dwarfs: Sagittarius, LMC, SMC, Ursa Minor, Sculptor, Draco, Sextans, Carina, Fornax, LeoII, and LeoI.
Ultra-faint dwarfs: Segue I, Ursa Major II, Segue II, Willman I, Coma Berenics, Bootes II, Bootes I, Pisces I, Ursa Major I, Hercules, Canes Venatici II, Leo IV, Leo V, Pisces II, Canes Venatici I.
We refer readers to Refs.~\cite{Walker:2009zp, Wolf:2009tu} for dynamical properties.
Note that $V_{\rm max} \geq V_{\rm 1/2} \simeq \sqrt{3} \sigma_{\rm l.o.s}$, where $V_{\rm 1/2}$ is the circular velocity at the half light radius and $\sigma_{\rm l.o.s}$ is the line-of-sight velocity dispersion~\cite{Wolf:2009tu}.
}
We multiply 3.5 by the number of ultra-faint satellites found in SDSS to take account of the SDSS limited sky coverage as in Refs.~\cite{Polisensky:2010rw, Kamada:2013sh, Lovell:2013ola, Kennedy:2013uta, Schneider:2014rda, Schneider:2016uqi, Murgia:2017lwo}.
$N_{\rm sat} > N_{\rm sat}^{\rm obs}$ places a lower bound on the conventional thermal WDM mass as $m_{\rm WDM} > 2.9$\,keV.%
\footnote{
We remark that we do not use the fitting function given by Eq.~(\refeq{eq:T2FitWDM}), but directly compute the linear matter power spectrum by using {\tt CLASS}~\cite{Blas:2011rf, Lesgourgues:2011rh}. 
}
As we see, $N_{\rm sat}$ implicitly depends on the lower bound on the satellite mass.
For example, once a number of smaller-size satellite galaxies are discovered in future, one has to repeat the above procedure by adjusting the lower bound on the satellite mass and scan model parameters on a model-by-model basis again.
This drives us to use the $\{ \alpha, \beta, \gamma \}$ parametrization.
Once the observational constraint on $\{ \alpha, \beta, \gamma \}$ is updated, one can easily update the constraint on models parameters by using a constructed map between model parameters and $\{ \alpha, \beta, \gamma \}$.

%%%%%%%%%%%%%%%%%%%%%%%%%%%%%%%%%%%%%%%%%%%%%%%%%%
\subsubsection*{Lyman-$\alpha$ forest}
%%%%%%%%%%%%%%%%%%%%%%%%%%%%%%%%%%%%%%%%%%%%%%%%%%

Another observable is the Lyman-$\alpha$ forest in high-resolution quasar spectra.
The flux power spectrum is a powerful probe of underlying galactic-scale structure, while the thermal history of the intergalactic medium has uncertainties.
The most stringent constraint seems to exclude the WDM solution to small-scale issues~\cite{Schneider:2013wwa}.

The procedure for the Lyman-$\alpha$ forest constraint is an example of mapping the reported lower bound on the conventional thermal WDM mass onto a given model.
We evaluate the impact of the WDM model on the Lyman-$\alpha$ forest data as follows.
This approach follows Ref.~\cite{Murgia:2017lwo}, which extends the approach of Ref.~\cite{Schneider:2016uqi}.
First, given a 3-dimensional linear matter power spectrum $P(k)$, we calculate the 1-dimensional power spectrum as
\begin{align}
  \label{eq:1d-pk}
  P_{\rm 1D} (k)
  &=
    \frac{1}{2 \pi} \int_k^{\infty} dk^{\prime} k^{\prime} P(k^{\prime}) \,.
\end{align}
Second, we normalize the 1D power spectrum by that in the CDM model:
\begin{align}
  r(k) = \frac{P_{\rm 1D} (k)}{P_{\rm 1D}^{\rm CDM} (k)} \,.
\end{align}
Third, we integrate $r(k)$ over the typical range of $k$ that a given Lyman-$\alpha$ forest spectrum probes:
\begin{align}
A = \int_{k_{\rm min}}^{k_{\rm max}} dk \, r(k) \,,
\end{align}
The dimensionless deviation of $A$ represents net suppression in the Lyman-$\alpha$ forest spectrum:
\begin{align}
  \label{fdeltaA}
  \delta A = \frac{A_{\rm CDM} - A}{A_{\rm CDM}} \,.
\end{align}
Finally, we compare $\delta A$ between our FIMP model and the conventional thermal WDM model with the reported lower bound on $m_{\rm WDM}$.
Note that one should use the typical range of $k$ for $A$ and the lower bound on $m_{\rm WDM}$ consistently from the same dataset or analysis.
If $\delta A > \delta A_{\rm WDM}$, then we regard our FIMP model is excluded.

Ref.~\cite{Murgia:2017lwo} suggests $k_{\rm min} = 0.5 \, h / {\rm Mpc}$ and $k_{\rm max} = 20 \, h / {\rm Mpc}$ for the MIKE/HIRES+XQ-100 combined dataset used in Ref.~\cite{Irsic:2017ixq}.
The dataset places the lower bound of $m_{\rm WDM} > 3.5$\,keV in the conventional thermal WDM model.
We find that $\delta A_{\rm WDM} = 0.46$ for $m_{\rm WDM} = 3.5$\,keV,%
\footnote{
We again remark that we do not use the fitting function given by Eq.~(\refeq{eq:T2FitWDM}), but directly compute the linear matter power spectrum by using {\tt CLASS}~\cite{Blas:2011rf, Lesgourgues:2011rh}.
This may be partially why our $\delta A_{\rm WDM} = 0.46$ is different from $\delta A_{\rm WDM} = 0.38$ in Ref.~\cite{Murgia:2017lwo}.
}
so we use $\delta A_{3.5 \, {\rm keV}}\equiv0.46$ as an upper bound of $\delta A$ of a given model.
As we see, $\delta A$ needs a data-dependent input $k_{\rm min}$ and $k_{\rm max}$ and thus one has to repeat the procedure for different dataset.
A more extendable procedure is presumable.
Our proposal is the $\{ \alpha, \beta, \gamma \}$ parameterization.
For a given new dataset, while one has to update constraints in terms of in terms of $\{ \alpha, \beta, \gamma \}$, one can use the constructed map between model parameters and $\{ \alpha, \beta, \gamma \}$ as it is.

%%%%%%%%%%%%%%%%%%%%%%%%%%%%%%%%%%%%%%%%%%%%%%%%%%
\subsection{$\{\alpha, \beta, \gamma\}$ parametrization of the transfer function}
\label{subsec:abcparam}
%%%%%%%%%%%%%%%%%%%%%%%%%%%%%%%%%%%%%%%%%%%%%%%%%%

As we described above, the thermal WDM model has been conventionally used to report observation constraints on the transfer function $T^{2} (k)$.
The single-parameter fitting function of $T^{2} (k)$ in the thermal WDM model is given by~\cite{Bode:2000gq, Hansen:2001zv, Viel:2005qj}%
\footnote{
We refer readers to Ref.~\cite{Abazajian:2005gj} for a fitting function of $T^2(k)$ in the resonantly produced sterile neutrino DM.
}
\begin{align}
T_{\rm WDM}^{2} (k) 
&= 
\left[ 1 + \left( \alpha k \right)^{2 \nu} \right]^{- 10 / \nu} \,.
\label{eq:T2FitWDM}
\end{align}
Here $\nu = 1.12$ and thus only $\alpha$ is a parameter related with the thermal WDM mass:
\begin{align}
\alpha = 0.049 \, {\rm Mpc} / h \left( \frac{m_{\rm WDM}}{\rm keV} \right)^{- 1.11} \left( \frac{\Omega_{\rm WDM}}{0.25} \right)^{0.11} \left( \frac{h}{0.7} \right)^{1.22}
\end{align}
from Ref.~\cite{Viel:2005qj}. 

However, the single-parameter ($m_{\mathrm{WDM}}$) characterization does not cover a wide range of beyond-WIMP models.
Ref.~\cite{Murgia:2017lwo} proposes characterizing the transfer function as
\begin{align}
T^{2} (k) 
&= 
\left[ 1 + \left( \alpha k \right)^{\beta} \right]^{2 \gamma} \,.
\label{eq:T2Fit}
\end{align}
This parametrization allow us to divide the procedure to place constraints on FIMPs into two with $\{ \alpha, \beta, \gamma \}$ being a ``common language''.
On one side, one calculates $\{\alpha, \beta, \gamma \}$ as a function of model parameters in a given model (corresponding to the left red flow in Fig.~\ref{fig:Schematic}).
On the other side, one reports a likelihood function from observations as a function of $\{ \alpha, \beta, \gamma \}$ (corresponding to the right red flow in Fig.~\ref{fig:Schematic}).
By combining these two, one can obtain the constraints on model parameters more easily.
This procedure is also very extendable.
Once a new observation date becomes available, what one has to do is just to update the latter, namely, constraints on $\{ \alpha, \beta, \gamma \}$.
One does not need to repeat the former.
One can use a constructed map between model parameters and $\{ \alpha, \beta, \gamma \}$ as it is.

A remaining challenge is how to share results from the two sides.
It is not apparent how to share 3-parameter results efficiently.
In this paper, we propose using neural network technique.
In the context of the paper, advantages of using a neural network are:
\begin{itemize}
\item[--] 
It expresses nonlinear relations quite efficiently.
\item[--] 
It learns nonlinearity without being explicitly taught.
\item[--] 
It provides us with a unified format in presenting results.
\end{itemize}
We indeed see these advantages in Sec.~\ref{sec:NN}.

%%%%%%%%%%%%%%%%%%%%%%%%%%%%%%%%%%%%%%%%%%%%%%%%%%
\section{Simplified FIMP model}
\label{sec:model}
\setcounter{equation}{0}
%%%%%%%%%%%%%%%%%%%%%%%%%%%%%%%%%%%%%%%%%%%%%%%%%%

In this work, we consider a simple setup.
The model contains a seemingly renormalizable interaction of Majorana DM $\chi$ with a heavy Dirac fermion $\Psi$ and a heavy scalar $\phi$:
\begin{align}
  \label{eq:lag_chi}
  {\cal L}_{\chi}
  &=
    y_{\chi} \phi \bar\Psi \chi + {\rm h.c.} \,,
\end{align}
with the Yukawa coupling $y_{\chi}$.
We assume the mass hierarchy of $m_{\Psi} > m_{\phi} \gg m_{\chi}$.%
\footnote{
The result will change only slightly for $m_{\phi} > m_{\Psi} \gg m_{\chi}$ and for different quantum statistics of particles~\cite{Bae:2017dpt}.
}

This simplified model virtually corresponds to a light axino FIMP model considered in Refs.~\cite{Bae:2017tqn, Bae:2017dpt}.
The axino FIMP model is based on a supersymmetric version of Dine-Fischler-Srednicki-Zhitnitsky axion model~\cite{Zhitnitsky:1980tq, Dine:1981rt}.
Axino is a fermionic supersymmetric partner of axion that dynamically explains why the strong interaction preserves $CP$ very precizely~\cite{Peccei:1977hh, Peccei:1977ur, Weinberg:1977ma, Wilczek:1977pj}.
One can identify $\chi$, $\Psi$, and $\phi$ as light axino, Higgsino (supersymmetric partner of Higgs), and Higgs in the axino FIMP model.

%%%%%%%%%%%%%%%%%%%%%%%%%%%%%%%%%%%%%%%%%%%%%%%%%%
\subsection{Freeze-in production}
\label{subsec:modelsetup}
%%%%%%%%%%%%%%%%%%%%%%%%%%%%%%%%%%%%%%%%%%%%%%%%%%

We assume that $\Psi$ is equilibrated in thermal plasma.
Freeze-in production of DM $\chi$ proceeds mainly through $2$-body decay of $\Psi \to \phi + \chi$.
The production process ceases (decouples) when the plasma temperature $T$ gets comparable with the mother particle mass; 
{\it i.e.}, the decoupling temperature is $T_{\rm dec} \sim m_{\Psi}$.

It is convenient to define a DM ``temperature'' as
\begin{align}
  \label{eq:nonthermal-temp}
  T_{\rm DM} = \left( \frac{g_{*}(T)}{g_{*}(T_{\rm dec})} \right)^{1/3} T \,,
\end{align}
with the effective number of massless degrees of freedom $g_{*}(T)$ and the decoupling temperature $T_{\rm dec}$.
This temperature scales as $T_{\rm DM} \propto 1 / a$ with the cosmic scale factor $a (t)$ and thus the dimensionless momentum $q = p / T_{\rm DM}$ is conserved after the decoupling.
In the following, we take $g_{*}(T_{\rm dec}) = g_{*}^{\rm SM} = 106.75$ (all the SM particles) as a baseline value.

%%%%%%%%%%%%%%%%%%%%%%%%%%%%%%%%%%%%%%%%%%%%%%%%%%
\subsubsection*{Case A: Decay with entropy production}
\label{subsubsec:CaseA}
%%%%%%%%%%%%%%%%%%%%%%%%%%%%%%%%%%%%%%%%%%%%%%%%%%

Meanwhile, we incorporate a different value of $g_{*}(T_{\rm dec})$ or entropy production after the decoupling, by introducing $\Delta$ as
\begin{align}
  \label{eq:nonthermal-temp-delta}
  T_{\rm DM} = \left( \frac{g_{*}(T)}{\Delta \times g_{*}^{\rm SM}} \right)^{1/3} T \,.
\end{align}
$\Delta>1$ takes account of entropy production after the decoupling, or lager degrees of freedom at the decoupling ({\it e.g.}, minimal supersymmetric standard model, where $g_{*}^{\rm MSSM} = 226.75$).
$\Delta<1$ is applied to the case of late decoupling, {\it i.e.}, $g_*(T_{\rm dec}) < g_{*}^{\rm SM}$.

We take into account only relevant model parameters to ``warmness'' of FIMP DM.
Note that warmness of FIMP DM depends on the phase space distribution $f(p)$ (equivalently, $f(q)$ and $T_{\rm DM}$) and the FIMP mass $m_{\chi}$.
The phase space distribution does not depend on an absolute scale of $m_{\phi}$ and $m_{\Psi}$, but is {\it sensitive} to the ratio $m_{\phi} / m_{\Psi}$ since the ratio determines the kinematic phase space of decay product, {\it i.e.}, $\chi$ in this case.
If the two masses are degenerate, the energy of $\chi$ in $\Psi \to \phi + \chi$ is suppressed and thus the resultant $\chi$'s are colder~\cite{Heeck:2017xbu, Bae:2017tqn, Bae:2017dpt}.

In this class of models, therefore, the relevant parameters are
\begin{align}
\frac{m_{2}}{m_{1}} \,,
~~~~
m_{\rm DM} \,,
~~~~
\Delta \,.
\end{align}
Hereafter we use the notation of $m_{1} = m_\Psi$, $m_{2}=m_{\phi}$, and $m_{\mathrm{DM}}=m_{\chi}$ for the sake of notational simplicity.
The Yukawa coupling $y_{\chi}$ is fixed by the observed DM abundance $\Omega_{\mathrm{DM}} = m_\chi s_0Y_\chi/\rho_c$.
While the colder phase space distribution is realized for a more degenerate mass spectrum, the larger Yukawa coupling or lighter $\Psi$ is necessary to obtain the observed DM abundance.

%%%%%%%%%%%%%%%%%%%%%%%%%%%%%%%%%%%%%%%%%%%%%%%%%%
\subsubsection*{Case B: Decay with scattering}
\label{subsubsec:CaseB}
%%%%%%%%%%%%%%%%%%%%%%%%%%%%%%%%%%%%%%%%%%%%%%%%%%

Generally a daughter particle $\phi$ has another interaction with a light Dirac fermion $f$:
\begin{align}
  \label{eq:lag_phi}
  {\cal L}_{\phi}
  &=
    y_{f} \phi {\bar f} f 
    + {\rm h.c.} \,,
\end{align}
with the Yukawa coupling $y_{f}$.
One can identify $f$ as top quark (again $\phi$ as Higgs) in the axino FIMP model~\cite{Bae:2017tqn, Bae:2017dpt}.
We assume the mass hierarchy of $m_{\Psi} > m_{\phi} \gg m_{f}$.
We also assume that $f$ is equilibrated in thermal plasma.
In this case, freeze-in production of $\chi$ occurs through $s$-channel scattering of $f {\bar f} \to \Psi \chi$ and $t$-channel scattering of $\Psi f \to \chi f$ as well as through 2-body decay of $\Psi \to \phi + \chi$.
The decoupling temperature is again $T_{\rm dec} \sim m_\Psi$.
$y_{f}$ determines the scattering contribution to the yield, $Y_{\rm scat}$.
Freeze-in production through scattering becomes more important for more degenerate $\phi$ and $\Psi$, since the partial decay width becomes smaller.

In summary, in this case, the relevant parameters are
\begin{align}
\frac{m_{2}}{m_{1}} \,,
~~~~
\frac{Y_{\rm scat}}{Y_{\rm total}} \,,
~~~~
m_{\rm DM} \,.
\end{align}
Again hereafter we use the notation of $m_{1} = m_\Psi$, $m_{2}=m_\phi$, and $m_{\mathrm{DM}}=m_{\chi}$ for the sake of notational simplicity.
$y_{\chi}$ is fixed by the observed DM abundance: $Y_{\rm dec} + Y_{\rm scat} = Y_{\rm total}$.
In this case, we do not vary $\Delta$ but take several values such as $\Delta = 0.3, 1$, and $3$.

%%%%%%%%%%%%%%%%%%%%%%%%%%%%%%%%%%%%%%%%%%%%%%%%%%
\subsection{Constraints}
\label{subsec:NNraints}
%%%%%%%%%%%%%%%%%%%%%%%%%%%%%%%%%%%%%%%%%%%%%%%%%%

We derive constraints from $N_{\rm sat}$ and from $\delta A$ through the conventional procedure described in Sec.~\ref{sec:rev} (corresponding to the blue flow in Fig.~\ref{fig:Schematic}).

\begin{figure}
\begin{center}
\includegraphics[width=0.4\columnwidth]{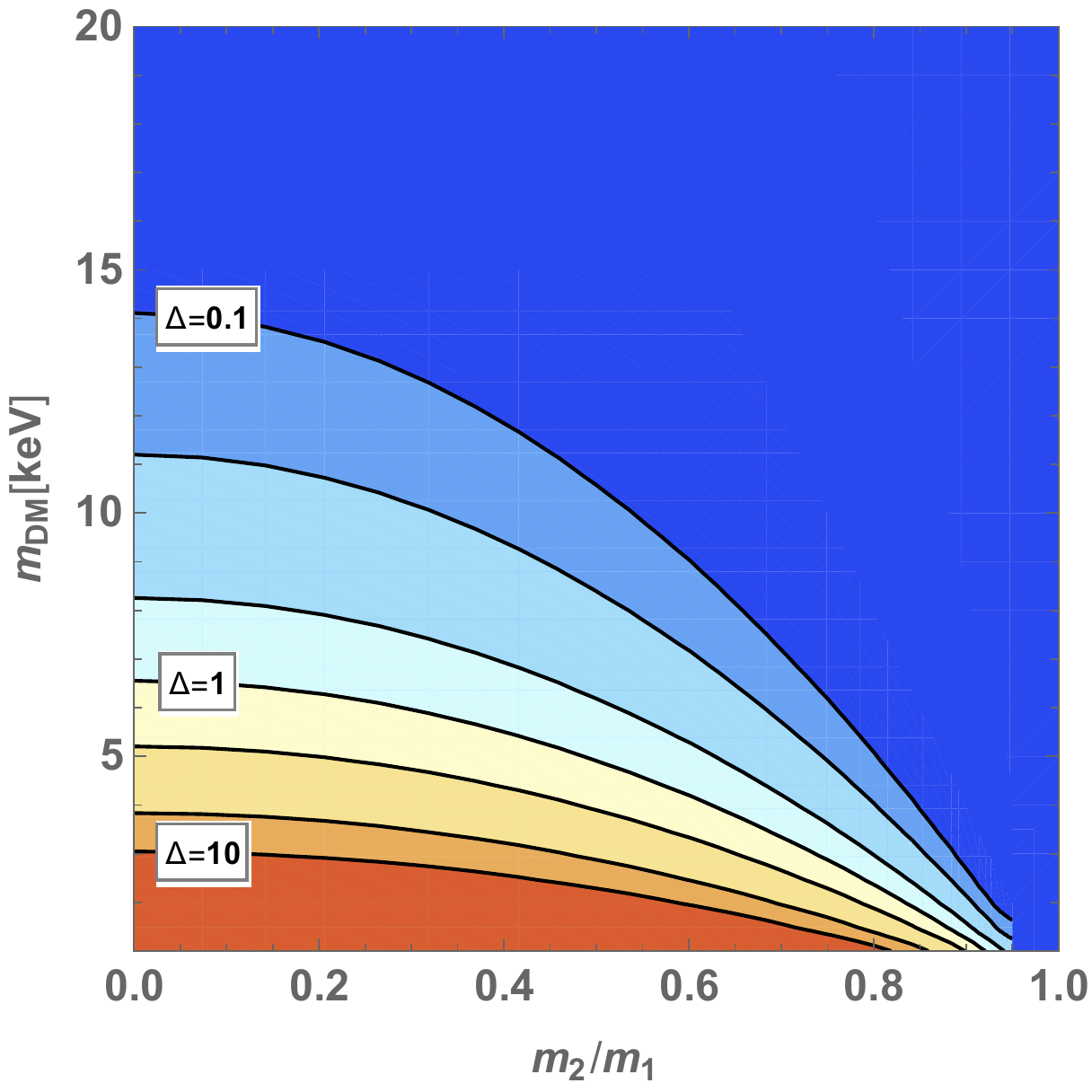}
\hskip 1cm
\includegraphics[width=0.4\columnwidth]{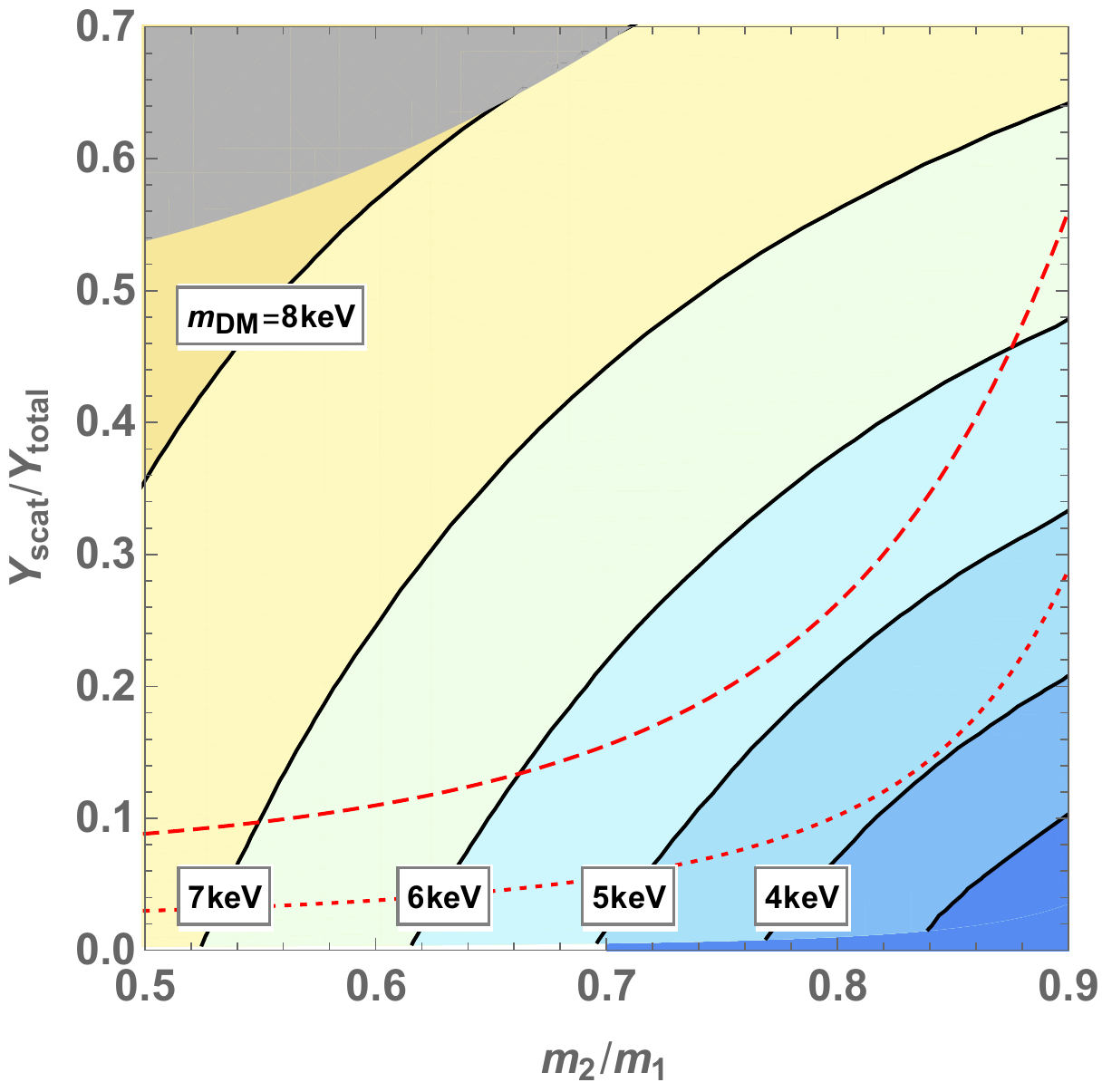}
\vskip 5mm
\includegraphics[width=0.4\columnwidth]{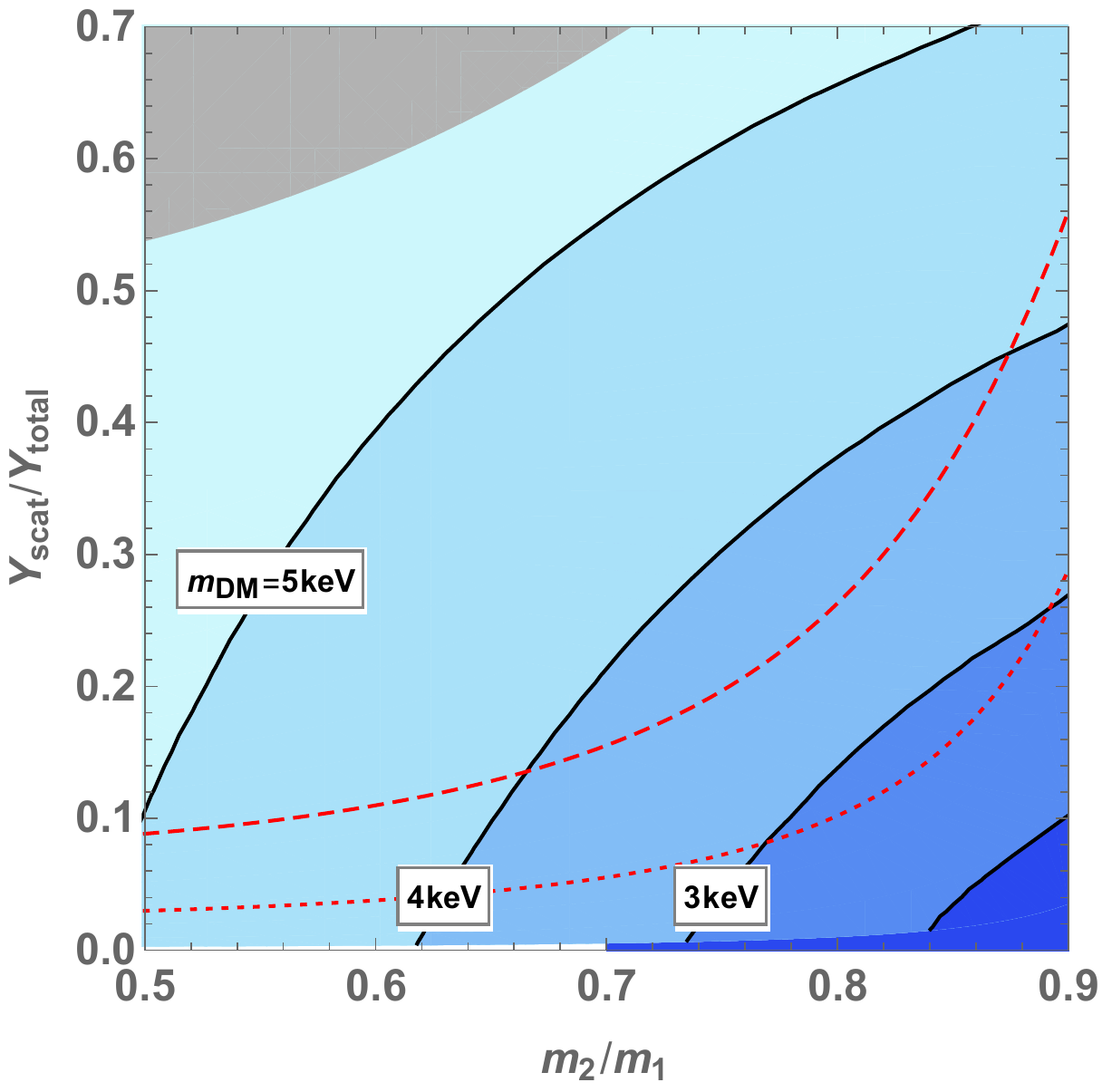}
\hskip 1cm
\includegraphics[width=0.4\columnwidth]{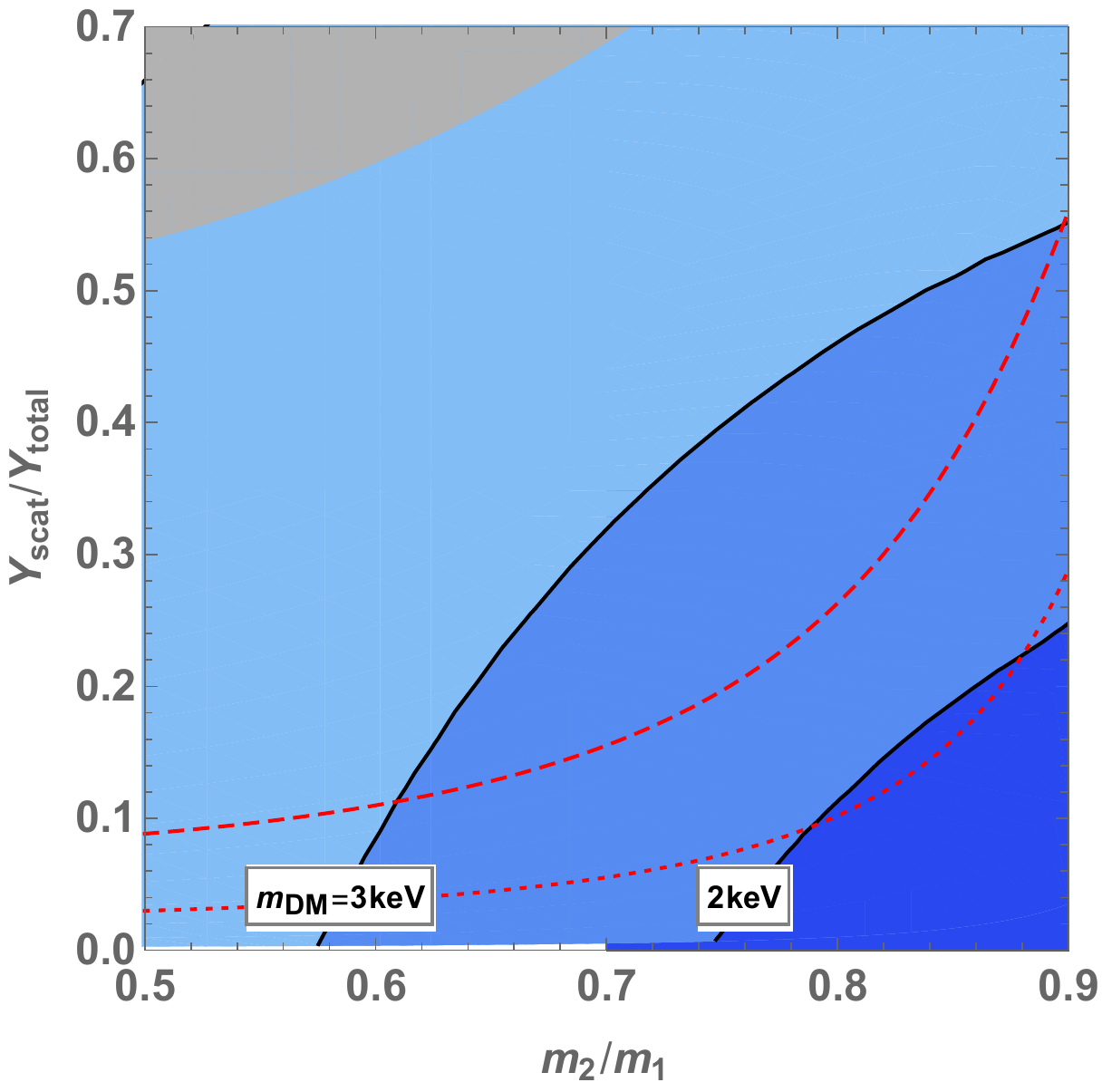}
\caption {\small
Constraints from $N_{\rm sat} > N_{\rm sat}^{\rm obs}$,
where $N_{\rm sat}^{\rm obs} = 63$ is the observed value explained in the main text.
Bluer regions are alive for each value of $m_{\rm DM}$.
The red contours are for $y_f = \sqrt{\pi/3}$ (dashed) and $\sqrt{1/3}$ (dotted).
{\bf Top-left:} Case A (Decay with entropy production).
{\bf Top-right:} Case B (Decay with scattering) with $\Delta = 0.3$.
{\bf Bottom-left:} Case B with $\Delta = 1$.
{\bf Bottom-right:} Case B with $\Delta = 3$.
}
\label{fig:NSat}
\end{center}
\end{figure}

First we present constrains from $N_{\rm sat}$ in Fig.~\ref{fig:NSat}.
The top-left panel is for Case A (Decay with entropy production), while the other panels are for Case B (Decay with scattering).
For Case A, bluer regions satisfy the condition $N_{\rm sat} > N_{\rm sat}^{\rm obs} = 63$ for each value of $\Delta$.
For Case B, the three panels correspond to $\Delta = 0.3$ (top-right),
$\Delta = 1$ (bottom-left), and $\Delta = 3$ (bottom-right), respectively.
As in Case A, bluer regions satisfy $N_{\rm sat} > N_{\rm sat}^{\rm obs}$ for each value of $m_{\rm DM}$.
We also display two lines corresponding to $y_f = \sqrt{\pi / 3}$ (red-dashed) and $y_f = \sqrt{1 / 3}$ (red-dotted), to depict a perturbative Unitarity limit.

\begin{figure}
\begin{center}
\includegraphics[width=0.4\columnwidth]{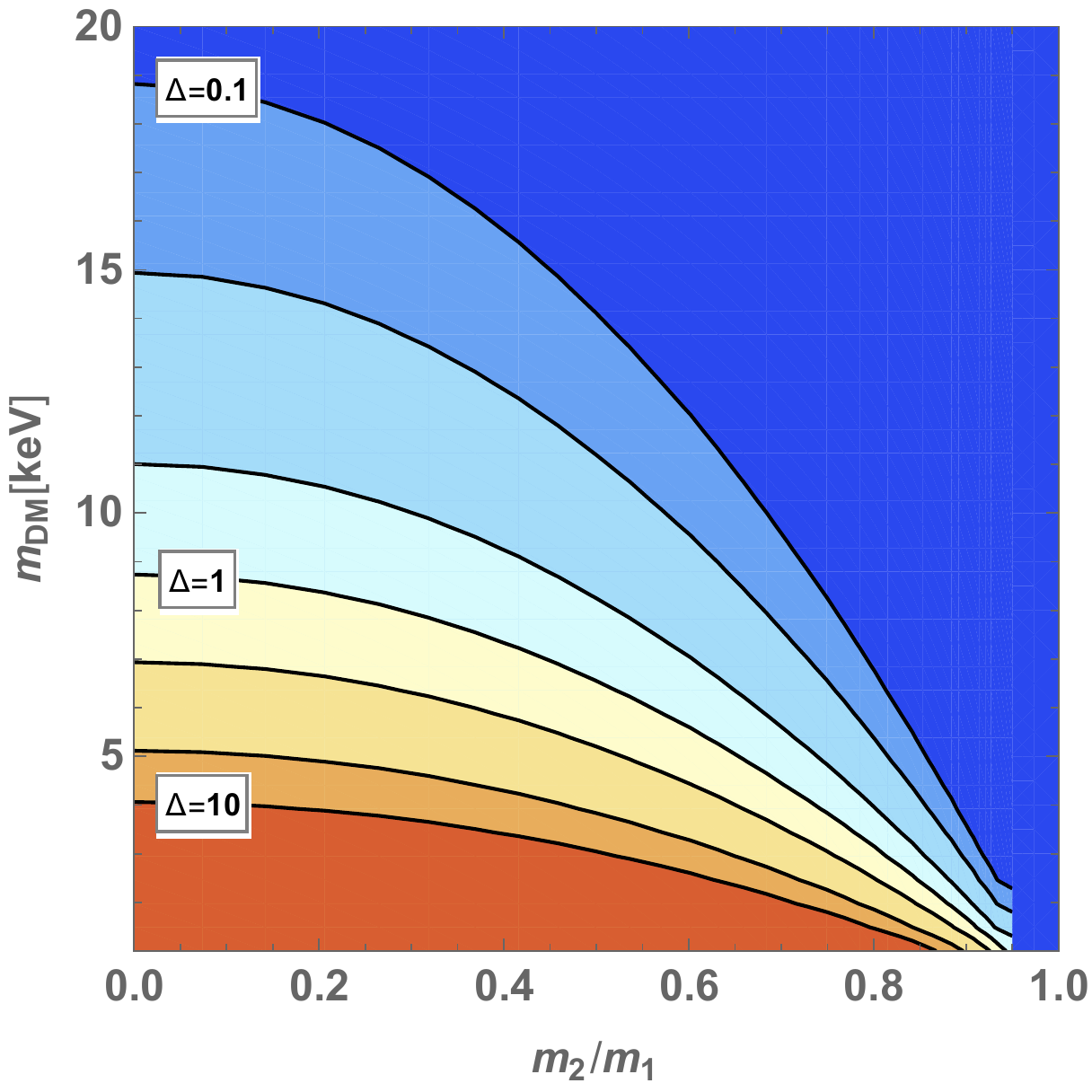}
\hskip 1cm
\includegraphics[width=0.4\columnwidth]{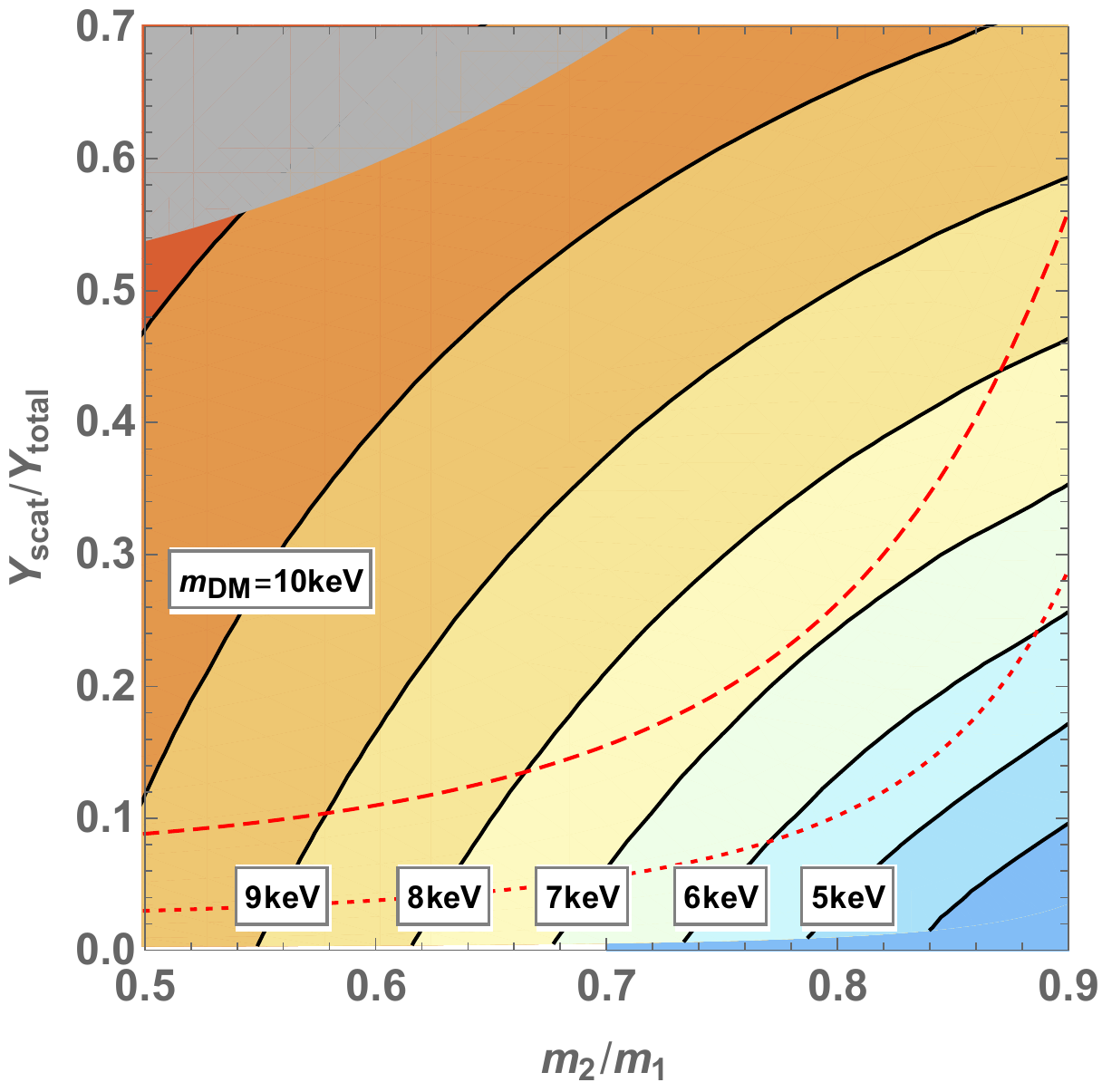}
\vskip 5mm
\includegraphics[width=0.4\columnwidth]{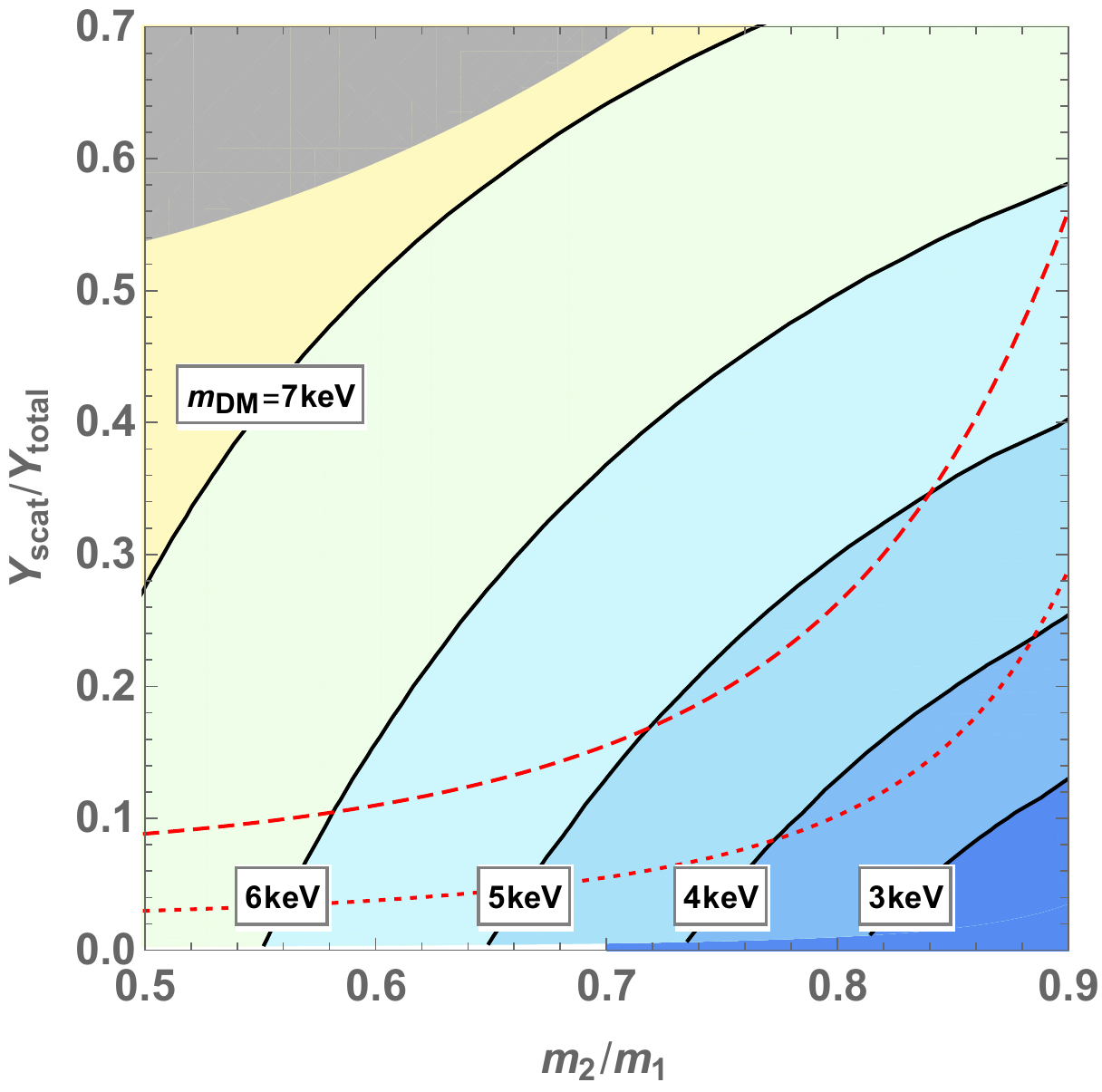}
\hskip 1cm
\includegraphics[width=0.4\columnwidth]{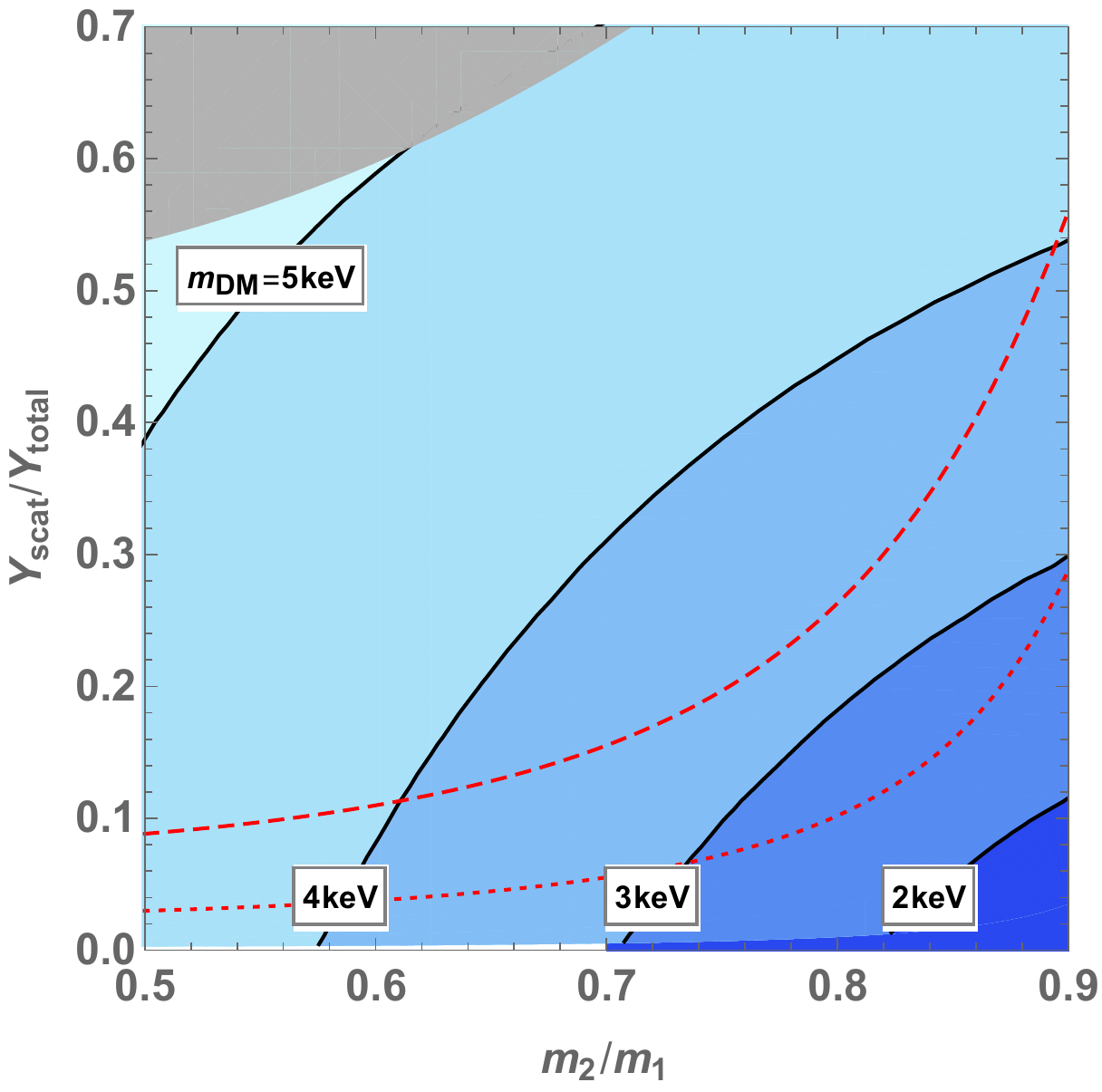}
\caption {\small
Constraints from $\delta A > \delta A_{3.5 \, {\rm keV}}$,
where $\delta A_{3.5 \, {\rm keV}} = 0.46$ is the value for $3.5$\,keV thermal WDM.
Bluer regions are alive for each contour.
{\bf Top-left:} Case A (Decay with entropy production).
{\bf Top-right:} Case B (Decay with scattering) with $\Delta = 0.3$.
{\bf Bottom-left:} Case B with $\Delta = 1$.
{\bf Bottom-right:} Case B with $\Delta = 3$.
}
\label{fig:deltaA}
\end{center}
\end{figure}

Next we show constraints from $\delta A$ in Fig.~\ref{fig:deltaA}.
The four panels are for Case A (top-left) and for Case B with $\Delta = 0.3$ (top-right),
$\Delta = 1$ (bottom-left), and $\Delta = 3$ (bottom-right), respectively.
For each parameter, bluer regions satisfy the condition $\delta A < \delta A_{3.5 \, {\rm keV}}$.
The red lines are the same as Fig.~\ref{fig:NSat}.
We see that $\delta A$ gives stronger constraints than $N_{\rm sat}$.

As repeatedly stated, constraints on the transfer function are often provided in terms of the conventional thermal WDM mass $m_{\rm WDM}$.
In Appendix~\ref{app:analytic} we convert $m_{\rm WDM} > 2.9$\,keV corresponding to $N_{\rm sat} > N_{\rm sat}^{\rm obs}$ and $m_{\rm WDM} > 3.5$\,keV corresponding to $\delta A < \delta A_{3.5 \, {\rm keV}}$ into constraints on our FIMP parameters.
We see that the constraints are qualitatively similar but quantitatively slightly different ($\sim 10\%$ in $m_{\rm DM}$) from those derived in this section.

%%%%%%%%%%%%%%%%
%%%%%%%%%%%%%%%%

%%%%%%%%%%%%%%%%
%%%%%%%%%%%%%%%%

%%%%%%%%%%%%%%%%%%%%%%%%%%%%%%%%%%%%%%%%%%%%%%%%%%
\section{Neural network approach}
\label{sec:NN}
\setcounter{equation}{0}
%%%%%%%%%%%%%%%%%%%%%%%%%%%%%%%%%%%%%%%%%%%%%%%%%%

%%%%%%%%%%%%%%%%
\begin{figure}
\centering
\includegraphics[width=0.6\columnwidth]{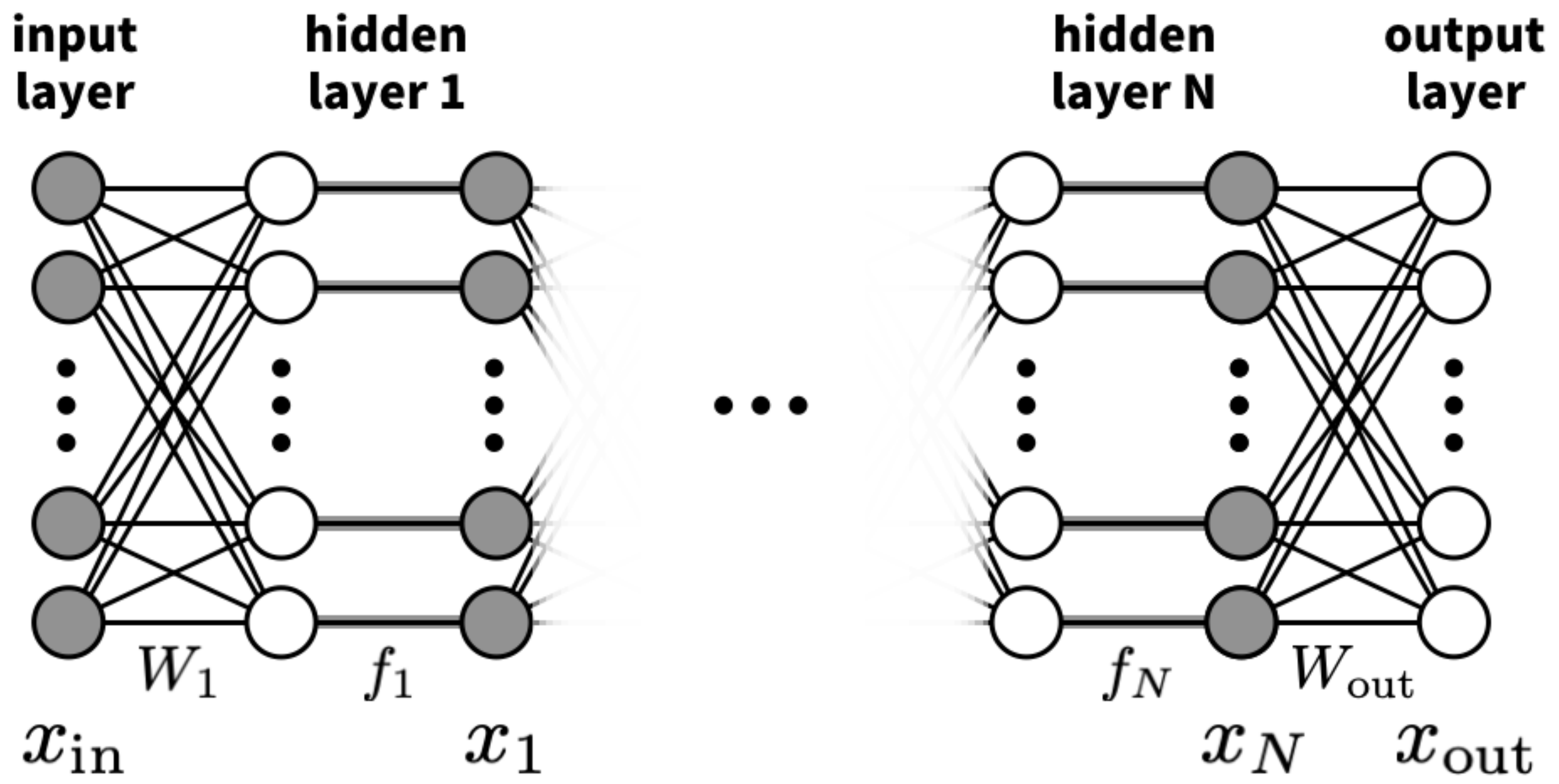}
\caption {\small
Schematic diagram of the neural network. Partly taken from Ref.~\cite{Jinno:2018dek}.
}
\label{fig:NN}
\end{figure}
%%%%%%%%%%%%%%%%

As stressed in Sec.~\ref{sec:Intro}, one of the main purposes of this paper is to provide ready-to-use maps
for ``Model parameters $\to$ $\{ \alpha, \beta, \gamma \}$" and also for ``$\{ \alpha, \beta, \gamma \}$ $\to$ Observables" (see the red flow in Fig.~\ref{fig:Schematic}).
Our proposal is to use a neural network for this purpose.
In the following we first explain our neural network setup in Sec.~\ref{subsec:setup}, and then construct concrete neural networks for ``Model parameters $\to$ $\{ \alpha, \beta, \gamma \}$" and ``$\{ \alpha, \beta, \gamma \}$ $\to$ Observables" in Sec.~\ref{subsec:NN1} and Sec.~\ref{subsec:NN2}, respectively.
Finally we combine the two neural networks to reproduce the constraints presented in Sec.~\ref{sec:model}
to demonstrate the precision of the neural networks.

%%%%%%%%%%%%%%%%%%%%%%%%%%%%%%%%%%%%%%%%%%%%%%%%%%
\subsection{Neural network setup}
\label{subsec:setup}
%%%%%%%%%%%%%%%%%%%%%%%%%%%%%%%%%%%%%%%%%%%%%%%%%%

The setup of our neural network is summarized in Fig.~\ref{fig:NN}.
We identify the input vector $\vec{x}_{\rm in}$ as the three model parameters for each of Case A and B.
As the layer proceeds, the original layer is operated by linear algebra and then multiplied by a nonlinear function $\vec{f}$. 
More concretely, the connections among the layers are given by
\begin{align}
\vec{x}_{1}
&= 
\vec{f} (W_{1} \vec{x}_{\rm in} + \vec{b}_{1}) \,,
\\
\vec{x}_{n}
&= 
\vec{f} (W_{n} \vec{x}_{n - 1} + \vec{b}_{n})
~~~~
(2 \leq n \leq N) \,,
\\
\vec{x}_{\rm out}
&= W_{\rm out} \vec{x}_{N} + \vec{b}_{\rm out} \,,
\end{align}
where $N$ is the number of hidden layers and $W$'s and $b$'s are called weight matrices and biases, respectively.
The nonlinear function $\vec{f}$ is understood as acting on each component:
\begin{align}
\vec{f} (\vec{y}) 
&\equiv 
\left(
\begin{matrix}
f (y_{1}) \\
f (y_{2}) \\
\vdots
\end{matrix}
\right) \,,
\end{align}
and we adopt a Rectified Linear Unit (ReLU)~\cite{Nair:2010:RLU:3104322.3104425} for the $f$ function:
\begin{align}
f (y)
&= 
\max (0, y) \,.
\end{align}

We train the neural network with supervised learning.
As we explain in the next subsections, 
we collect $\Order (10,000)$ combinations of the input $\vec{x}_{\rm in}$ 
and the true value (from direct calculations) of the output $\vec{x}_{\rm out}^{\rm (true)}$.
Note that, with such a large number of data points,
it is much more efficient to recast the obtained data onto the neural network and share the neural network parameters than to provide the data itself.
Training of the neural network is performed through the updates of the weight matrices and biases 
so that the output of the neural network $\vec{x}_{\rm out}$ gets closer 
to the true value $\vec{x}_{\rm out}^{\rm (true)}$.
The closeness is measured by the loss function $E$, which we take as
\begin{align}
E
&= 
\sum_i \left| \left( \vec{x}_{\rm out}^{\rm (true)} \right)_i - \left( \vec{x}_{\rm out} \right)_i \right| \,,
\end{align}
where $(\vec{x})_i$ denotes the $i$-th component of $\vec{x}$.

For the number of hidden layers, we use $N = 2$ in this paper.
Then the relation between the input and output reduces to
\begin{align}
\vec{x}_{\rm out}
&= 
W_{\rm out}~\vec{f} \left( W_2~\vec{f} \left( W_1 \vec{x}_{\rm in} + \vec{b}_1 \right) + \vec{b}_2 \right) + \vec{b}_{\rm out} \,.
\label{eq:xinxout}
\end{align}
We construct the neural network using the public code {\tt TensorFlow}~\cite{TensorFlow},\footnote{
We use the version {\tt r1.1.7}.
}
and train it for $\Order (10^5)$ epochs.
The whole dataset is split into training ($90\%$) and test ($10\%$) subsets,
and the former is used to train the neural network, 
while the latter is used to monitor the training process and avoid possible overfitting.
We also apply a 10\% dropout~\cite{Srivastava:2014:DSW:2627435.2670313} to avoid overfitting.
We use Adam Optimizer~\cite{Kingma:2014vow} with a learning rate of 0.001.

%%%%%%%%%%%%%%%%%%%%%%%%%%%%%%%%%%%%%%%%%%%%%%%%%%
\subsection{Model parameters $\to$ $\{\alpha, \beta, \gamma\}$}
\label{subsec:NN1}
%%%%%%%%%%%%%%%%%%%%%%%%%%%%%%%%%%%%%%%%%%%%%%%%%%

We first construct a neural network connecting model parameters and the transfer function parameters $\{ \alpha, \beta, \gamma \}$.
Before moving on, however, we remark that parameter degeneracy often appears when we fit $\{ \alpha, \beta, \gamma \}$ to the resulting power spectrum in the simplified FIMP model in Sec.~\ref{sec:model}.
Indeed Ref.~\cite{Murgia:2017lwo} also notices this parameter degeneracy (see Appendix.~A of Ref.~\cite{Murgia:2017lwo}).
Meanwhile, Ref.~\cite{Murgia:2018now} reports that the combination of $|\beta \times \gamma|$ is well constrained by observational data (see Fig.~4 of Ref.~\cite{Murgia:2018now}), while the orthogonal direction $\beta/\gamma$ is not very sensitive.
Therefore, in this paper, we fix this orthogonal direction by the relation
\begin{align}
\gamma
&= 
- \beta \,.
\end{align}
As a result, the output $\vec{x}_{\rm out}$ becomes a two-component vector.

%%%%%%%%%%%%%%%%%%%%%%%%%%%%%%%%%%%%%%%%%%%%%%%%%%
\subsubsection*{Case A: Decay with entropy production}
%%%%%%%%%%%%%%%%%%%%%%%%%%%%%%%%%%%%%%%%%%%%%%%%%%

Let us first take Case A (see Sec.~\ref{subsec:modelsetup}).
We identify the output $\vec{x}_{\rm out}$ and input $\vec{x}_{\rm in}$ as
\begin{align}
\vec{x}_{\rm in}
&= 
\left(
\begin{matrix}
\displaystyle 
\frac{\displaystyle \log_{10} \left( 1 - \frac{m_{2}}{m_{1}} \right) - (\vec{x}_{{\rm in},0})_1}{(\vec{\sigma}_{\rm in})_1} 
\\[2.5ex]
\displaystyle 
\frac{\log_{10} m_{\rm DM}\,{\rm [keV]} - (\vec{x}_{{\rm in},0})_2}{(\vec{\sigma}_{\rm in})_2} 
\\[2.5ex]
\displaystyle 
\frac{\log_{10} \Delta - (\vec{x}_{{\rm in},0})_3}{(\vec{\sigma}_{\rm in})_3}
\end{matrix}
\right) \,,
~~~~
\vec{x}_{\rm out}
= 
\left(
\begin{matrix}
\displaystyle 
\frac{\log_{10} \alpha - (\vec{x}_{{\rm out},0})_1}{(\vec{\sigma}_{\rm out})_1} 
\\[2.5ex]
\displaystyle 
\frac{\log_{10} \beta - (\vec{x}_{{\rm out},0})_2}{(\vec{\sigma}_{\rm out})_2} 
\end{matrix}
\right) \,.
\end{align}
Here $\vec{x}_{{\rm in},0}$ and $\vec{x}_{{\rm out},0}$ are the means of the input and output data, respectively,
while $\vec{\sigma}_{\rm in}$ and $\vec{\sigma}_{\rm out}$ are the standard deviations.
These are constant vectors introduced to normalize the data and make learning more efficient.

For the dataset, we sample about $20,000$ data points 
from $0 < m_2 / m_1 < 1$, $1 \leq m_{\rm DM}\,{\rm [keV]} \leq 20$, and $0.1 \leq \Delta \leq 10$.
We exclude data points in the gray-shaded regions of Figs.~\ref{fig:NSatNN} and \ref{fig:deltaANN},
and thus the resulting neural networks cannot be used for the input parameters in these regions.%
\footnote{
The reason for excluding the gray-shaded regions is as follows.
For Case A, the right-top corner of the parameter space corresponds to the CDM limit.
Since the transfer function approaches unity in this region, 
the parameter set $\{ \alpha, \beta, \gamma \}$ are not uniquely determined by fitting even after $\gamma = -\beta$ is imposed.
For Case B, the left-top corner corresponds to the large Yukawa coupling limit and thus the perturbative Unitarity violation problem arises.
}

%%%%%%%%%%%%%%%%%%%%%%%%%%%%%%%%%%%%%%%%%%%%%%%%%%
\subsubsection*{Case B: Decay with scattering}
%%%%%%%%%%%%%%%%%%%%%%%%%%%%%%%%%%%%%%%%%%%%%%%%%%

Next let us take Case B (see Sec.~\ref{subsec:modelsetup}).
We identify the output $\vec{x}_{\rm out}$ and input $\vec{x}_{\rm in}$ as
\begin{align}
\vec{x}_{\rm in}
&= 
\left(
\begin{matrix}
\displaystyle 
\frac{\displaystyle \log_{10} \left( 1 - \frac{m_{2}}{m_{1}} \right) - (\vec{x}_{{\rm in},0})_1}{(\vec{\sigma}_{\rm in})_1} 
\\[2.5ex]
\displaystyle 
\frac{\displaystyle \log_{10} \frac{Y_{\rm scat}}{Y_{\rm total}} - (\vec{x}_{{\rm in},0})_2}{(\vec{\sigma}_{\rm in})_2} 
\\[2.5ex]
\displaystyle 
\frac{\log_{10} m_{\rm DM} - (\vec{x}_{{\rm in},0})_3}{(\vec{\sigma}_{\rm in})_3}
\end{matrix}
\right) \,,
~~~~
\vec{x}_{\rm out}
= 
\left(
\begin{matrix}
\displaystyle 
\frac{\log_{10} \alpha - (\vec{x}_{{\rm out},0})_1}{(\vec{\sigma}_{\rm out})_1} 
\\[2.5ex]
\displaystyle 
\frac{\log_{10} \beta - (\vec{x}_{{\rm out},0})_2}{(\vec{\sigma}_{\rm out})_2} 
\end{matrix}
\right) \,.
\end{align}
For the dataset, we sample about $20,000$ data points 
from $0.5 < m_2 / m_1 < 0.9$, $0 \leq Y_{\rm scat} / Y_{\rm total} \leq 0.7$, and $1 \leq m_{\rm DM}\,{\rm [keV]} \leq 15$.
We again exclude data points in the gray-shaded regions of Figs.~\ref{fig:NSatNN} and \ref{fig:deltaANN}.

%%%%%%%%%%%%%%%%%%%%%%%%%%%%%%%%%%%%%%%%%%%%%%%%%%
\subsection{$\{\alpha, \beta, \gamma\}$ $\to$ Observables}
\label{subsec:NN2}
%%%%%%%%%%%%%%%%%%%%%%%%%%%%%%%%%%%%%%%%%%%%%%%%%%

We next construct a neural network that maps $\{ \alpha, \beta, \gamma \}$
onto the observables, more specifically, $N_{\rm sat}$ and $\delta A$ introduced in Sec.~\ref{sec:rev}. 
We identify the input and output as
\begin{align}
\vec{x}_{\rm in}
&= 
\left(
\begin{matrix}
\displaystyle 
\frac{\log_{10} \alpha - (\vec{x}_{{\rm in},0})_1}{(\vec{\sigma}_{\rm in})_1} 
\\[2.5ex]
\displaystyle 
\frac{\log_{10} \beta - (\vec{x}_{{\rm in},0})_2}{(\vec{\sigma}_{\rm in})_2} 
\\[2.5ex]
\displaystyle 
\frac{\log_{10} (- \gamma) - (\vec{x}_{{\rm in},0})_3}{(\vec{\sigma}_{\rm in})_3} 
\end{matrix}
\right) \,,
\end{align}
\begin{align}
\vec{x}_{\rm out}
&= 
\left(
\begin{matrix}
\displaystyle 
\frac{\log_{10} N_{\rm sat} - (\vec{x}_{{\rm out},0})_1}{(\vec{\sigma}_{\rm out})_1} 
\end{matrix}
\right)
~~~~{\rm or}~~~~
\left(
\begin{matrix}
\displaystyle 
\frac{\log_{10} \delta A - (\vec{x}_{{\rm out},0})_1}{(\vec{\sigma}_{\rm out})_1} 
\end{matrix}
\right).
\end{align}
Note that we do not assume $\gamma = -\beta$ in contrast to the previous subsection,
and thus $\vec{x}_{\rm in}$ is a three-component vector.
This is to accommodate broader class of models than the models we adopt in this paper.
Also note that $\vec{x}_{\rm out}$ is a one-component vector,
which means that we construct neural networks for ``$\{ \alpha, \beta, \gamma\} \to N_{\rm sat}$"
and for ``$\{ \alpha, \beta, \gamma\} \to \delta A$" separately.

For the dataset, we sample about $70,000$ points from 
$0.001 \leq \alpha \leq 0.1$, $0.1 \leq \beta \leq 10$, and $0.1 \leq \gamma \leq 10$.

%%%%%%%%%%%%%%%%%%%%%%%%%%%%%%%%%%%%%%%%%%%%%%%%%%
\subsection{Combined results}
\label{subsec:results}
%%%%%%%%%%%%%%%%%%%%%%%%%%%%%%%%%%%%%%%%%%%%%%%%%%

Before combining the two neural networks constructed in the previous subsections, we remark that we discuss details about the precision of the neural network in Appendix~\ref{app:precision}.
We provide the resultant neural network parameters through the {\tt arXiv} website.
See Appendix~\ref{app:howto} for further explanation of the data files.
We also provide a {\tt Mathematica} file ({\tt freeze-in.nb}) for illustration.

\begin{figure}
\begin{center}
\includegraphics[width=0.4\columnwidth]{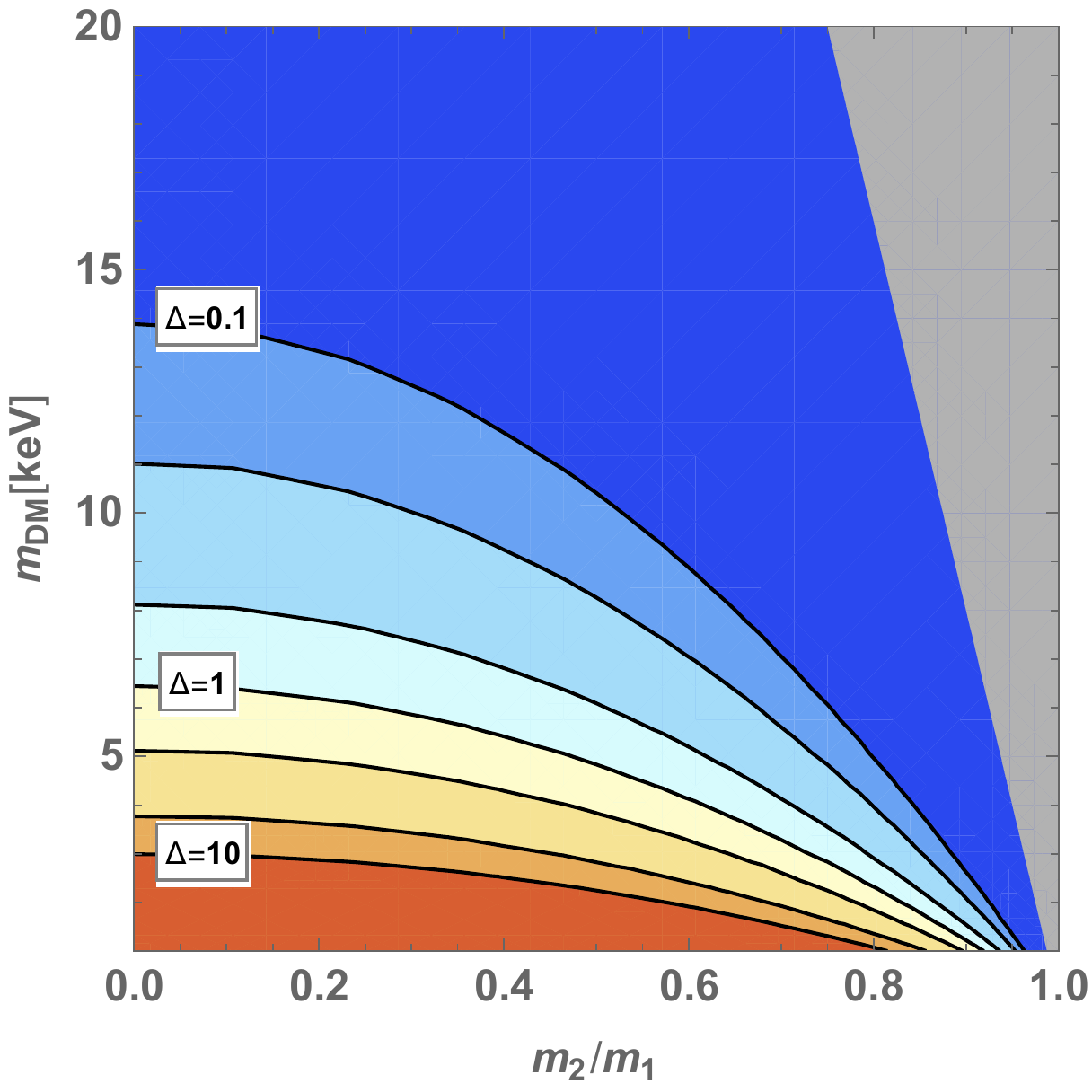}
\hskip 1cm
\includegraphics[width=0.4\columnwidth]{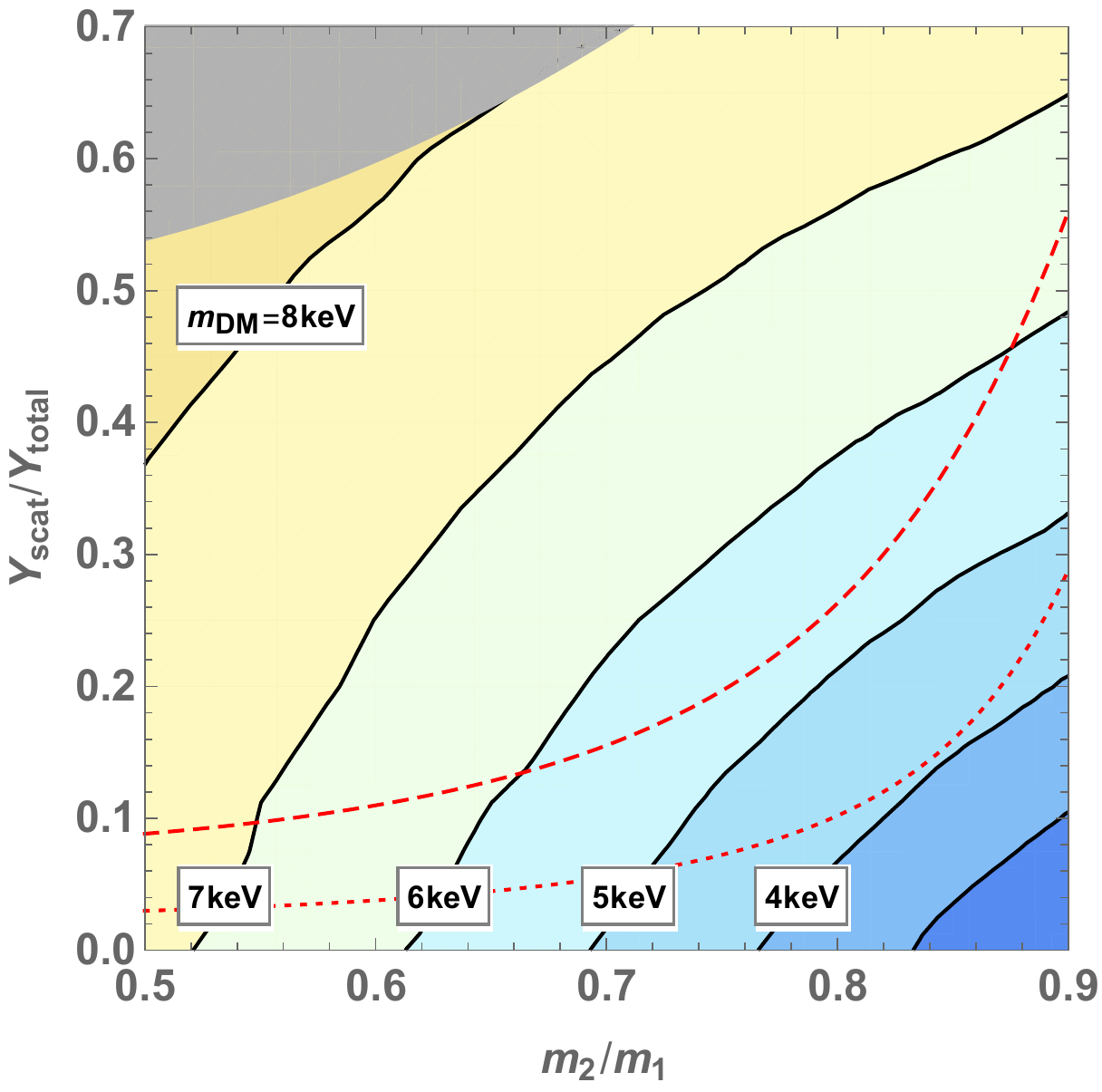}
\vskip 5mm
\includegraphics[width=0.4\columnwidth]{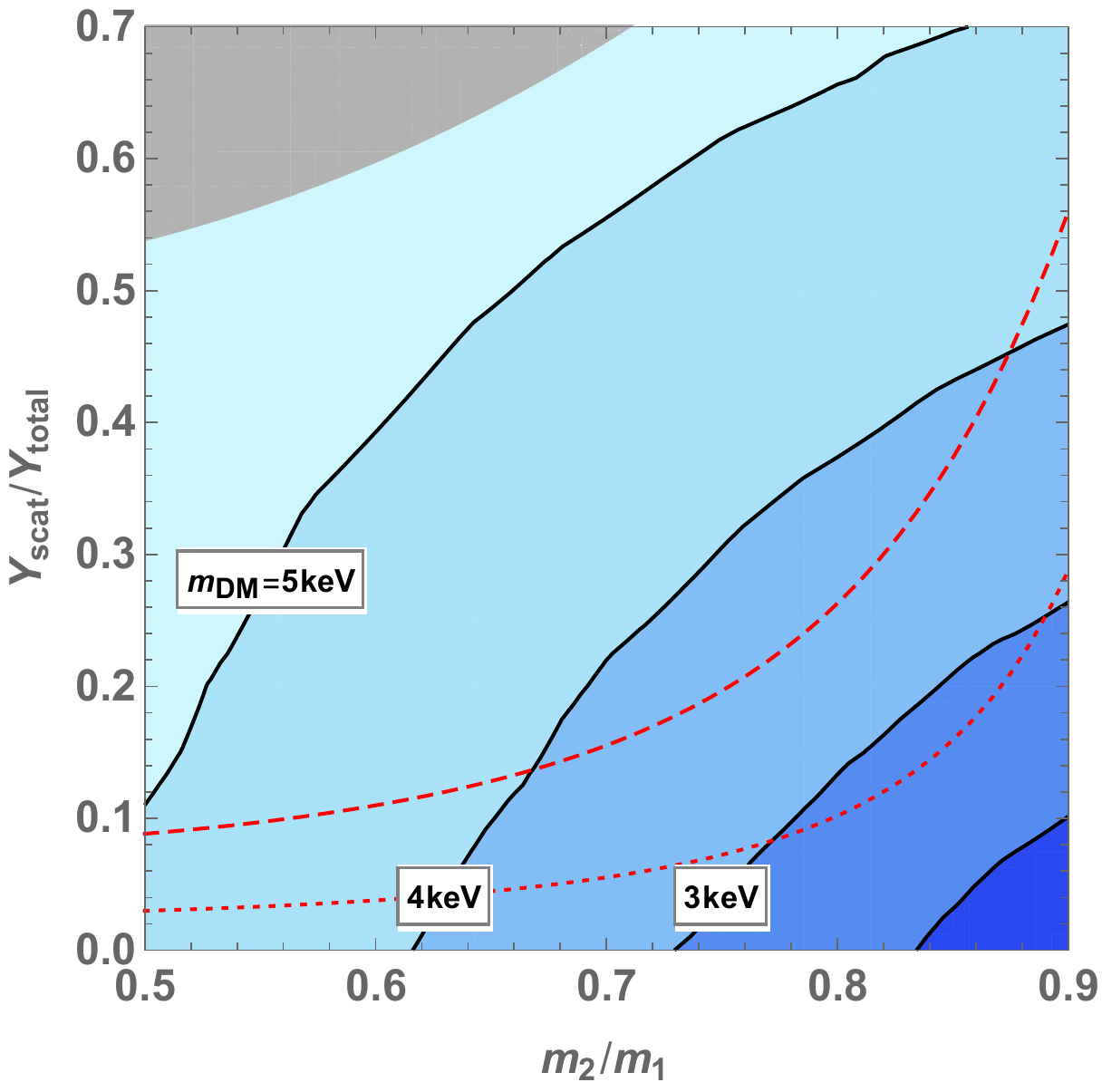}
\hskip 1cm
\includegraphics[width=0.4\columnwidth]{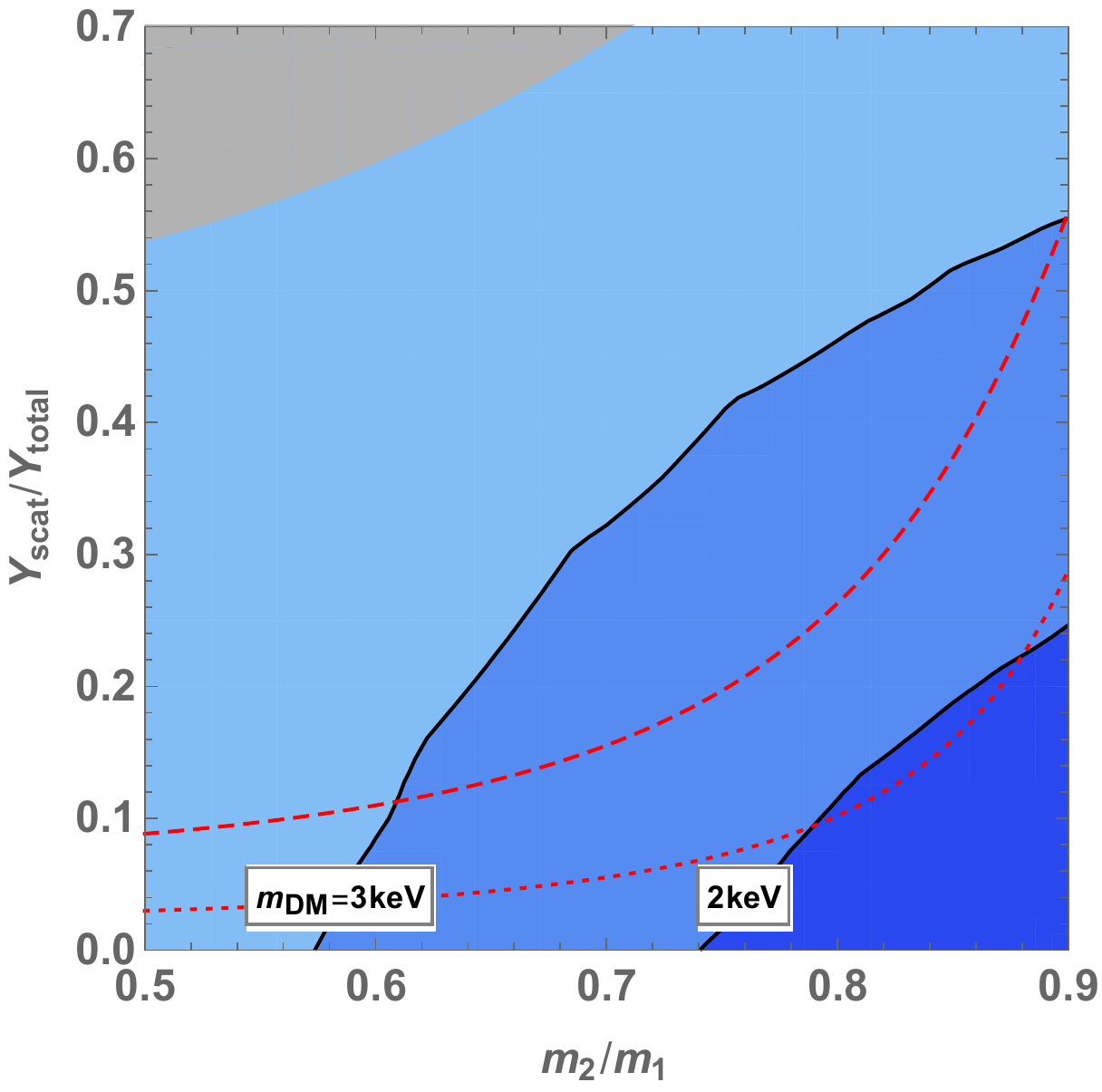}
\caption {\small
Constraints from $N_{\rm sat} > N_{\rm sat}^{\rm obs}$ reproduced by the neural network.
Compare this figure with Fig.~\ref{fig:NSat}.
}
\label{fig:NSatNN}
\end{center}
\end{figure}

\begin{figure}
\begin{center}
\includegraphics[width=0.4\columnwidth]{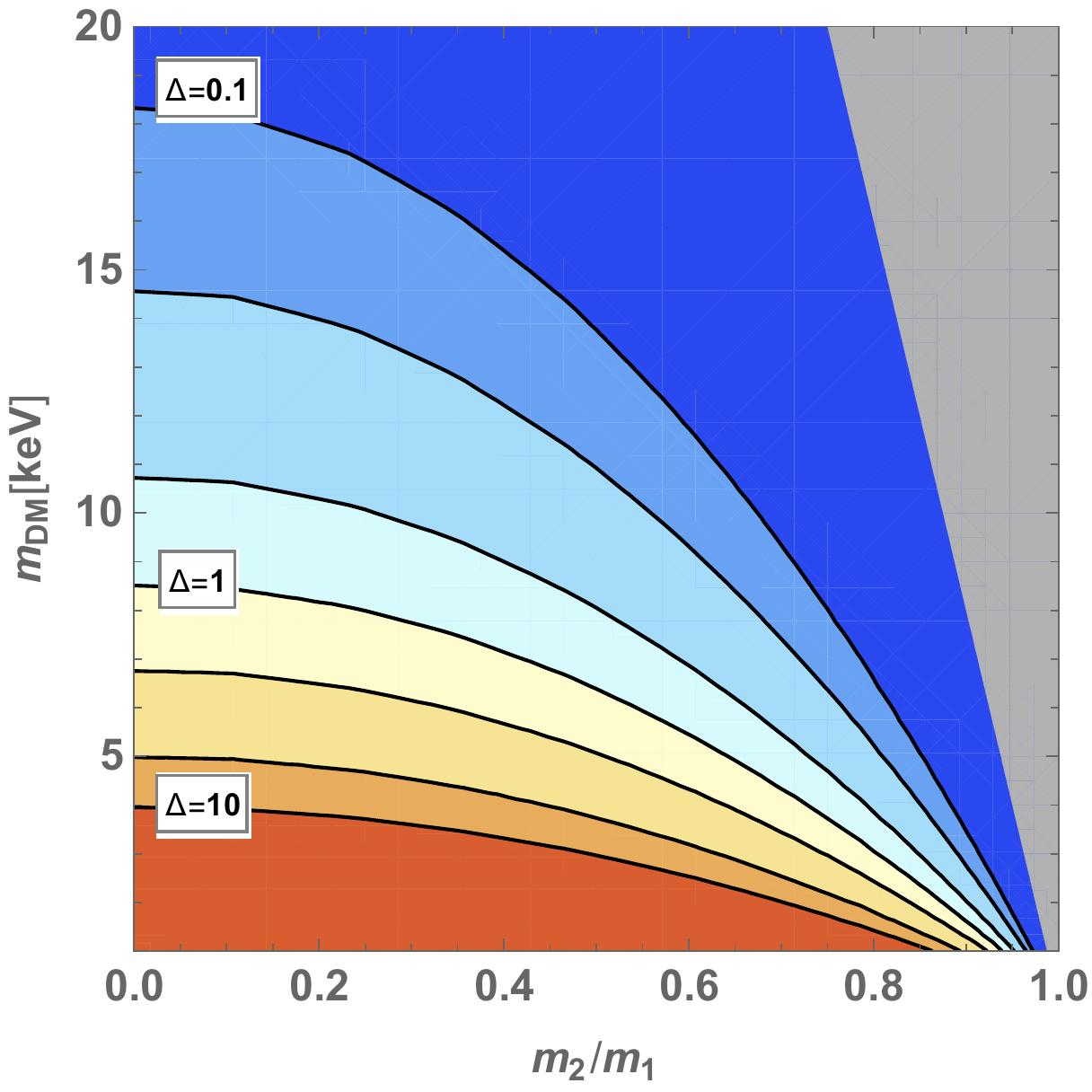}
\hskip 1cm
\includegraphics[width=0.4\columnwidth]{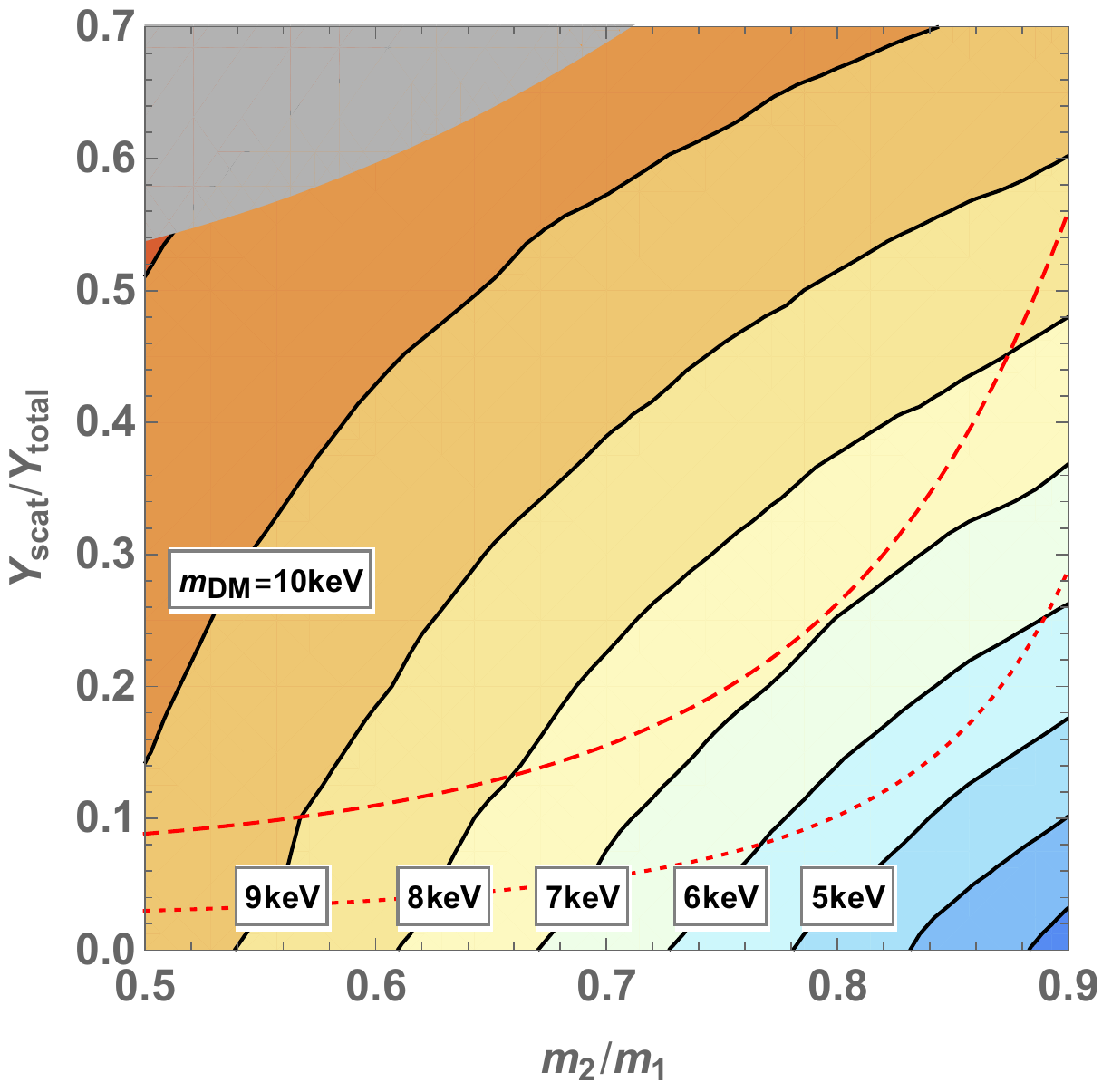}
\vskip 5mm
\includegraphics[width=0.4\columnwidth]{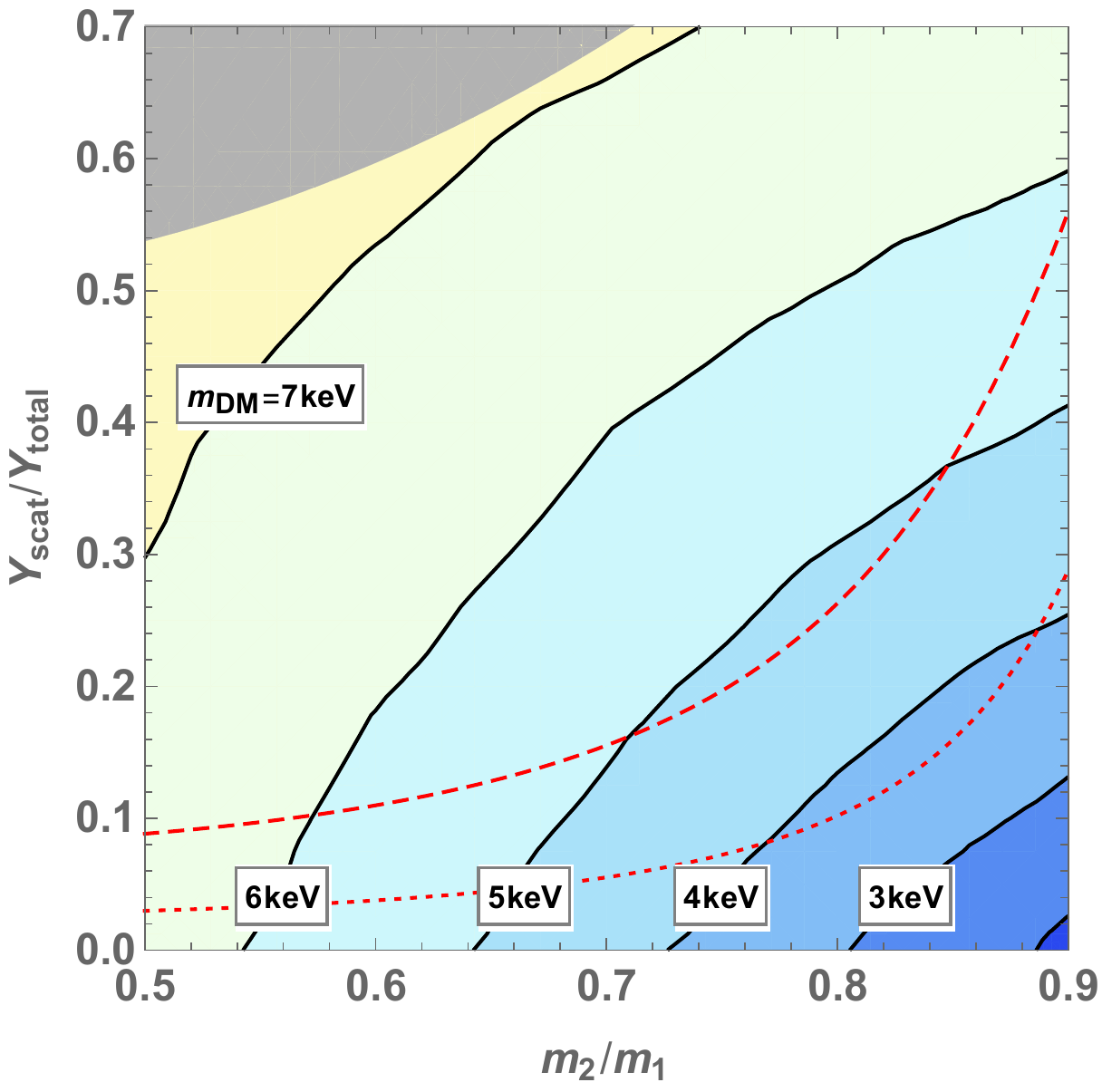}
\hskip 1cm
\includegraphics[width=0.4\columnwidth]{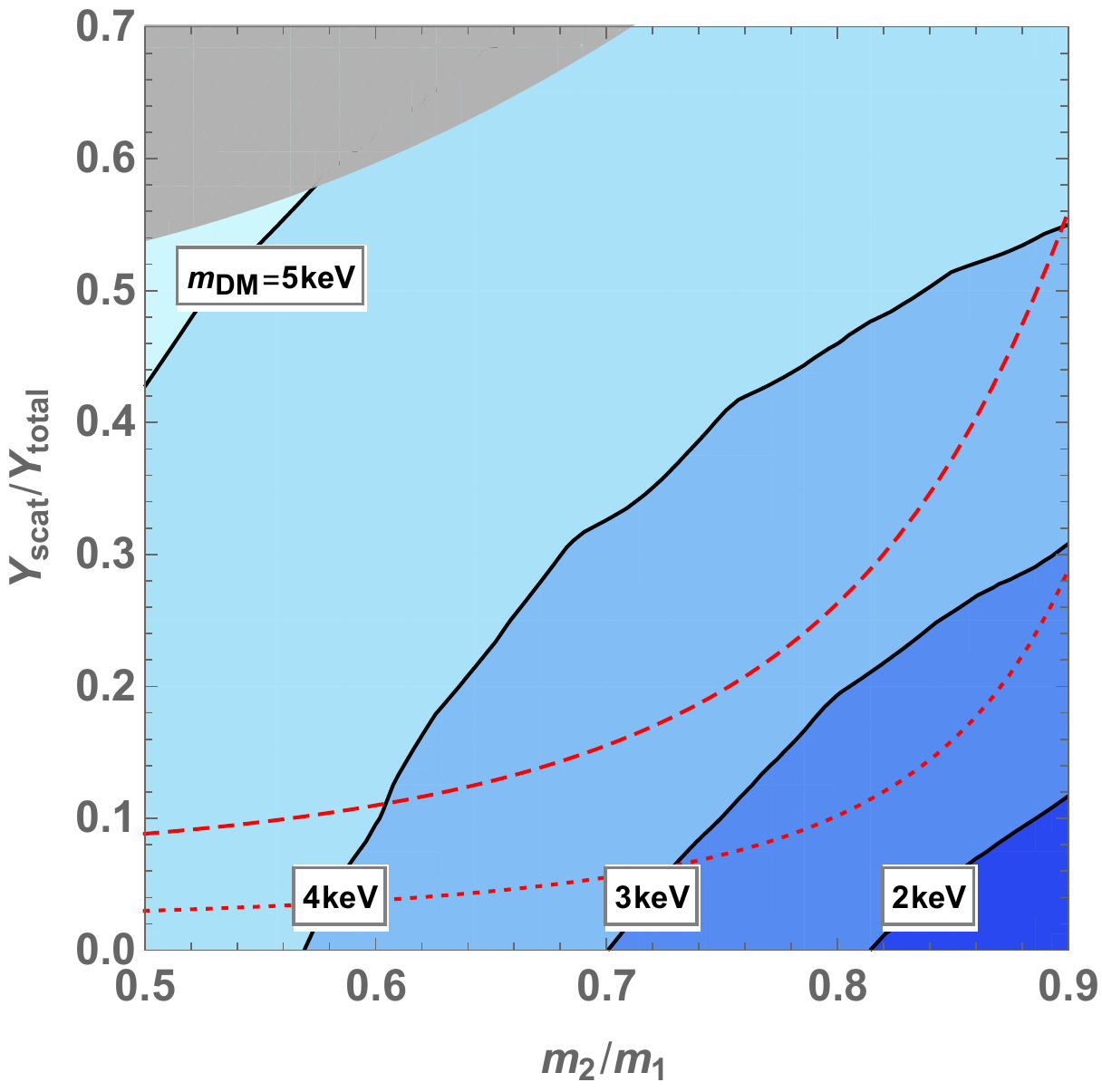}
\caption {\small
Constraints from $\delta A <  \delta A_{3.5 \, {\rm keV}}$ reproduced by the neural network. 
Compare this figure with Fig.~\ref{fig:deltaA}.
}
\label{fig:deltaANN}
\end{center}
\end{figure}

Now let us check the precision of the neural network by combining the two neural networks.
The results should coincide with the constraints obtained in Sec.~\ref{sec:model} as long as the neural networks work well.
Figs.~\ref{fig:NSatNN} and \ref{fig:deltaANN} are the constraints from $N_{\rm sat} > N_{\rm sat}^{\rm obs}$ and $\delta A <  \delta A_{3.5 \, {\rm keV}}$ derived through the combination of the two neural networks and thus should be compared with Figs.~\ref{fig:NSat} and \ref{fig:deltaA}, respectively.
We see that the neural networks nicely reproduce the original constraints.

We again stress constructing nonlinear maps for
``Model parameters $\to$ Linear matter power spectrum" and for ``Linear matter power spectrum$\to$ Observables" separately is very useful and time-saving:
given the common language of $\{ \alpha, \beta, \gamma \}$,
those interested in particle physics models can provide $\{ \alpha, \beta, \gamma \}$ as functions of model parameters,
while those who reports observational constraints can update the constraints in terms of $\{ \alpha, \beta, \gamma \}$.
Neural network technique provides us with a ready-to-use format for this procedure.

%%%%%%%%%%%%%%%%
%%%%%%%%%%%%%%%%

%%%%%%%%%%%%%%%%
%%%%%%%%%%%%%%%%

%%%%%%%%%%%%%%%%%%%%%%%%%%%%%%%%%%%%%%%%%%%%%%%%%%
\section{Summary}
\label{sec:sum}
\setcounter{equation}{0}
%%%%%%%%%%%%%%%%%%%%%%%%%%%%%%%%%%%%%%%%%%%%%%%%%%

Galactic-scale structure formation of the Universe is of particular interest in DM research.
Beyond-WIMP scenarios alter galactic-scale structure formation, while conventional WIMP DM behaves as CDM.
Precise measurement of galactic-scale structure in near-future observations may hint beyond-WIMP scenarios.
On the other hand, here is a practical bottleneck.
Impacts of beyond-WIMP scenarios on galactic-scale structure vary model by model.
In principle, one has to repeat the two-step procedure on a model-by-model basis:
\begin{itemize}
\item[]
\begin{center}
Model $\to$ Linear matter power spectrum $\to$ Observables,
\end{center}
\end{itemize}
which is sketched by the blue flow in Fig.~\ref{fig:Schematic}.
Each step requires different disciplines and dedicated computations.
Following this procedure in the model-by-model basis is very time-consuming.

We may improve the situation by characterizing the transfer function ({\it i.e.}, the linear matter power spectrum) with some parameter.
One (likely particle physicist) calculates the transfer function parameter as a function of model parameters.
Another reports observational constraints in terms of the transfer function parameter.
Now we can get constraints on the model parameters very easily by combining the two results.
Although a single-parameter characterization (the thermal WDM mass $m_{\rm WDM}$) has been conventionally used, 3-parameter characterization is proposed to cover a wide range of beyond-WIMP scenarios.
Our main stress is that neural network technique facilitates sharing results from one side to another by providing the results in a ready-to-use format.

We devoted this paper to demonstrating how we can actually work out with $\{ \alpha, \beta, \gamma \}$ and a neural network.
To be specific, we considered a simplified model of light (keV-scale) FIMP DM
Freeze-in production from $2$-body decay gives a main contribution to the relic abundance.
We also took into account entropy production after the decoupling and freeze-in production from scattering.
We constructed first a map between the FIMP model parameters and $\{ \alpha, \beta, \gamma \}$ and next a map between $\{ \alpha, \beta, \gamma \}$ and the observables, {\it i.e.}, the number of satellite galaxies and Lyman-$\alpha$ forest, by adopting neural network technique.
We provided the constructed maps in a ready-to-use format through the {\tt arXiv} website.
Meanwhile, we performed the conventional procedure to derive the direct constraints on the FIMP model parameters.
The constraints derived through $\{ \alpha, \beta, \gamma \}$ and a neural network are in good agreement with those derived through the conventional procedure.

Although we focused on a simplified model of FIMP DM in this paper, it is worth performing a similar study in other FIMP models such as sterile neutrino DM and superWIMP DM and also in other alternatives to CDM such as Fuzzy DM and late kinetic decoupling of DM.
Our suggestion will facilitate comparison between beyond-WIMP models and future updates of constraints on galactic-scale structure formation, {\it e.g.}, from redshifted 21cm surveys.

%%%%%%%%%%%%%%%%%%%%%%%%%%%%%%%%%%%%%%%%%%%%%%%%%%
\section*{Acknowledgments}
%%%%%%%%%%%%%%%%%%%%%%%%%%%%%%%%%%%%%%%%%%%%%%%%%%

The work of KJB, RJ, and AK was supported by IBS under the project code, IBS-R018-D1.
The work of RJ was supported by Grants-in-Aid for JSPS Overseas Research Fellow (No. 201960698).
The work of RJ was supported by the Deutsche Forschungsgemeinschaft 
under Germany's Excellence Strategy -- EXC 2121 ,,Quantum Universe`` -- 390833306.
The work of KY was supported by JSPS KAKENHI Grant Number JP18J10202.

\clearpage

\appendix

%%%%%%%%%%%%%%%%%%%%%%%%%%%%%%%%%%%%%%%%%%%%%%%%%%
\section{Comparison with an analytic map}
\label{app:analytic}
%%%%%%%%%%%%%%%%%%%%%%%%%%%%%%%%%%%%%%%%%%%%%%%%%%

In this appendix we derive constraints on our FIMP model parameters by converting the thermal WDM mass $m_{\rm WDM}$.
Proposed ways of converting $m_{\rm WDM}$ onto a given model are as follows:
\begin{itemize}
\item[--]
One compares the characteristic quantity such as the free-streaming length~\cite{Bond:1982uy, Bond:1983hb} 
and Jeans length~\cite{Bond:1980ha, Davis:1981yf, Peebles:1982ib} between a given WDM model 
and the conventional thermal WDM model.
If the free-streaming length in the given model is larger than that in the conventional thermal WDM model with an observational lower bound on $m_{\rm WDM}$, the given model is regarded as disfavored by the same observation.
See Ref.~\cite{Kamada:2013sh} for comparison of the transfer function in different WDM models with the Jeans length fixed.

\item[--]
One compares the transfer function $T^{2} (k)$ below some critical wavenumber between a given model and the conventional thermal WDM model.
If $T^{2} (k)$ in the given model is smaller in amplitude than that in the conventional thermal WDM model with a n observational lower bound on $m_{\rm WDM}$, the given model is regarded as disfavored by the same observation.
A suggested choice of the critical wavenumber is the half mode $k_{1/2}$ 
where $T^2_{\rm WDM} (k_{1/2}) = 1/2$~\cite{Konig:2016dzg}.
\end{itemize}

In this appendix, we adopt a ``warmness" quantity (equivalently, the Jeans length) calculated from a DM phase space distribution~\cite{Kamada:2013sh}:
\begin{align}
  \label{eq:warmness}
  \sigma \equiv \frac{\sqrt{\langle{p^2}\rangle}}{m_\chi} = \tilde{\sigma} \times \frac{T_{\rm DM}}{m_\chi}\,,
\end{align}
where $\langle{p^2}\rangle$ is the 2nd moment of the DM phase space distribution and thus
\begin{align}
  \label{eq:sigmatilde}
  \tilde\sigma^2 =
  \frac{\int d^3q\, q^2 f(q)}{\int d^3q\,f(q)}\,.
\end{align}
$\tilde\sigma$ depends on the shape of the phase space distribution.
For a given observational lower bound on $m_{\mathrm{WDM}}$, 
a WDM model is regarded as disfavored by the same observation, if $\sigma > \sigma_{m_{\mathrm{WDM}}}$.
Using the definition of DM temperature given by Eq.~\eqref{eq:nonthermal-temp}, we obtain the constraint on a FIMP as
\begin{align}
  \label{eq:m-chi-limit}
  m > 7 \, {\rm keV} \left(\frac{m_{\mathrm{WDM}}}{2.5 \, {\rm keV}}\right)^{4/3}\left(\frac{\tilde\sigma}{3.6}\right)\left(\frac{106.75}{g_* (T_{\rm dec})}\right)^{1/3}\,.
\end{align}
Note that in the conventional thermal WDM model, WDM particles follows the Fermi-Dirac distribution, and thus  $\tilde{\sigma}_{\mathrm{WDM}} \simeq 3.6$.

In our simplified FIMP model, the phase space distribution can be expressed analytically~\cite{KY}, and thus $\tilde\sigma$ is also analytically derivable.
As a result, we can construct an analytic map between $m_{\rm WDM}$ and the model parameters.
The total $\tilde\sigma$ is calculated from each production process as
\begin{align}
  \label{eq:sigmatile-total}
  \tilde\sigma^2
  =
  \frac{Y_{\rm dec}}{Y_{\rm total}}\tilde\sigma^2_{\rm dec}
  +
  \frac{Y_{\text{scat,\,t-ch}}}{Y_{\rm total}}\tilde\sigma^2_{\text{scat,\,t-ch}}
  +
  \frac{Y_{\text{scat,\,s-ch}}}{Y_{\rm total}}\tilde\sigma^2_{\text{scat,\,s-ch}}\,,
\end{align}
where each $\tilde\sigma^2$ is calculated analytically as
\begin{align}
  \tilde\sigma^2_{\rm dec}
  &=
  \frac{35}{4}(1-r^2)^2\,,
  \label{eq:sigmatilde-2body-1}
  \\
  \tilde\sigma^2_{\text{scat,\,t-ch}}
  &=
  \frac{35}{4}\,,
  \label{eq:sigmatilde-tch}
  \\
  \tilde\sigma^2_{\text{scat,\,s-ch}}
  &=
    \frac{7(105r-265r^3+191r^5-15r^7-15(1-r^2)^3(7+r^2)\tanh^{-1}r)}{12r^4(r(3-r^2)+(-3+2r^2+r^4)\tanh^{-1}r)}\,,
  \label{eq:sigmatilde-sch}
\end{align}
and each FIMP yield is also obtained as
\begin{align}
   Y_{\rm dec}
  &= 2\times\frac{3y_\chi^2 M_0}{32\pi^2 m_\Psi}
  \left( 1 -r^2 \right)^2\,,
   \label{eq:yield-2body-anal}\\
  Y_{\text{scat,\,t-ch}}
  &=
  4\times \frac{3N_fy_\chi^2y_f^2M_0}{128\pi^4m_\Psi}\cdot
  \frac{(2-r^2)\tanh^{-1}\sqrt{1-r^2} -\sqrt{1-r^2}}{3(1-r^2)^{3/2}}\,,
  \label{eq:yield-tch-anal}\\
  Y_{\text{scat,\,s-ch}}
  &=
  2\times \frac{3N_fy_\chi^2y_f^2M_0}{128\pi^4m_\Psi}\cdot
  \frac{r(3-r^2)+(-3+2r^2+r^4)\tanh^{-1}(r)}{2r^5}\,.
    \label{eq:yield-s-anal}
\end{align}
Here $r \equiv m_2 / m_1$, prefactors count a number of particle spieces ($\Psi/{\bar \Psi}$ and $f/\bar f$), and $M_{0}$ is a dimensionful constant whose expression is not relevant in this appendix.

\begin{figure}
\begin{center}
\includegraphics[width=0.4\columnwidth]{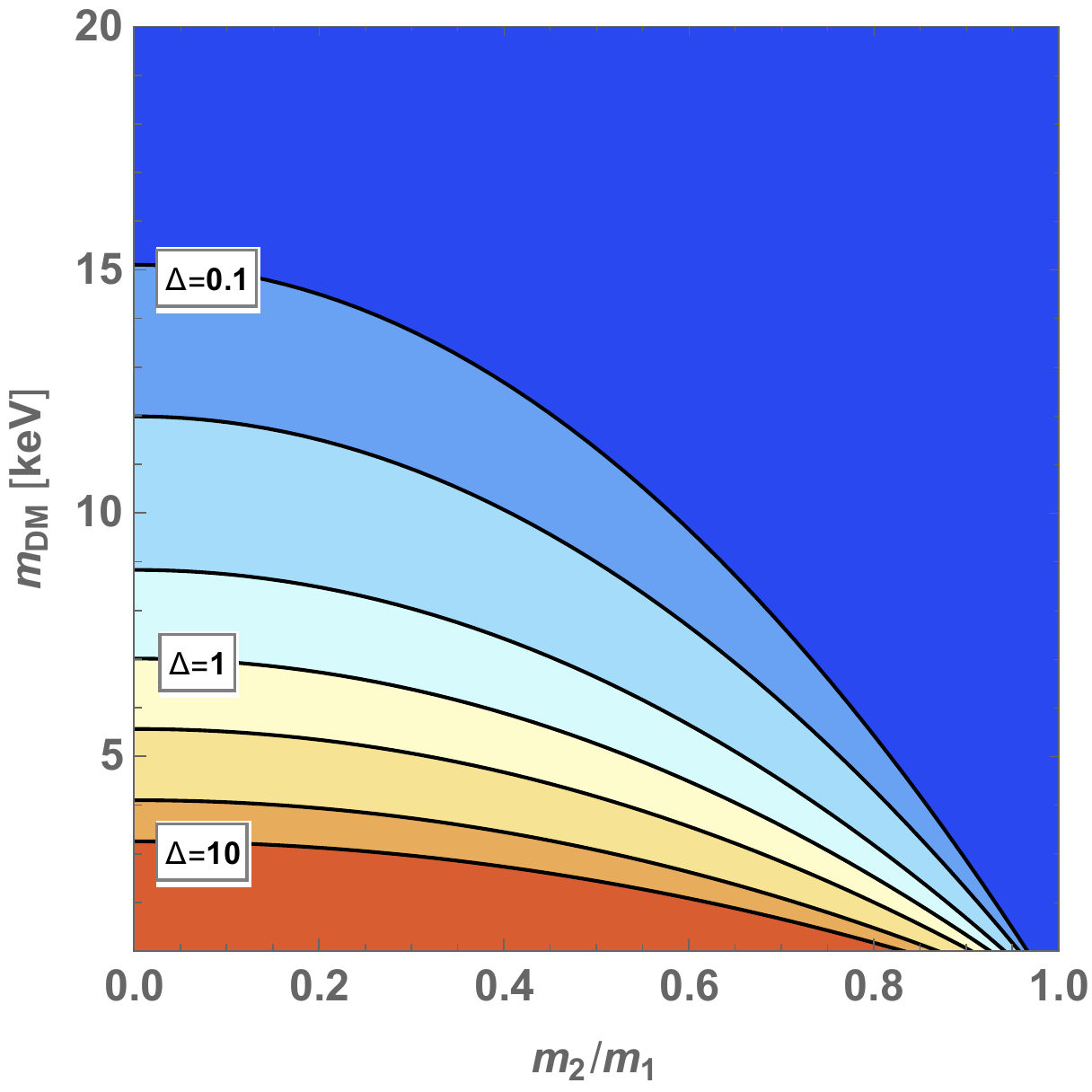}
\hskip 1cm
\includegraphics[width=0.4\columnwidth]{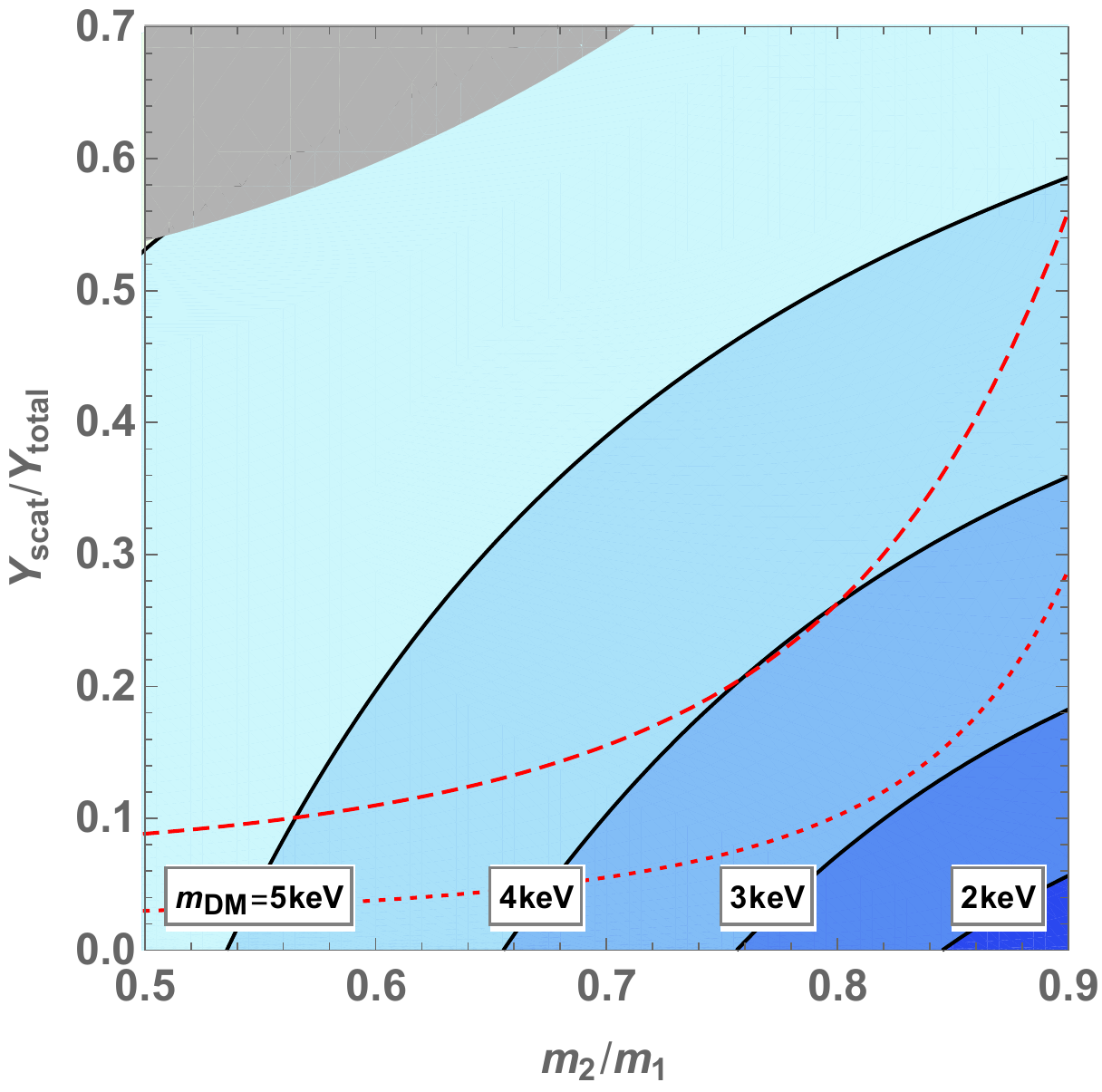}
\caption {\small
Constraints on Case A (left) and Case B with $\Delta = 1$ (right) from $\sigma < \sigma_{2.9\,{\rm keV}}$, where $\sigma_{2.9\,{\rm keV}}$ denotes the warmness of 2.9\,keV conventional WDM.
The left and right panels should be compared with the top-left and bottom-left panels of Fig.~\ref{fig:NSat}, respectively.
}
\label{fig:sigma2.9}
\end{center}
\vskip 1cm
\begin{center}
\includegraphics[width=0.4\columnwidth]{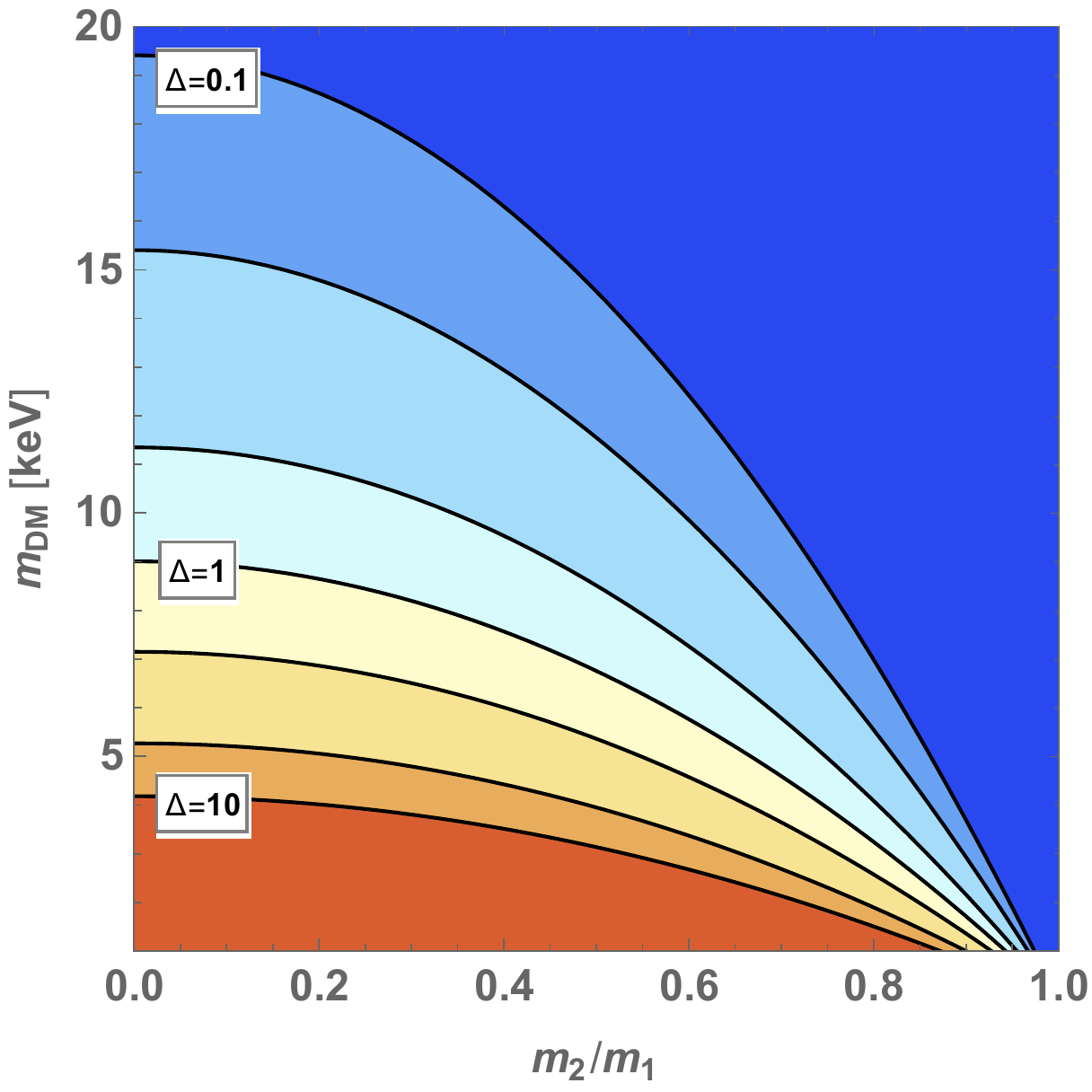}
\hskip 1cm
\includegraphics[width=0.4\columnwidth]{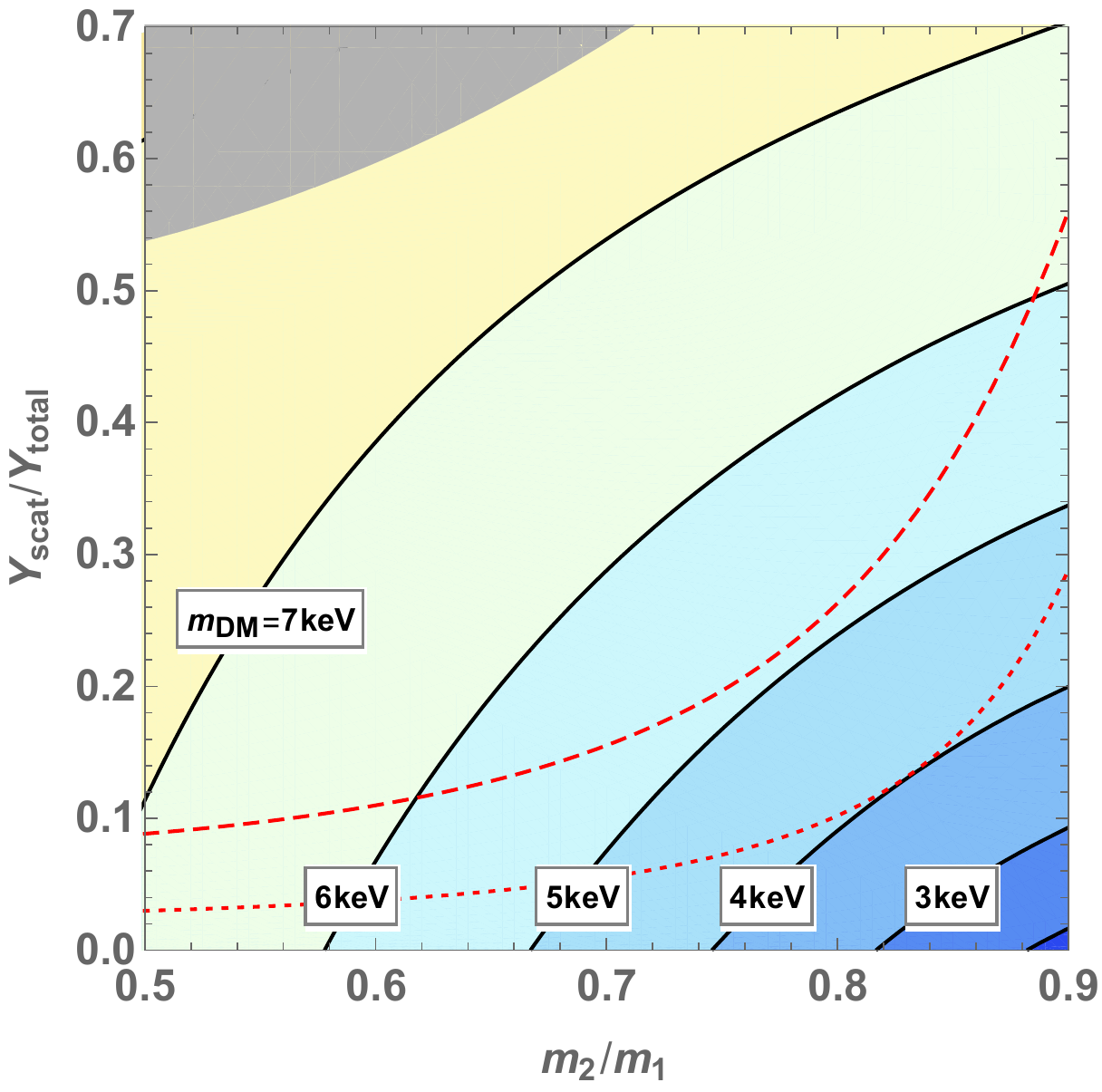}
\caption {\small
Constraints on Case A (left) and Case B with $\Delta = 1$ (right) from $\sigma < \sigma_{3.5\,{\rm keV}}$, where $\sigma_{3.5\,{\rm keV}}$ denotes the warmness of 3.5\,keV conventional WDM.
This figure should be compared with the top-left and bottom-left panels of Fig.~\ref{fig:deltaA}.
}
\label{fig:sigma3.5}
\end{center}
\end{figure}

In this way, we derive the constraints on our simplified FIMP model from $m_{\rm WDM}$ through warmness.
First, Fig.~\ref{fig:sigma2.9} shows constraints from the observed number of Milky Way satellites, $\sigma > \sigma_{2.9 \, {\rm keV}}$.
$m_{\rm WDM} > 2.9$\,keV corresponds to $N_{\rm sat} > N_{\rm sat}^{\rm obs}$.
The left and right panels are for Case A (Decay with entropy production) and Case B (Decay with scattering) with $\Delta = 1$ and should be compared with the top-left and top-right panels of Fig.~\ref{fig:NSat}, respectively.
We see the results are $\sim 10\%$ different with each other, while are qualitatively equivalent.

Next, Fig.~\ref{fig:sigma3.5} shows constraints from the Lyman-$\alpha$ forest data, $\sigma > \sigma_{3.5 \, {\rm keV}}$.
$m_{\rm WDM} > 3.5$\,keV corresponds to $\delta A <  \delta A_{3.5 \, {\rm keV}}$.
The left and right panels are for Case A and Case B with $\Delta = 1$ and should be compared with the top-left and top-right panels of Fig.~\ref{fig:deltaA}, respectively.
We see the derived results are $\sim 10\%$ different from those from the direct modeling in Sec.~\ref{subsec:NNraints}, as for $N_{\rm sat} > N_{\rm sat}^{\rm obs}$.%
\footnote{
We also derive the constraints through $k_{1/2}$.
The derived constraints are again $\sim 10\%$ different from those from the direct modeling.
}

%%%%%%%%%%%%%%%%
%%%%%%%%%%%%%%%%

%%%%%%%%%%%%%%%%%%%%%%%%%%%%%%%%%%%%%%%%%%%%%%%%%%
\section{Further check: original data vs neural network}
\label{app:precision}
%%%%%%%%%%%%%%%%%%%%%%%%%%%%%%%%%%%%%%%%%%%%%%%%%%

In this appendix, we take a closer look at the difference between the original data and neural network.

\begin{figure}
\begin{center}
\includegraphics[width=0.48\columnwidth]{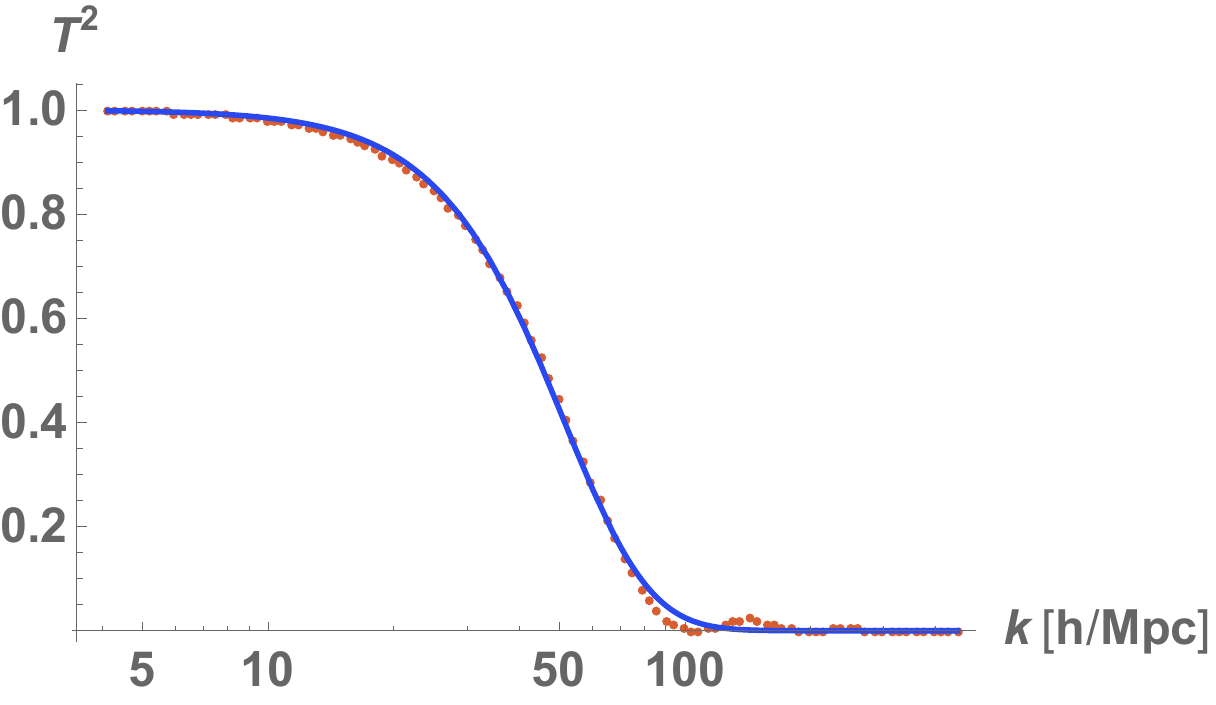}
\includegraphics[width=0.48\columnwidth]{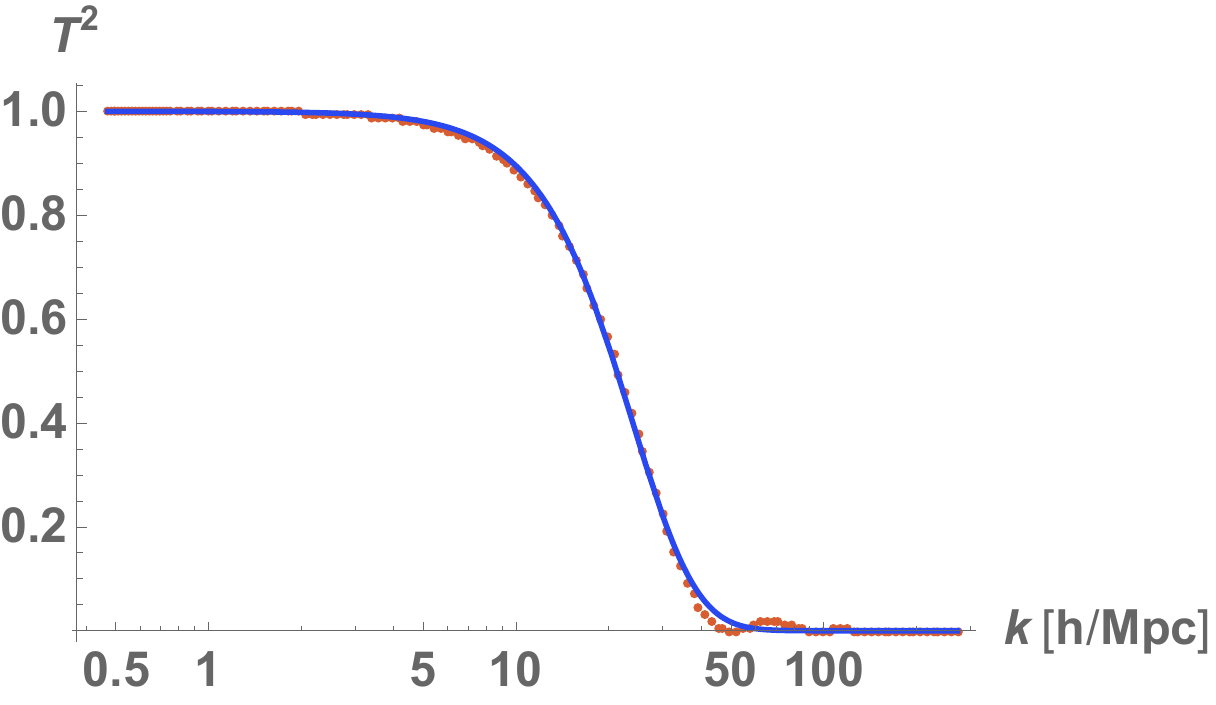}
\caption {\small
Comparison between original data (red) and fitting function with $\alpha$ and $\beta$ given by the neural network (blue).
We show two benchmark points.
{\bf Left:} Case A (Decay with entropy production) with 
$m_2 / m_1 = 0.499$, $m_{\rm DM} = 10.256$\,keV, and $\Delta = 1$.
{\bf Right:} Case B (Decay with scattering) with 
$m_2 / m_1 = 0.701$, $Y_{\rm scat} / Y_{\rm tot} = 0.367$, $m_{\rm DM} = 4$\,keV, and $\Delta = 1$.
}
\label{fig:T2}
\end{center}
\end{figure}

First we check validity of the $\{ \alpha, \beta, \gamma \}$ parametrization itself 
(which is irrelevant to the precision of the neural network).
Fig.~\ref{fig:T2} shows the transfer function for Case A with $m_2 / m_1 = 0.499$, $m_{\rm DM} = 10.256$\,keV, and $\Delta = 1$ (left panel) and Case B with $m_2 / m_1 = 0.701$, $Y_{\rm scat} / Y_{\rm tot} = 0.367$, $m_{\rm DM} = 4$\,keV, and $\Delta = 1$ (right panel).
The red points are data points, while the blue lines are $T^{2} (k)$ given by Eq.~\eqref{eq:T2Fit} with the fitted values of $\{ \alpha, \beta, \gamma \}$ ($\gamma = - \beta$ as explained in Sec.~\ref{subsec:NN1}).
We see that the $\{ \alpha, \beta, \gamma \}$ parametrization nicely reproduces the original data.

\begin{figure}
\begin{center}
\includegraphics[width=0.44\columnwidth]{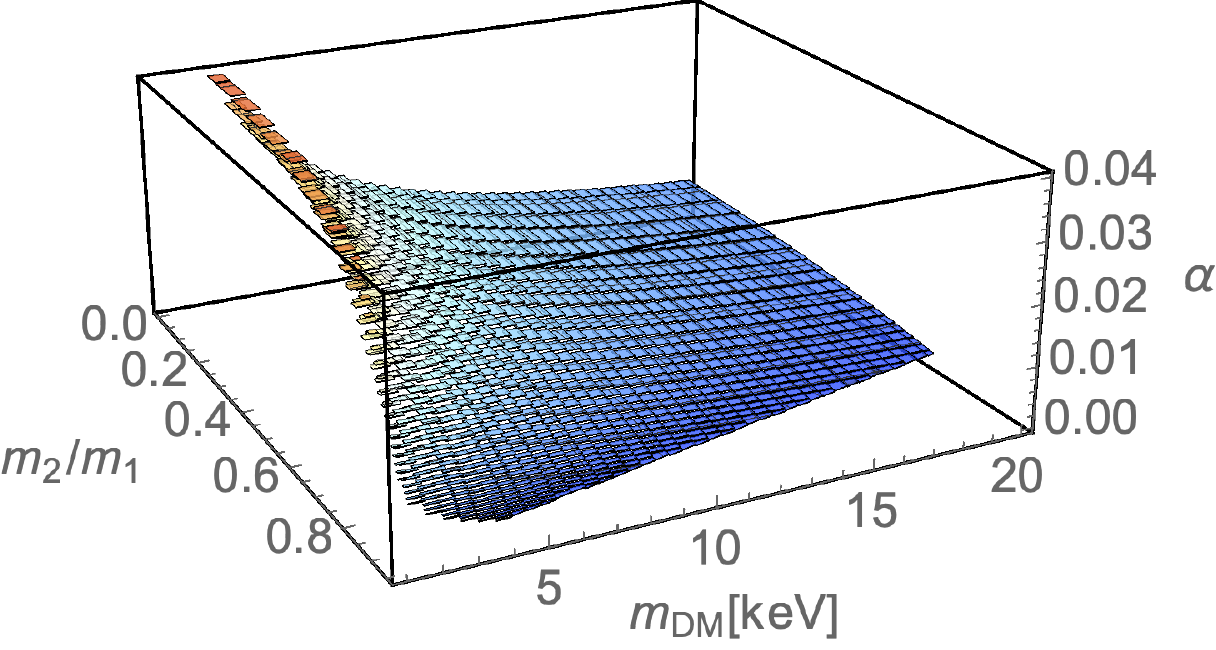}
\hskip 0.2cm
\vline
\hskip 0.4cm
\includegraphics[width=0.41\columnwidth]{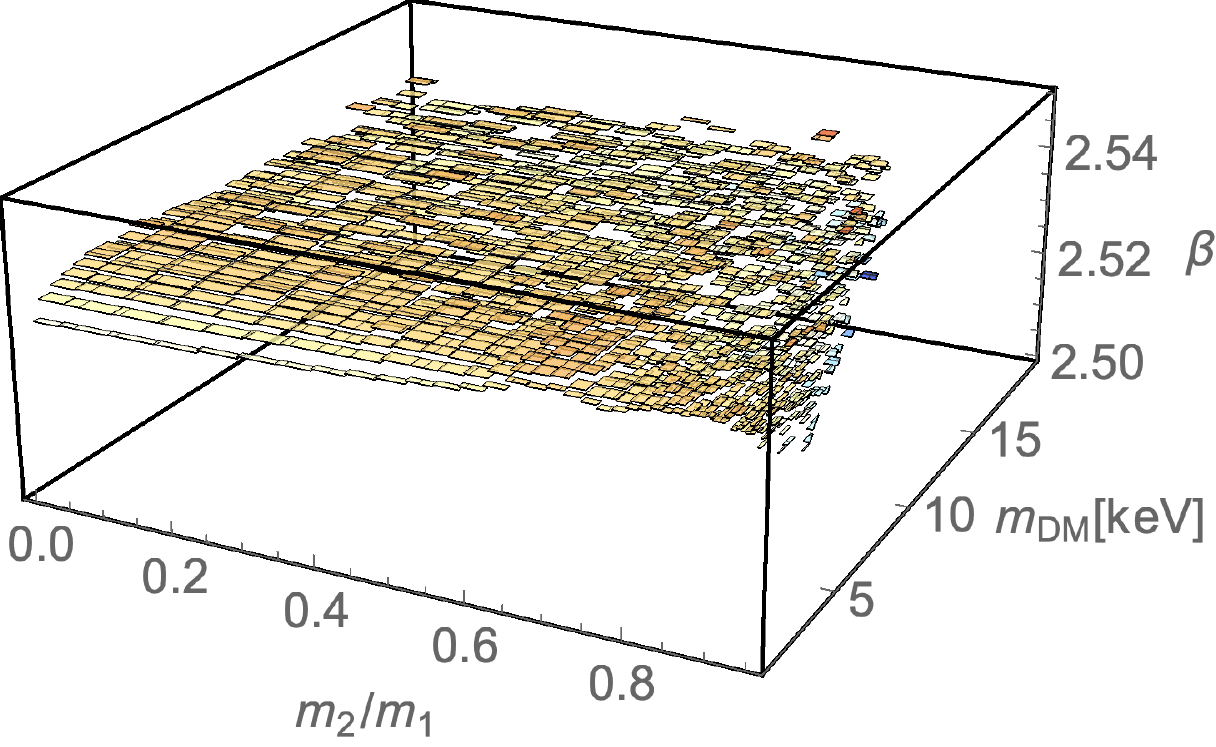}
\vskip 0.4cm
\includegraphics[width=0.44\columnwidth]{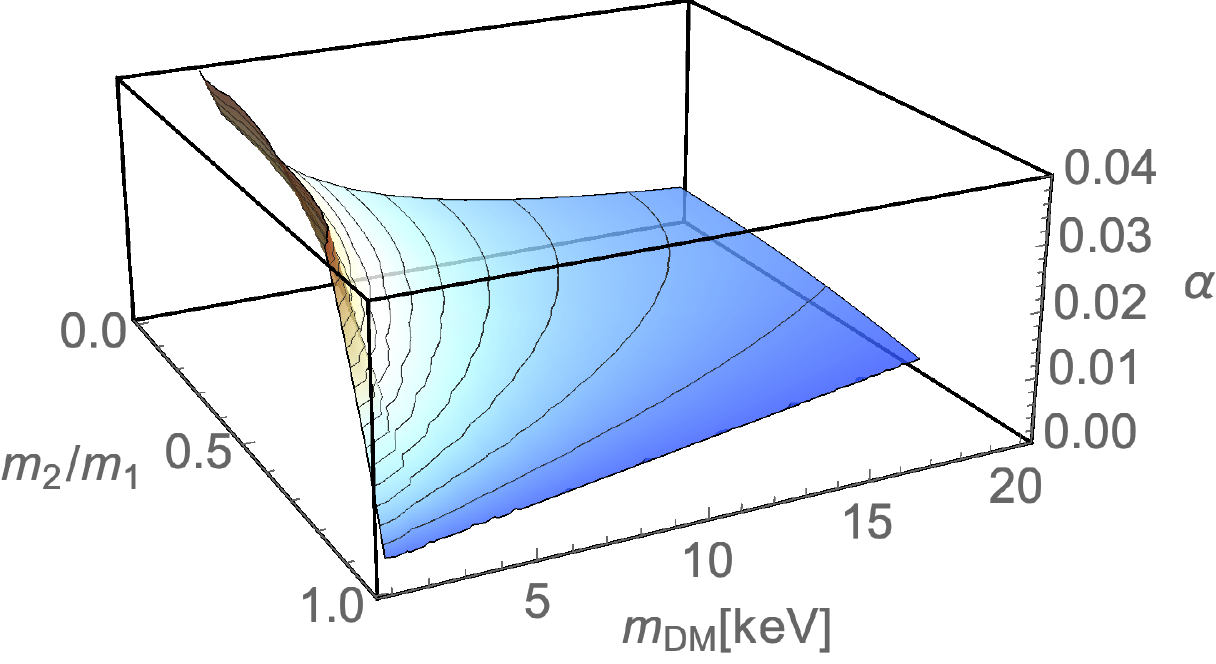}
\hskip 0.2cm
\vline
\hskip 0.4cm
\includegraphics[width=0.41\columnwidth]{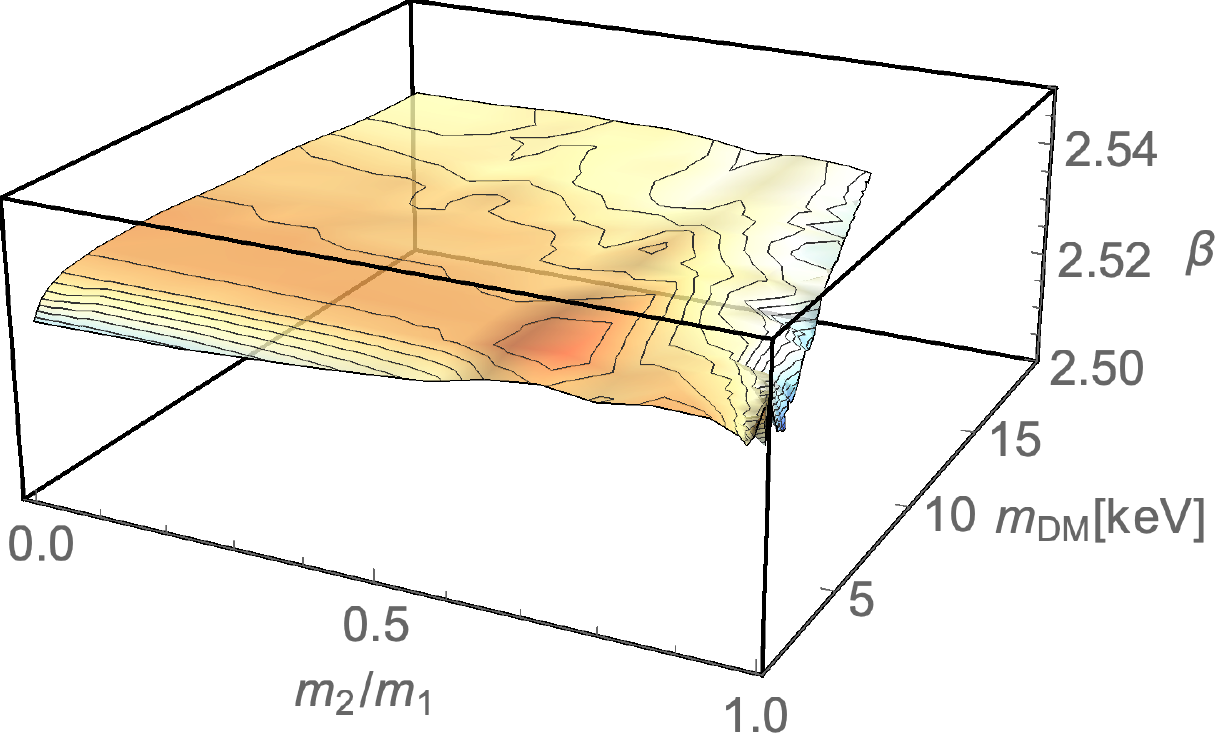}
\caption {\small
Original data (upper) and functional forms learned by the neural network (lower) for $\Delta = 1$.
The left and right columns correspond to $\alpha$ and $\beta$, respectively.
}
\label{fig:Decay_DataNN}
\end{center}
\vskip 1cm
\begin{center}
\includegraphics[width=0.44\columnwidth]{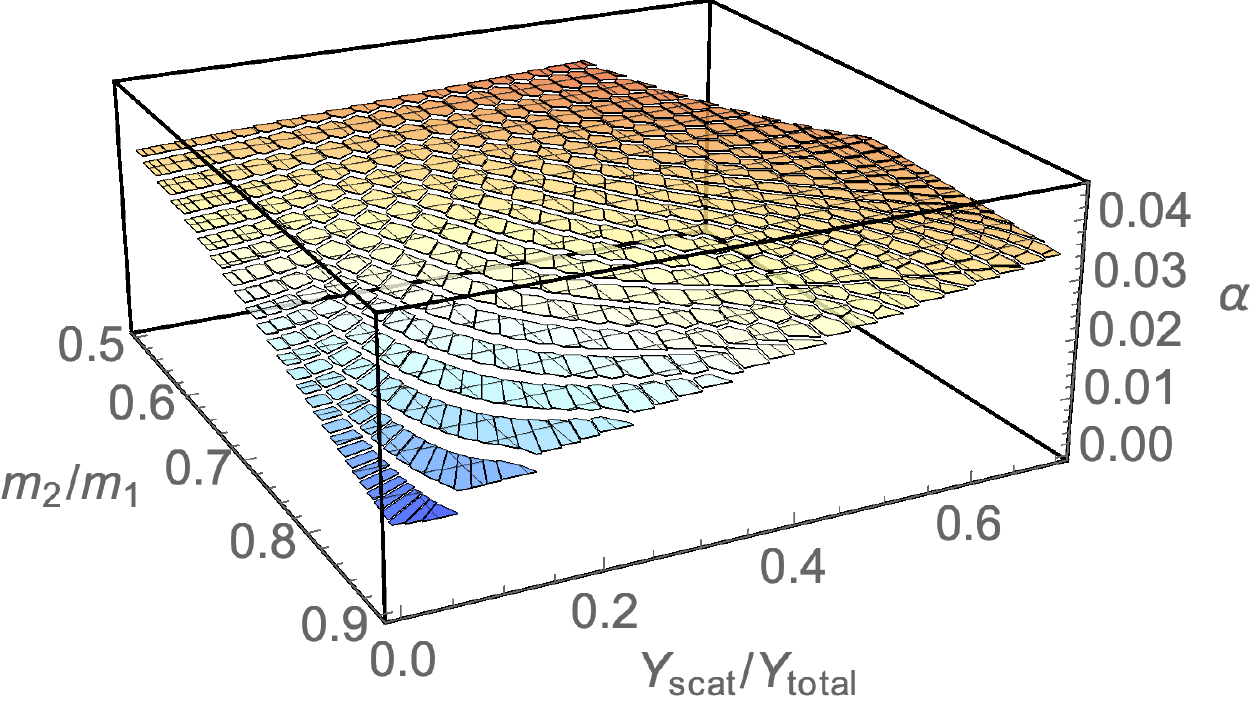}
\hskip 0.2cm
\vline
\hskip 0.4cm
\includegraphics[width=0.41\columnwidth]{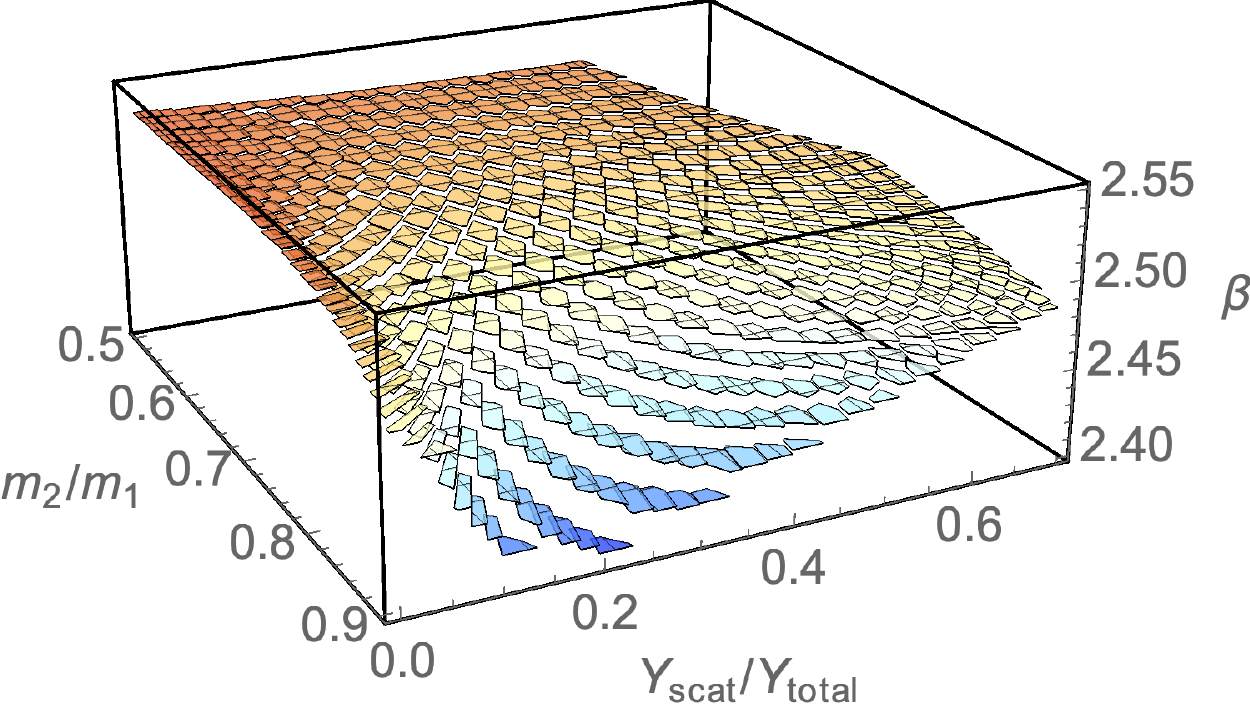}
\vskip 0.4cm
\includegraphics[width=0.44\columnwidth]{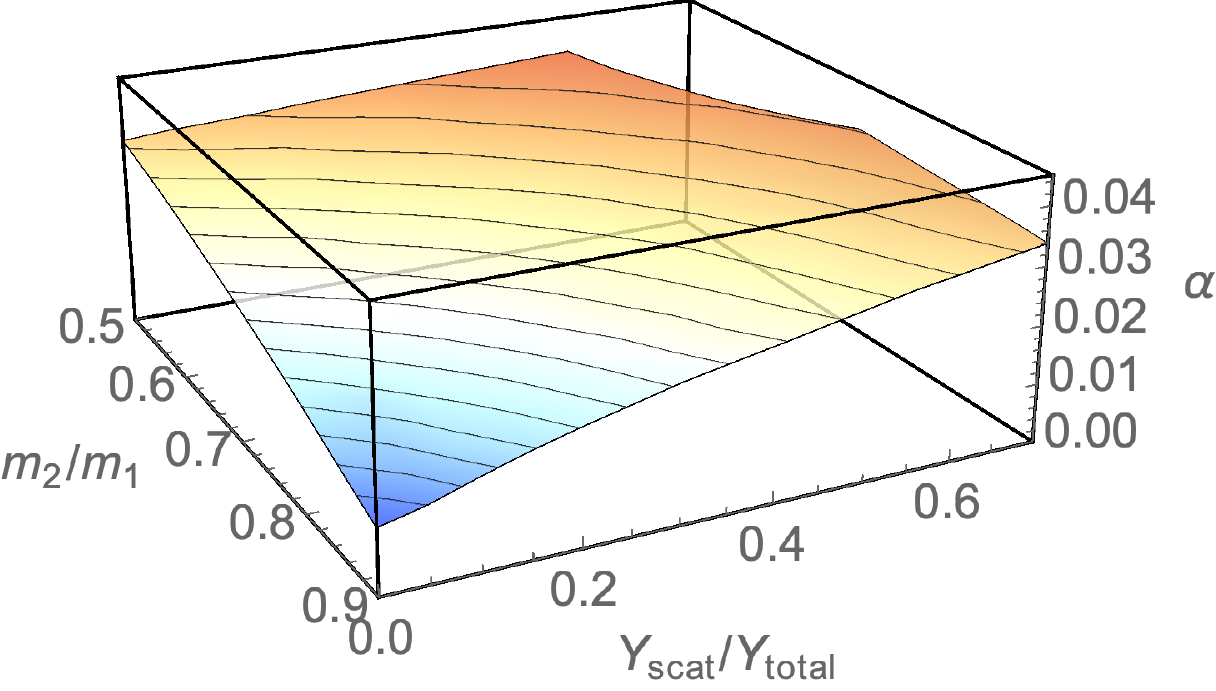}
\hskip 0.2cm
\vline
\hskip 0.4cm
\includegraphics[width=0.41\columnwidth]{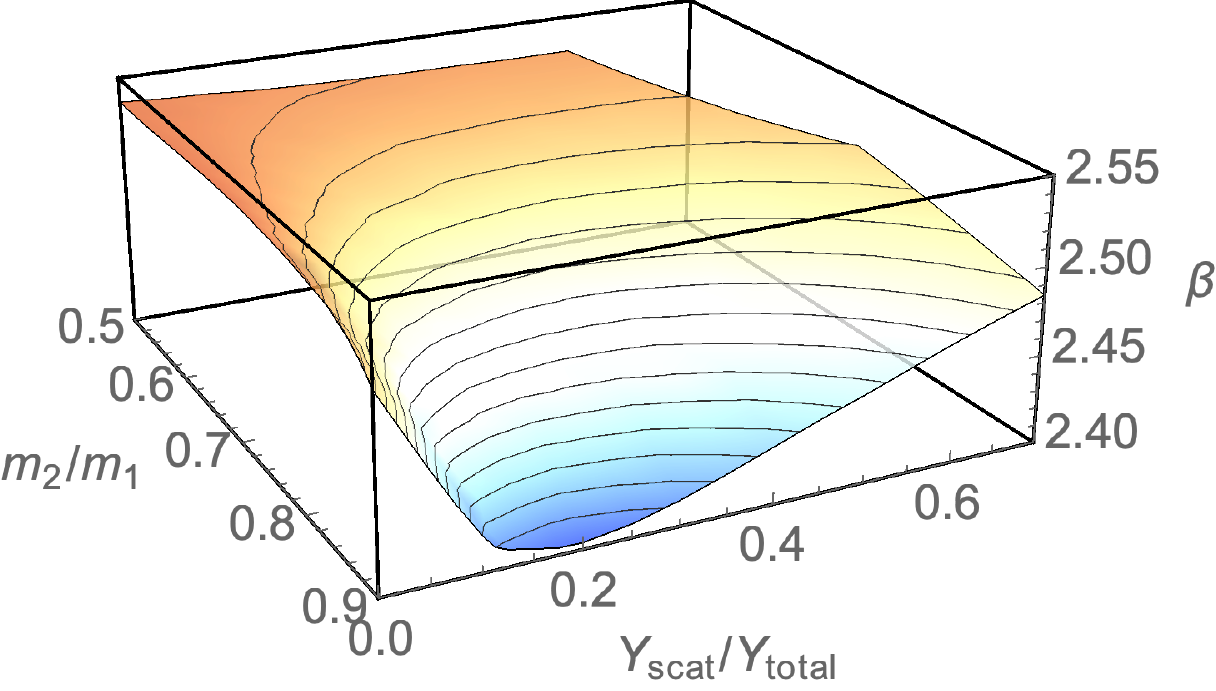}
\caption {\small
Original data (upper) and functional forms learned by the neural network (lower) for $m_{\rm DM} = 4$\,keV.
The left and right columns correspond to $\alpha$ and $\beta$, respectively.
}
\label{fig:Scattering_DataNN}
\end{center}
\end{figure}

Next we examine the precision of the neural network.
Fig.~\ref{fig:Decay_DataNN} compares the original values of $\alpha$ and $\beta$ (upper panels)
and the fit from the neural network (lower panels) for Case A with $\Delta = 1$.
Similarly, Fig.~\ref{fig:Scattering_DataNN} is for Case B with $m_{\rm DM} = 4$\,keV.
We see that the neural network not only reproduces the original data quite well, but also somewhat smoothens artificial fluctuations in the original data.

\begin{figure}
\begin{center}
\includegraphics[width=0.45\columnwidth]{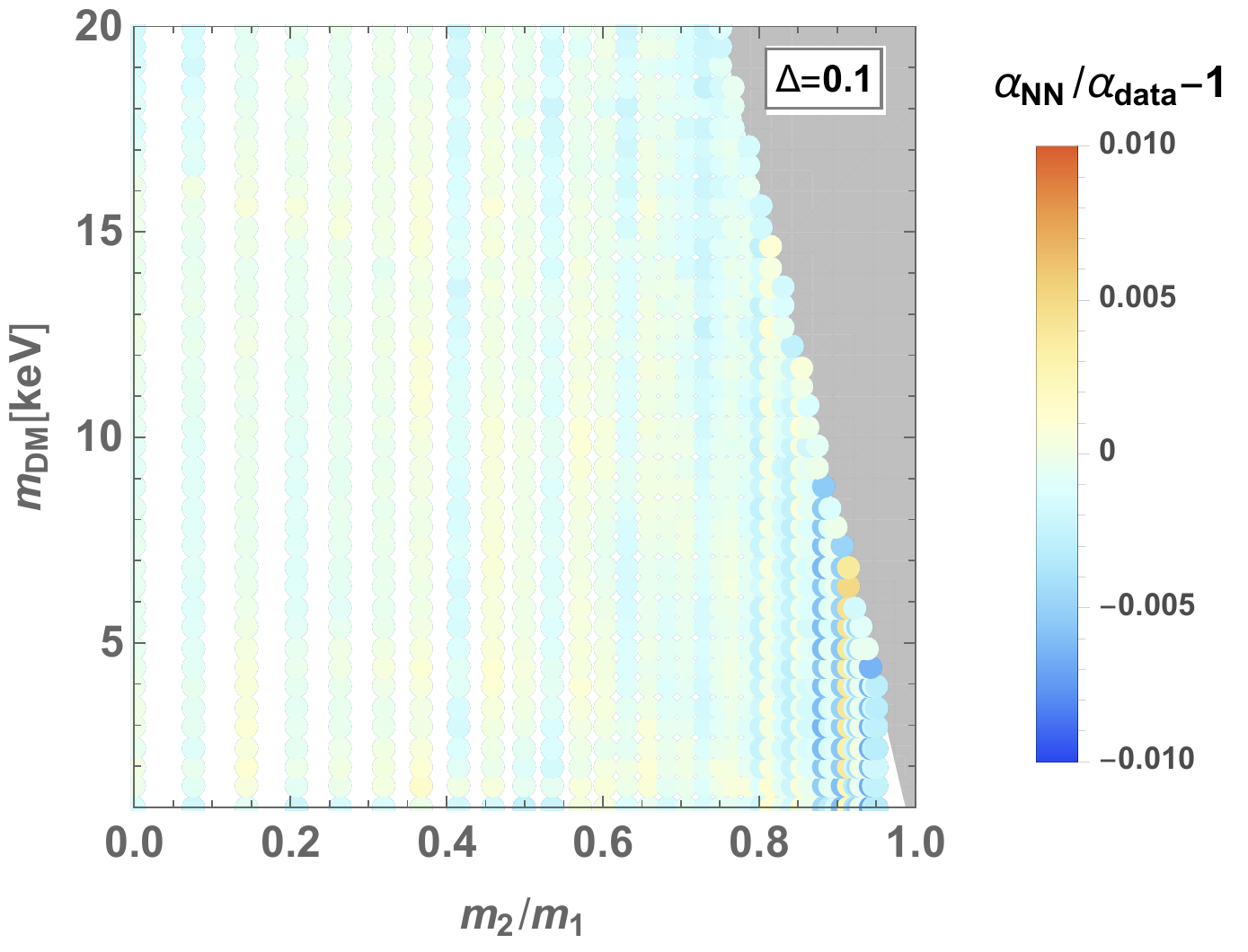}
\hskip 0.5cm
\includegraphics[width=0.45\columnwidth]{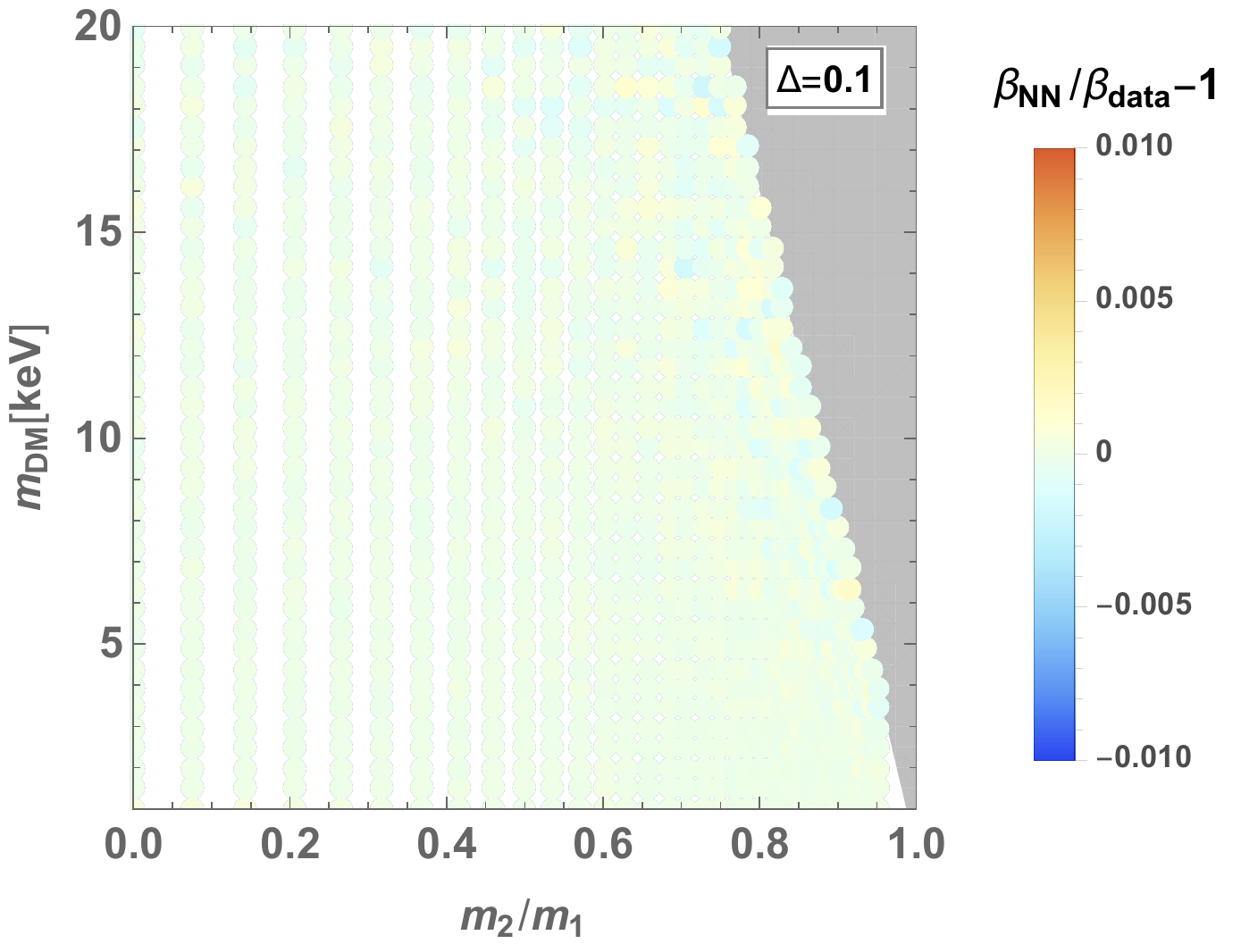}
\vskip 0.5cm
\includegraphics[width=0.45\columnwidth]{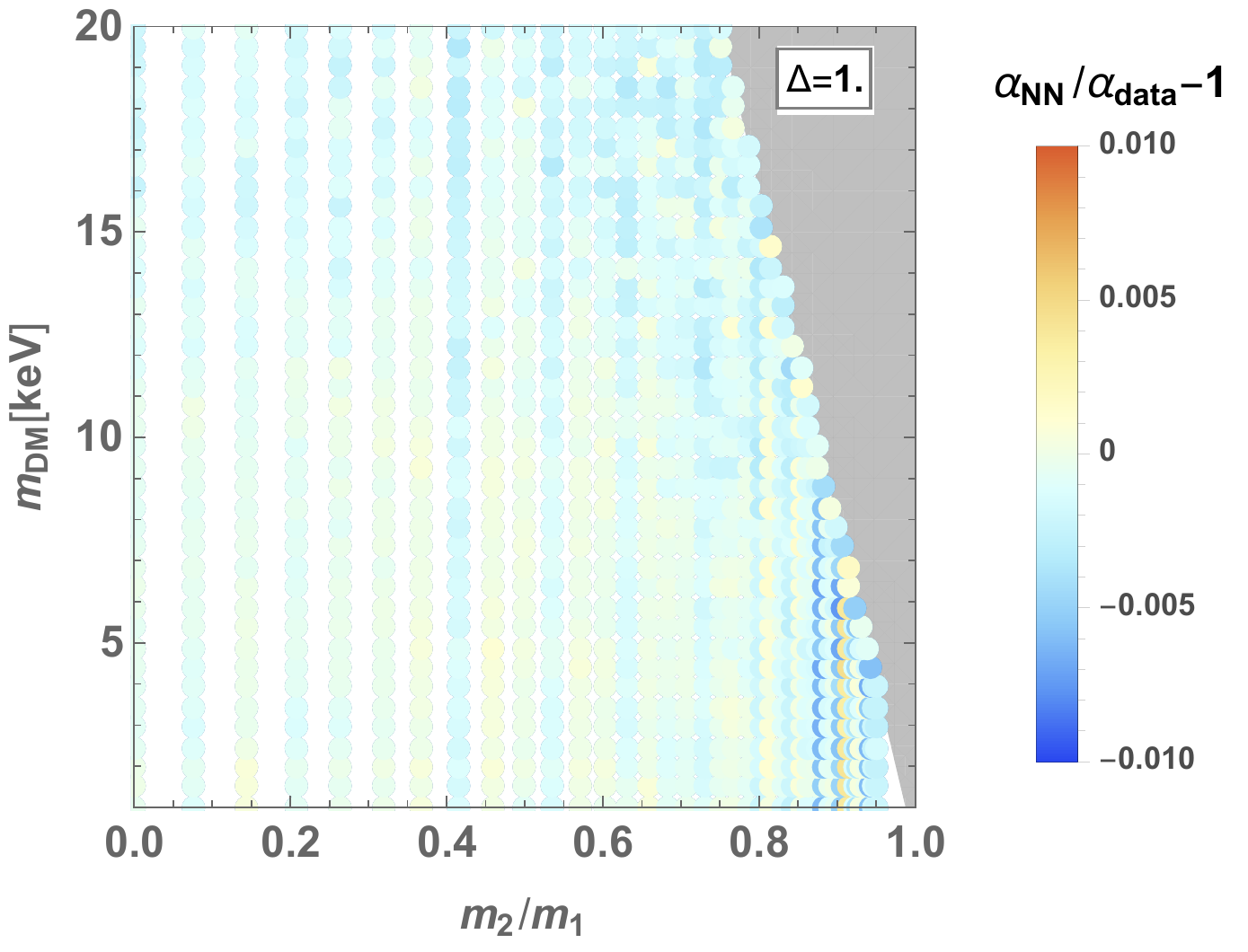}
\hskip 0.5cm
\includegraphics[width=0.45\columnwidth]{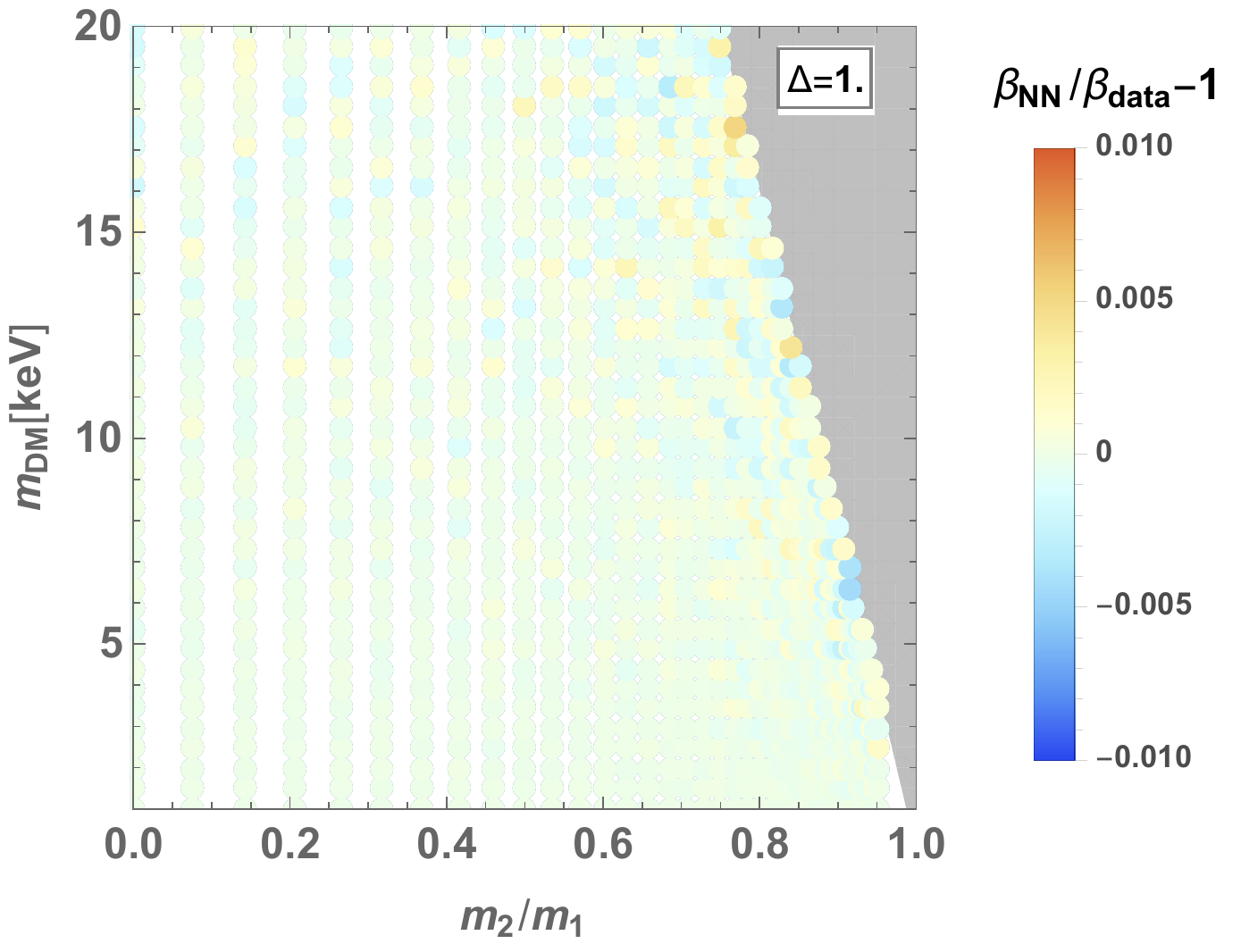}
\vskip 0.5cm
\includegraphics[width=0.45\columnwidth]{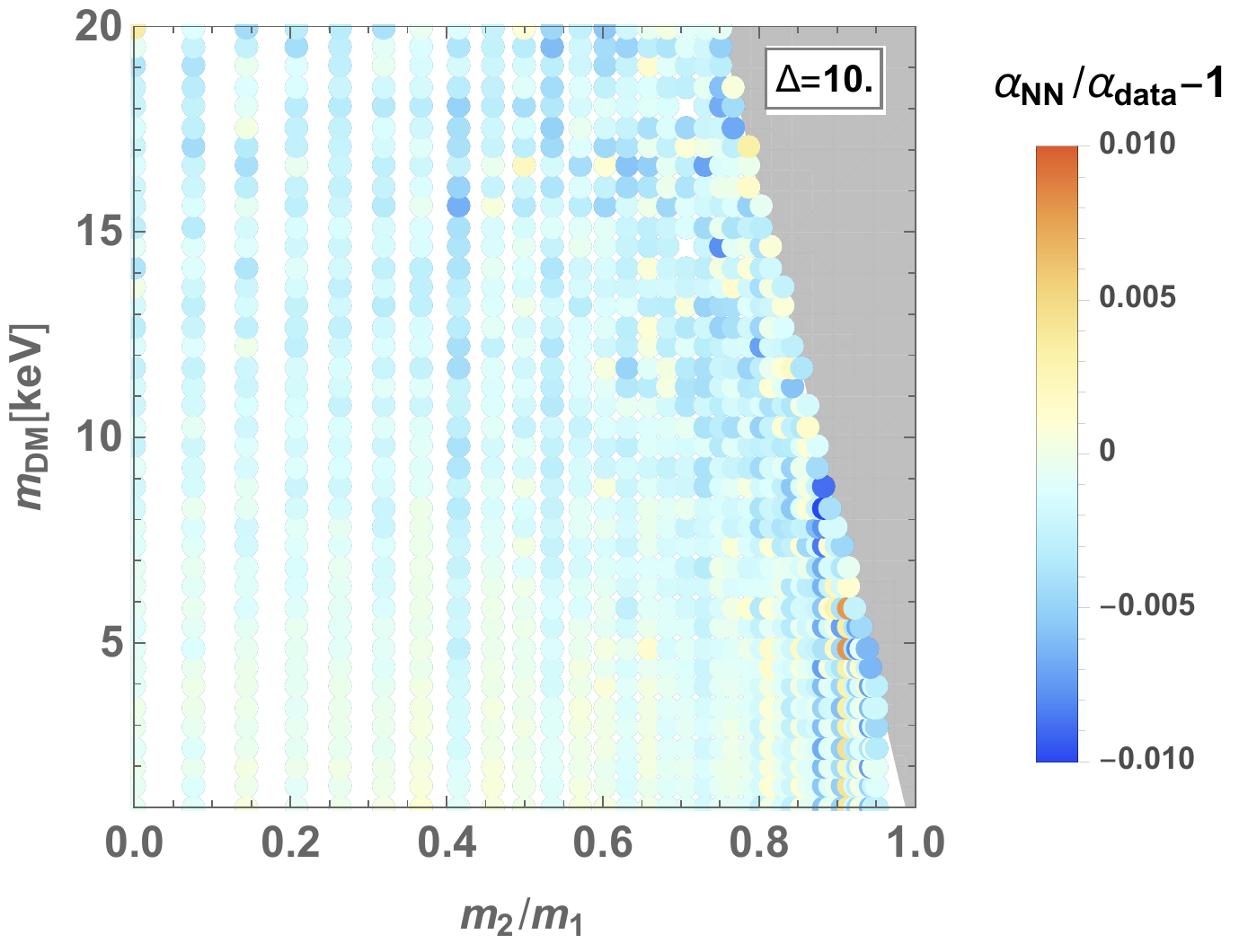}
\hskip 0.5cm
\includegraphics[width=0.45\columnwidth]{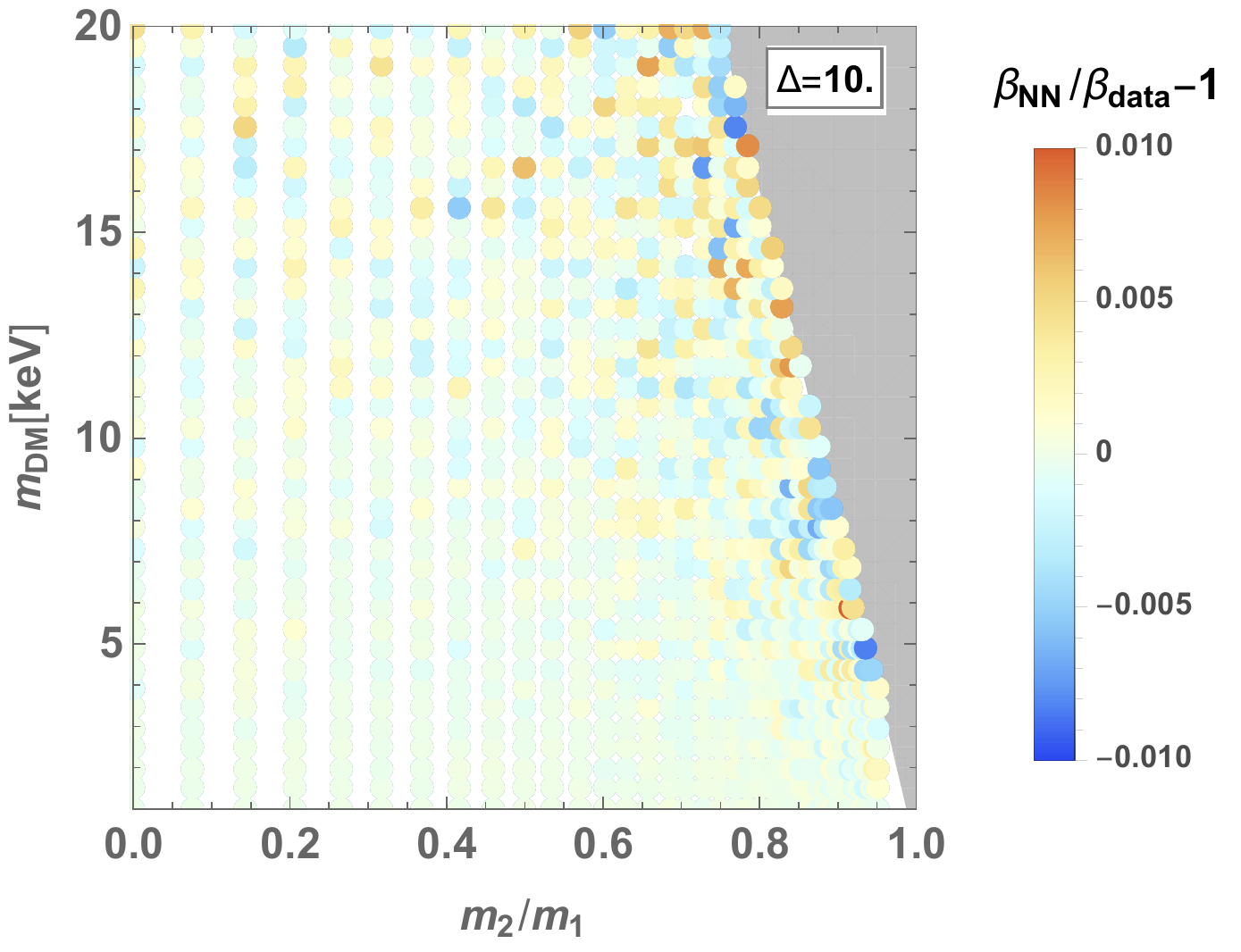}
\caption {\small
Relative error between $\alpha$ (left) or $\beta$ (right) obtained from the original data 
and those learned by the neural network.
This figure is for Case A (Decay with entropy production).
We show $\Delta = 0.1$ (top), $1$ (middle), and $10$ (bottom).
}
\label{fig:Decay_Error}
\end{center}
\end{figure}

\begin{figure}
\begin{center}
\includegraphics[width=0.45\columnwidth]{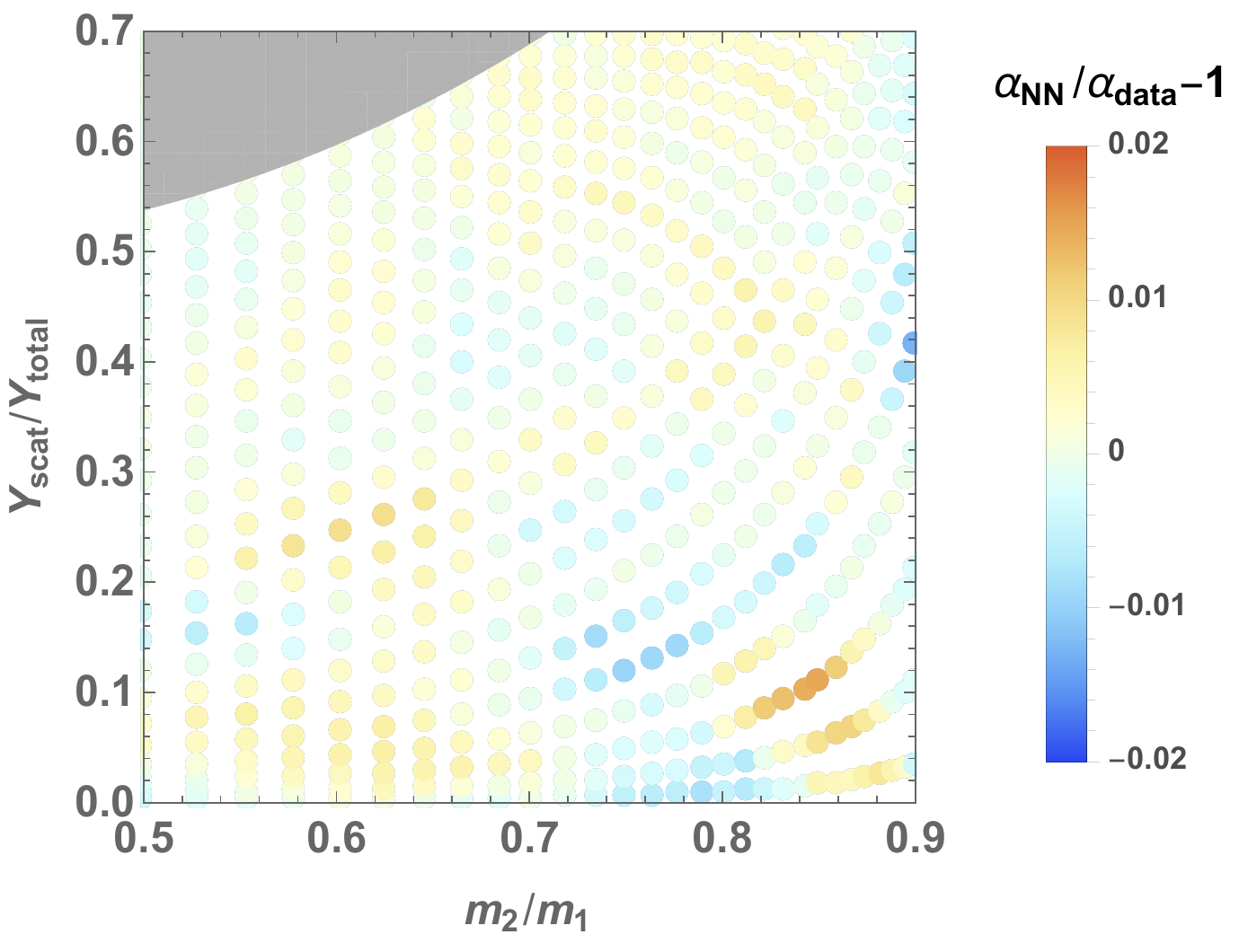}
\hskip 0.5cm
\includegraphics[width=0.45\columnwidth]{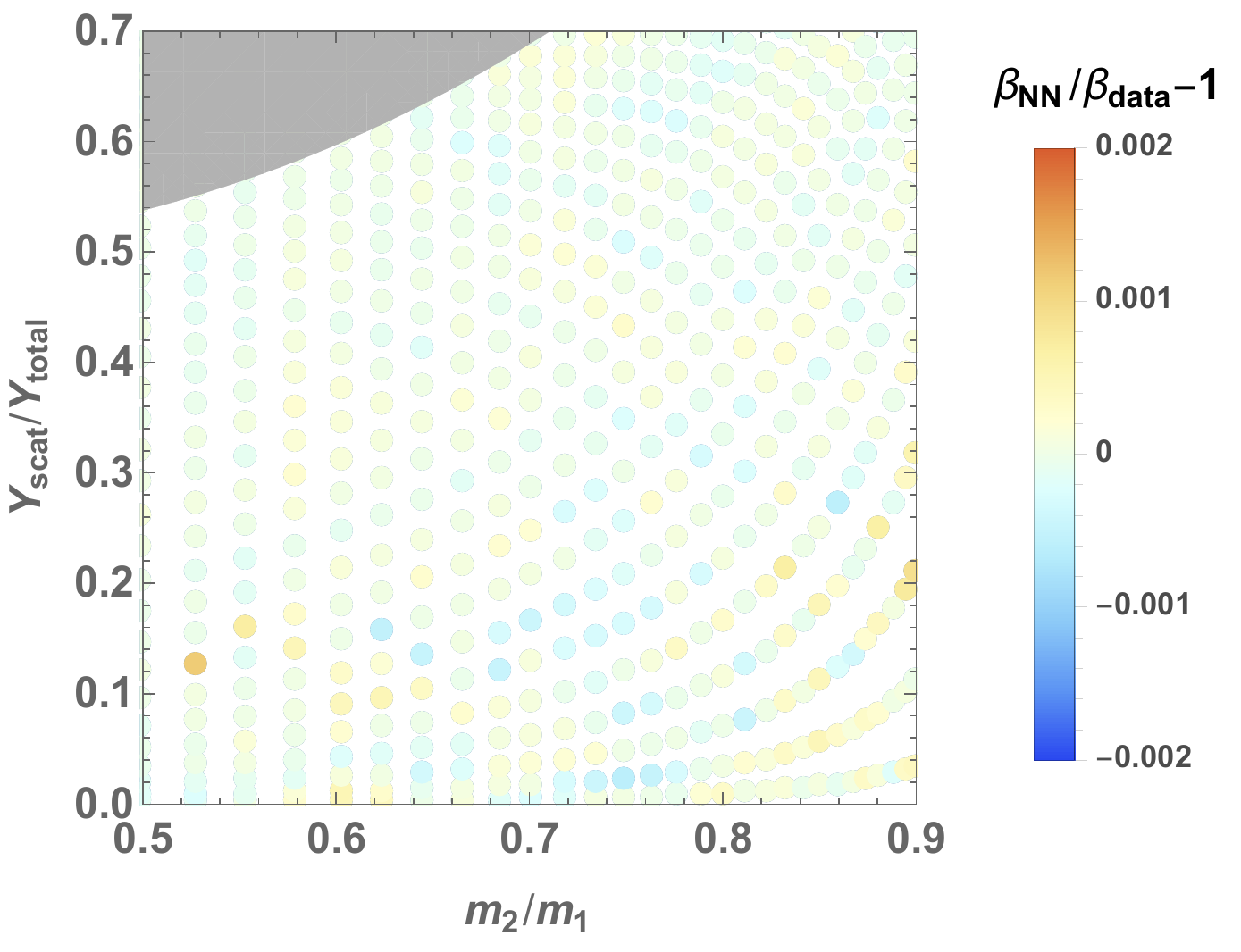}
\vskip 0.5cm
\includegraphics[width=0.45\columnwidth]{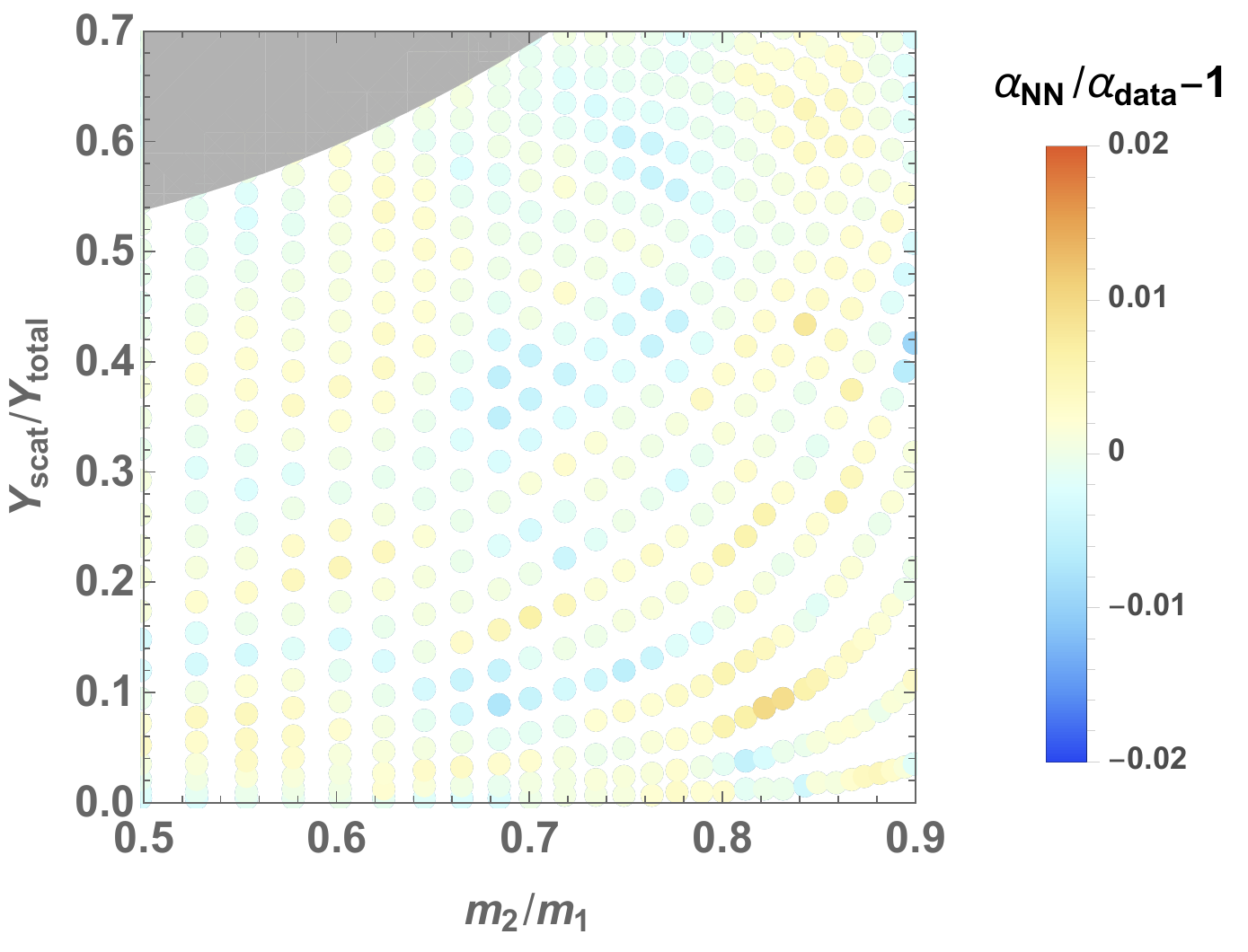}
\hskip 0.5cm
\includegraphics[width=0.45\columnwidth]{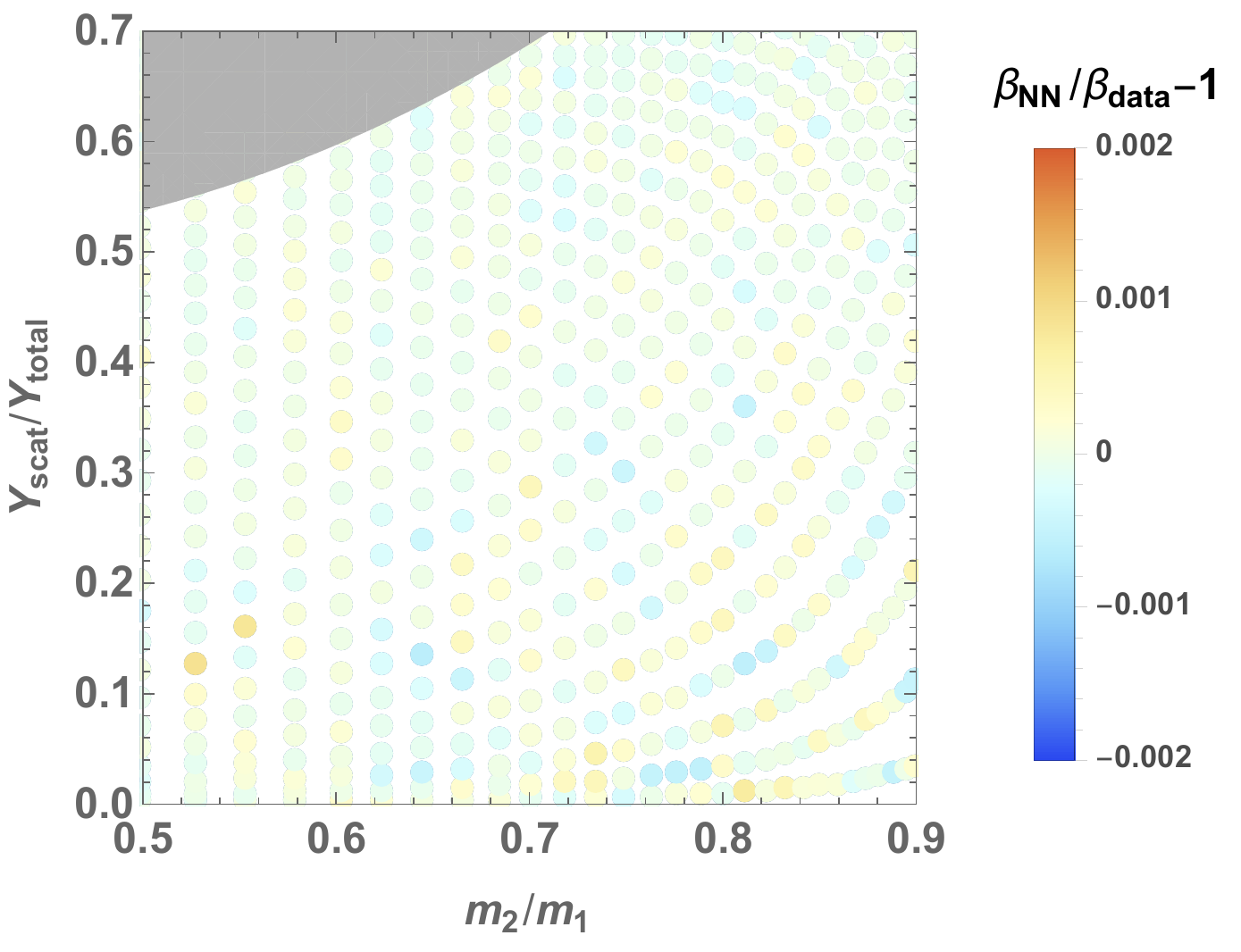}
\vskip 0.5cm
\includegraphics[width=0.45\columnwidth]{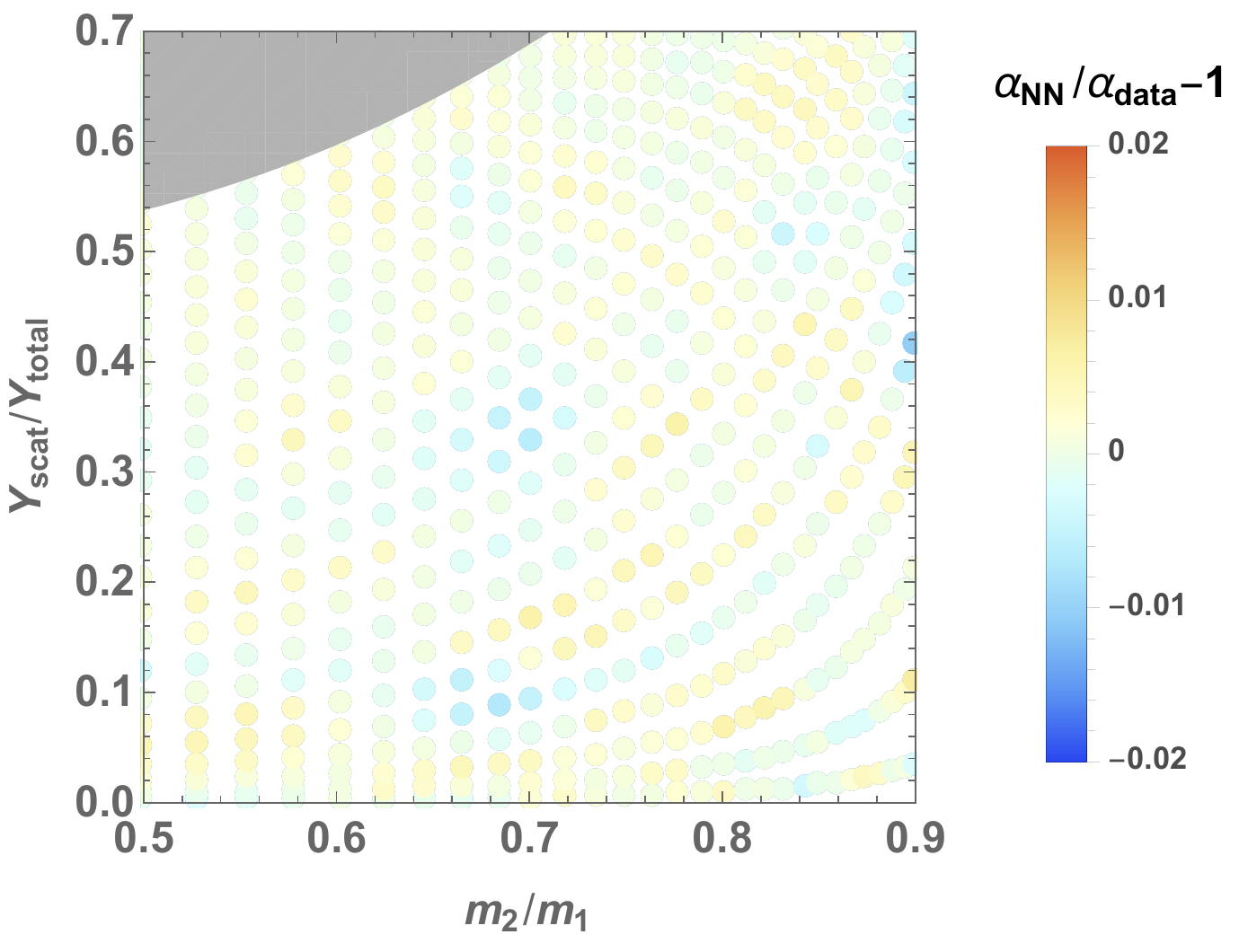}
\hskip 0.5cm
\includegraphics[width=0.45\columnwidth]{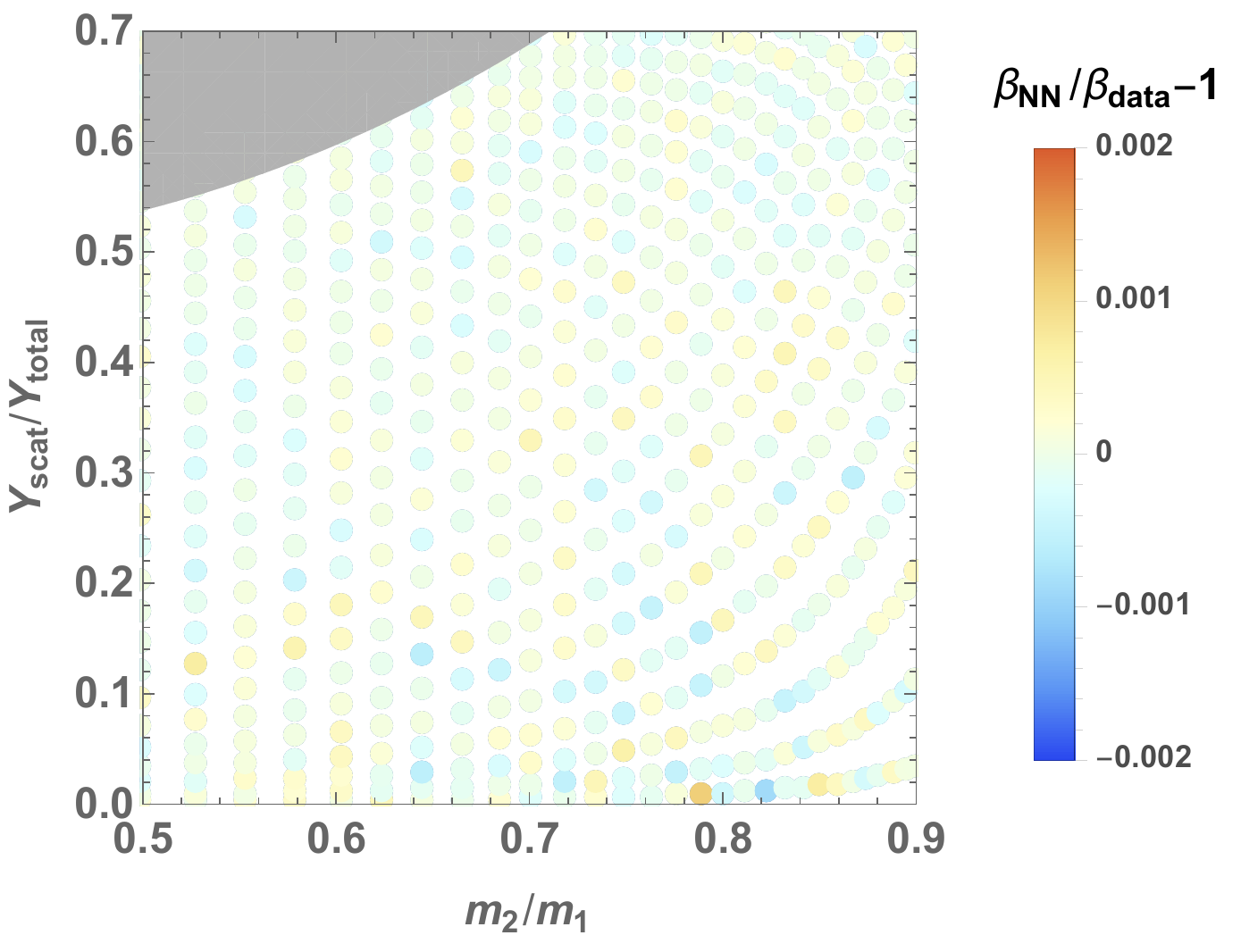}
\caption {\small
Relative error between $\alpha$ (left) or $\beta$ (right) obtained from the original data 
and those learned by the neural network.
This figure is for Case B (Decay with scattering).
We show $\Delta = 1$ with $m_{\rm DM} = 2$ (top), $4$ (middle), and $6$\,keV (bottom).
}
\label{fig:Scattering_Error}
\end{center}
\end{figure}

Figs.~\ref{fig:Decay_Error} and \ref{fig:Scattering_Error} are color plots for the relative error between the original data and neural network for $\alpha$ (left columns) and $\beta$ (right columns), respectively.
Fig.~\ref{fig:Decay_Error} is for Case A with $\Delta = 0.1$, $1$, and $10$ from top to bottom, 
while Fig.~\ref{fig:Scattering_Error} is for Case B with $m_{\rm DM} = 2$, $4$, and $6$\,keV 
for $\Delta = 1$ from top to bottom.
The relative error for Case A is at most 1$\%$ in $\alpha$ and $\beta$, while for Case B the error is at most $2\%$ and $0.2\%$ in $\alpha$ and $\beta$, respectively.

We finally comment that the error of the neural network for ``$\{ \alpha, \beta, \gamma \} \to$ Observables''
is much smaller than that for ``Model parameters $\to \{ \alpha, \beta, \gamma \}$''.

%%%%%%%%%%%%%%%%
%%%%%%%%%%%%%%%%

%%%%%%%%%%%%%%%%
%%%%%%%%%%%%%%%%

%%%%%%%%%%%%%%%%
%%%%%%%%%%%%%%%%

%%%%%%%%%%%%%%%%
%%%%%%%%%%%%%%%%

%%%%%%%%%%%%%%%%%%%%%%%%%%%%%%%%%%%%%%%%%%%%%%%%%%
\section{How to use the neural network data}
\label{app:howto}
%%%%%%%%%%%%%%%%%%%%%%%%%%%%%%%%%%%%%%%%%%%%%%%%%%

In this appendix we explain how to use the data provided through the {\tt arXiv} website.
The datafile we provide are
\begin{itemize}
\item
{\tt mean.tsv}, {\tt std.tsv},
\item
{\tt b1.tsv}, {\tt b2.tsv}, {\tt bout.tsv},
\item
{\tt w1.tsv}, {\tt w2.tsv}, {\tt wout.tsv}.
\end{itemize}
The first items are the means, $\vec{x}_0 \equiv (\vec{x}_{{\rm in},0}^{\rm T}, \vec{x}_{{\rm out},0}^{\rm T})^{\rm T}$, and standard deviations, $\vec{\sigma} \equiv (\vec{\sigma}_{\rm in}^{\rm T}, \vec{\sigma}_{\rm out}^{\rm T})^{\rm T}$, which shift and normalize the neural network input and output.
The second items are the biases, $\vec{b}_1$, $\vec{b}_2$, and $\vec{b}_{\rm out}$, while the last items are the weight matrices, $W_1$, $W_2$, and $W_{\rm out}$.

The data files for ``Model parameters $\to \{ \alpha, \beta, \gamma \}$" are in the directory of {\tt freeze-in/CaseA} for Case A, and in {\tt freeze-in/CaseB/Delta=...} for Case B, respectively.
The data files for ``$\{ \alpha, \beta, \gamma \} \to$ Observables" are in {\tt freeze-in/NSat} and {\tt freeze-in/deltaA} for $N_{\rm sat}$ and $\delta A$, respectively.

%%%%%%%%%%%%%%%%%%%%%%%%%%%%%%%%%%%%%%%%%%%%%%%%%%
\small
\bibliography{ref}
%%%%%%%%%%%%%%%%%%%%%%%%%%%%%%%%%%%%%%%%%%%%%%%%%%

%%%%%%%%%%%%%%%%%%%%%%%%%%%%%%%%%%%%%%%%%%%%%%%%%%
\end{document}